\documentstyle[12pt]{article}

\input mssymb

 1
\font\supbscriptsizebbfont=msym7 scaled \magstep 1
 1
 3
\font\bbfont=msym10 scaled \magstep1  

\def\supbscriptsizeBbb#1{\hbox{\supbscriptsizebbfont #1}}

\def\Bbb#1{\hbox{\bbfont #1}}

\input psfig
\psfiginit

\newcommand{\longrightrightarrow}{\mbox{\raisebox{.4ex}{$\longrightarrow$}}
  \hspace{-19pt}\mbox{\raisebox{-.4ex}{$\longrightarrow$}} }
\newcommand{\pr}{\mbox{\rm pr}}

\begin{document}

\begin{titlepage}

$ $

\vspace{-2.5cm}

\noindent\hspace{-.5cm}
\parbox{4cm}{February 1996}\
  \hspace{9cm}\
  \parbox{5cm}{{\sc umtg} -- {\small 190}   \newline
  {\tt hep-th/9603198}  }

\vspace{2cm}

\centerline{\large\bf Lorentz Surfaces and Lorentzian CFT}
\vspace{0.2cm}
\centerline{\small\bf ------  with an appendix on
                         quantization of string phase space}

\vspace{1.5cm}

\centerline{\large
 Chien-Hao Liu\footnote{e-mail: chienliu@phyvax.ir.miami.edu}}

\vspace{1em}

\centerline{\it Department of Physics}
\centerline{\it University of Miami}
\centerline{\it P.O. Box 248046}
\centerline{\it Coral Gables, FL.\ 33124}

\vspace{1cm}

\centerline{({\sl Dedicated to {\it Ling-Miao Chou},
                                whom I can never thank enough.})}

\vspace{1cm}

\begin{quotation}
\centerline{\bf Abstract}
\vspace{0.3cm}

\baselineskip 13pt

{\small The interest in string Hamiltonian system has recently been
 rekindled due to its application to target-space duality. In this
 article, we explore another direction it motivates. In Sec.\ 1,
 conformal symmetry and some algebraic structures of the system that
 are related to interacting strings are discussed. These lead one
 naturally to the study of Lorentz surfaces in Sec.\ 2. In contrast
 to the case of Riemann surfaces, we show in Sec.\ 3 that there are
 Lorentz surfaces that cannot be conformally deformed into Mandelstam
 diagrams. Lastly in Sec.\ 4, we discuss speculatively the prospect
 of Lorentzian conformal field theory.

 Additionally, to have a view of what quantum picture a string
 Hamiltonian system may lead to, we discuss independently in the
 Appendix a formal geometric quantization of the string phase space.
} 
\end{quotation}

\vspace{1em}

\noindent
MSC number 1991: 05C90, 53C50, 53Z05, 57M50, 81S10, 81T40.

\bigskip

\baselineskip 12pt

{\footnotesize
\noindent{\bf Acknowledgements.}
This work follows from numerous discussions with Orlando Alvarez,
who helped me clarify all sorts of premature ideas and generously
proofread and commented the draft. It is also under the shadow of
Bill Thurston, whose insight on geometry greatly influences me. To
both of them I am deeply indebted. I would also like to express my
gratitude to J$\phi$rgen E.\ Anderson, Martin Halpern,
Duong Phong, and Alan Weinstein for their courses, from which I had
my first contact with CFT and quantization. Besides, I want to thank
Hung-Wen Chang, Thom Curtright, Marco Monti, Rafael Nepomechie, and
Radu T\v{a}tar for discussions and assistance.}

\end{titlepage}

\baselineskip 15pt

\newpage
\pagenumbering{arabic}

\begin{flushleft}
{\Large\bf 0. Introduction and Outline.}
\end{flushleft}
\vspace{-3cm}
\centerline{\sc Lorentz Surfaces and CFT --- with appendix}
\vspace{2cm}
\begin{flushleft}
{\bf Introduction.}
\end{flushleft}
The interest in string Hamiltonian system has recently been rekindled
due to its application to target-space duality
(e.g.\ [C-Z], [A-AG-B-L]).
In this article, we explore another direction it provides.

Interacting strings can be realized as collections of partially
ordered integral filaments in the string Hamiltonian system
$LT^{\ast}M$. They can be regarded as maps from Lorentzian world-sheets
$\Sigma$ into the target-space $M$. Together with the conformal symmetry
in the system, one is motivated to the study of Lorentz surfaces.
Depending on the role singularities on Lorentz surfaces play, there are
coarse and fine Lorentz surfaces. We discuss only coarse ones due to
technical reasons. Like pants decompositions for Riemann surfaces, one
has rompers decompositions for Lorentz surfaces. Such decompositions
provide a way to study their moduli spaces. In contrast to Riemannian
case, we show that there are Lorentz surfaces that cannot be rectified
into Mandelstam diagrams. As theory of Riemann surfaces to conformal
field theory (CFT), theory of Lorentz surfaces should lead to a
Lorentzian counterpart of CFT. We explore this prospect
speculatively at the end.

Additionally, to have a view of what quantum picture the string
Hamiltonian system may lead to, we discuss independently in the
Appendix a formal geometric quantization of the string phase space.

Readers are referred to [AG-G-M-V], [At1-2], [F-S], [G-S-W], [Ka1-2],
[Mo-S1-2], [L-T], [Se1-3], [Thor], and [Zw] for strings, CFT, and string
fields; [B-E], [H-E] and [Pe] for Lorentzian manifolds; [C-B] and [St]
for surface theory; [Bo], [K\"{o}] for graph theory; [Br], [G-S], [Mi],
[\'{S}n] and [Wo] for geometric quantization.

\bigskip

\begin{flushleft}
{\bf Outline.}
\end{flushleft}
{\small

\baselineskip 13pt

\begin{quote}
 {\bf 1. Beginning with string Hamiltonian systems.}

 {\bf 2. Lorentz surfaces and their moduli.}
    \begin{quote}
      2.1 Lorentz surfaces and Mandelstam diagrams.

      2.2 Basic structures and coarse conformal groups.

      2.3 Rompers decompositions and coarse moduli spaces.
    \end{quote}

 {\bf 3. Rectifiability into Mandelstam diagrams.}
    \begin{quote}
      3.1 \parbox[t]{11cm}{Branched coverings, positive cones, and
               rectifiability.}

      3.2 \parbox[t]{11cm}{Electrical circuits and examples of
               unrectifiability.}
    \end{quote}

 {\bf 4. Toward Lorentzian conformal field theories.}

 {\bf Appendix. Quantization of string phase space.}
\end{quote}
} 

\baselineskip 15pt

\newpage

\section{Beginning with string Hamiltonian systems.}
For introduction of notations, let us recall first the following basic
objects:

\bigskip

\centerline{
 \begin{minipage}[b]{14.5cm}{
 \begin{tabular}{cccc}
   \multicolumn{1}{c}{physical object}\rule[-2ex]{0ex}{1ex}
           & & & \multicolumn{1}{c}{mathematical presentation}\\ \hline
   \rule{0ex}{4ex}\parbox[t]{3.5cm}{target-space }        & & &
       \parbox[t]{9cm}{$M=(M,ds^2,B)$ with the metric $ds^2$, also
         denoted by $\langle\;,\;\rangle$, either Riemannian or
         Lorentzian, and $B$ a 2-form, (a $B$-field in physicists'
         terminology)}  \\
   \rule{0ex}{4ex}\parbox[t]{3.5cm}{particle}            & & &
       \parbox[t]{9cm}{a smooth map
         $\phi:S^1\rightarrow M$} \\
   \rule{0ex}{4ex}\parbox[t]{3.5cm}{configuration space }     & & &
       \parbox[t]{9cm}{$LM$ = the loop space of $M$, which contains
         all $\phi$ }\\
   \rule{0ex}{4ex}\parbox[t]{3.5cm}{(momentum) phase space}     & & &
        \parbox[t]{9cm}{$LT^{\ast}M$; it has a canonical symplectic
         structure {\boldmath $\omega$} given by
       $$
        \mbox{\boldmath $\omega$}_{\gamma}(\eta,\xi)\;=\;
            \int_{S^1}\:d\sigma\,
         \omega\left(\eta_{\gamma(\sigma)},\xi_{\gamma(\sigma)}\right),
       $$
         where $\omega$: the canonical symplectic structure on
        $T^{\ast}M$, $\gamma\in LT^{\ast}M$, and $\eta$, $\xi$:
         tangent vectors at $\gamma$  }\\
   \rule{0ex}{4ex}\parbox[t]{3.5cm}{Lagrangian $\cal L$ on
           \newline $T_{\ast}LM$ ($=LT_{\ast}M$)}  & & &
        \parbox[t]{9cm}{
       ${\cal L}(\phi,X)\:
       =\:\int_{S^1}\,d\sigma\,{\cal L}(\phi,X;\sigma)$
                            with ${\cal L}(\phi,X;\sigma)$ being
       $$\frac{1}{2}\left(\langle X(\sigma),X(\sigma)\rangle\,-\,
         \langle\phi_{\ast}\partial_{\sigma},
              \phi_{\ast}\partial_{\sigma}\rangle\right)\:+\:
                B\left(X(\sigma),\phi_{\ast}\partial_{\sigma}\right)
       $$                          }\\
   \rule{0ex}{4ex}\parbox[t]{3.5cm}{Legendre transformation }     & & &
       \parbox[t]{9cm}{the map
       $$
       \begin{array}{ccc}
           LT_{\ast}M    & \longrightarrow   & LT^{\ast}M   \\
           (\phi,X)     & \longmapsto  & (\phi,\pi)
       \end{array}
       $$
       with
       $$
        \pi(\sigma)\;=\;\frac{\delta{\cal L}}{\delta X}(\sigma)\;=\;
                     \langle\;\cdot\;,X(\sigma)\rangle\:+\:
                B\left(\;\cdot\;,\phi_{\ast}\partial_{\sigma}\right)
       $$      } \\
   \rule{0ex}{4ex}\parbox[t]{3.5cm}{Hamiltonian $\cal H$ on
                          \newline $LT^{\ast}M$ } & & &
       \parbox[t]{9cm}{ the push-forward of $\cal L$ under the Legendre
        transformation; explicitly ${\cal H}(\phi,\pi;\sigma)$ equals
         $$
          \frac{1}{2}
          \langle\pi-B_{\phi}(\sigma),\pi-B_{\phi}(\sigma)\rangle^{\sim}
            \;+\;\frac{1}{2}\langle\phi_{\ast}\partial_{\sigma}\:,\:
                      \phi_{\ast}\partial_{\sigma}\rangle,
         $$
         where $B_{\phi}(\sigma)
                   =B(\;\cdot\;,\phi_{\ast}\partial_{\sigma})$ and
         $\langle\;,\;\rangle^{\sim}$ is the induced metric on
              fibers of $T^{\ast}M$ from $\langle\;,\;\rangle$}
 \end{tabular}   }
 \end{minipage}
} 

\vspace{1cm}

\noindent
In principle, the Hamiltonian system
$(LT^{\ast}M, \mbox{\boldmath $\omega$}, {\cal H})$ contains all the
classical information of a free closed bosonic string. The Liouville
1-form $\mbox{\boldmath $\theta$}$ on $LT^{\ast}M$ is given similarly by
$$
 \mbox{\boldmath $\theta$}_{\gamma}(\eta)\;
   =\; \int_{S^1}d\sigma\,
              \theta_{\gamma(\sigma)}(\eta_{\gamma(\sigma)})\,,
$$
where $\eta$ is a tangent vector at $\gamma$; and it satisfies
$d\mbox{\boldmath $\theta$}=\mbox{\boldmath $\omega$}$.

\bigskip

\begin{flushleft}
{\bf Completeness and symmetry of string system.}
\end{flushleft}
For a general target-space $M$, the string Hamiltonian flow
$\rho_t$ may not be complete. The integral trajectories of the
Hamiltonian vector field $X_{\cal H}$ may not be all definable to the
whole $\Bbb R$. Such incompleteness arises for two reasons: (1) $M$
itself may not be complete. (2) Singularities (i.e.\ nonsmoothness) of
string could arise when evolving toward either future or past; and
hence $X_{\cal H}$ is no longer defined there. The collection of
singular loops that appear in Case (2) forms by definition the
{\it s-boundary} $\partial_{\rm s}LT^{\ast}M$ of
$(LT^{\ast}M, X_{\cal H})$. The union
$LT^{\ast}M\cup\partial_{\rm s}LT^{\ast}M$ shall be denoted by
$\overline{LT^{\ast}M}$. One can define the {\it life-span map}
$(\tau^-,\tau^+)$ from $LT^{\ast}M$ to  $[-\infty,0]\times [0,+\infty]$
by assigning to a $\gamma\in LT^{\ast}M$ the pair
$(\inf\,\{t\},\sup\,\{t\})$ where $t$ is such that $\rho_t(\gamma)$ is
well-defined.

Now recall that in the Lagrangian formalism the string action $I$ is
defined on the path-space of $LM$. Let $A_{ab}=S^1\times (a,b)$,
$-\infty\leq a<0<b \leq\infty$, be a standard Lorentzian cylinder with
the metric $dt^2-d\sigma^2$. Then $I$ is a functional on the space
$\amalg_{ab}\,\mbox{\it Map}\,(A_{ab},M)$ (with the
$C^{\infty}$-topology) of all smooth maps $f$ from some $A_{ab}$ to $M$
defined by
$$
I(f)\;
  =\;\int_{A_{ab}}\,
         \left\{\,\frac{1}{2}|df|^2\ast 1\,+\,f^{\ast}B\,\right\}\;
  =\;\int_{A_{ab}}\,
      \left\{\,\frac{1}{2}\mbox{tr}\left(f^{\ast}ds^2\right)\ast 1\,
                            +\,f^{\ast}B\,\right\}\,,
$$
where $\ast 1$ is the metric-induced area-form $d\sigma\wedge dt$ on
$A_{ab}$. This action is invariant under conformal changes of metrics
on $A_{ab}$.

Let $S_I$ be the space of {inextendable} extrema of $I$. As maps from
intervals into $LM$, these extrema are naturally lifted to
$LT_{\ast}M$; and the map
$$
\begin{array}{ccccc}
\kappa_0 &: & S_I  & \longrightarrow   & LT_{\ast}M \\
    & & f & \longmapsto
             & f_{\ast}\left(\partial_t|_{S^1\times\{0\}}\right)
\end{array}
$$
gives an identification of $S_I$ to $LT_{\ast}M$. The Legendre
transformation now takes them further to $LT^{\ast}M$. The image
$\cal J$ consists of parametrized maximal integral trajectories of
$X_{\cal H}$. The map
$$
\begin{array}{ccccc}
 \iota_0 &: &{\cal J} &\longrightarrow &LT^{\ast}M \\
         & & j & \longmapsto  &j(0)
\end{array}
$$
identifies $\cal J$ with $LT^{\ast}M$; and the interval
$(\tau^-(j(0)),\tau^+(j(0)))$ parametrizes $j$.

Let $\mbox{\it Conf}\,(S^1\times{\Bbb R})$ be the group of conformal
diffeomorphisms of $(S^1\times{\Bbb R}, dt^2-d\sigma^2)$. Recall that
it has four components:
 $\mbox{\it Conf}^{\:(+,\uparrow)}(S^1\times{\Bbb R})$,
 $\mbox{\it Conf}^{\:(-,\uparrow)}(S^1\times{\Bbb R})$,
 $\mbox{\it Conf}^{\:(-,\downarrow)}(S^1\times{\Bbb R})$, and
 $\mbox{\it Conf}^{\:(+,\downarrow)}(S^1\times{\Bbb R})$, where $\pm$
(resp.\ $\uparrow\hspace{-.2em}\downarrow$) indicates whether it
preserves the space- (resp.\ time-)orientation of $S^1\times{\Bbb R}$.
The identity component
$\mbox{\it Conf}^{\:(+,\uparrow)}(S^1\times{\Bbb R})$
is a $\Bbb Z$-covering of
$\mbox{\it Diff}^{\,+}(S^1)\times\mbox{\it Diff}^{\,+}(S^1)$.
One can define an action of $\mbox{\it Conf}\,(S^1\times{\Bbb R})$ on
$S_I$ by precomposition: Let
$h\in\mbox{\it Conf}\,(S^1\times{\Bbb R})$ and
$$
S_I^h\;=\;\{f\in S_I\,|\,
            h(S^1\times\{0\})\subset\mbox{\rm Dom}\,(f)\}\,.
$$
For $f\in S_I$, one defines $h\cdot f$ by $f\circ h$. This
{\it right action} is in general {\it pseudo} since the defined domain
$S_I^h$ of $h$ may not be the whole $S_I$. Conjugated to $LT^{\ast}M$
by $\kappa_0$ and Legendre transformation, it then becomes a
(pseudo-)action on $LT^{\ast}M$. Notice that this action leaves
$X_{\cal H}$ invariant; thus it commutes with the flow $\rho_t$
whenever defined. In fact, $\rho_t$ is the restriction of this action
to the subgroup ${\Bbb R}$ of pure translations on $S^1\times{\Bbb R}$
along the $t$-direction.

With these prerequisites, we now demonstrate the following well-known
folklore (e.g.\ [Se3]).

\bigskip

\noindent {\bf Proposition 1.1 [symplecticity].} {\it The symplecticity
 of the $\mbox{\it Conf}\,(S^1\times{\Bbb R})$-action on $LT^{\ast}M$
 depends on its components as follows.
 $$
  \begin{array}{ccl}
   \mbox{\it Conf}^{\:(+,\uparrow)}(S^1\times{\Bbb R}) &:
         &\mbox{symplectic}\,,\\
   \mbox{\it Conf}^{\:(-,\uparrow)}(S^1\times{\Bbb R}) &:
         &\mbox{symplectic when $B=0$}\,,\\
   \mbox{\it Conf}^{\:(-,\downarrow)}(S^1\times{\Bbb R}) &:
        &\mbox{anti-symplectic}\,,\\
   \mbox{\it Conf}^{\:(+,\downarrow)}(S^1\times{\Bbb R}) &:
        &\mbox{anti-symplectic when $B=0$}\;.\\
  \end{array}
 $$     }

\bigskip

\noindent {\it Proof.} Assume first that
$h\in\mbox{\it Conf}^{\:(+,\uparrow)}(S^1\times{\Bbb R})$ and that
$S^1\times\{0\}$ and $h(S^1\times\{0\})$ are disjoint, say,
$h(S^1\times\{0\})$ lies in the chronological future domain of
$S^1\times\{0\}$. Together they bound a compact annulus $\Sigma$ in
$S^1\times{\Bbb R}$. Let $I_{\Sigma}$ be the functional on $S_I$
defined by
$$
 I_{\Sigma}(f)\;
   =\;\int_{\Sigma}\,
     \left\{\,\frac{1}{2}\mbox{tr}\left(f^{\ast}ds^2\right)\ast 1\,
                           +\,f^{\ast}B\,\right\}\,,
$$
where $\Sigma$ is endowed with the standard submanifold metric from
$S^1\times{\Bbb R}$. This functional depends only on the conformal
class of the metric on $\Sigma$. Consider a new metric on $\Sigma$
that is conformal to the standard metric and satisfies the following
two properties: (1) It is smoothly extendable to a metric on
$S^1\times{\Bbb R}$ that is conformal to the standard one; (2) it is
the standard metric in a neighborhood of $S^1\times\{0\}$ and
coincides with the push-forward metric $h_{\ast}(dt^2-d\sigma^2)$ in
a neighborhood of $h(S^1\times\{0\})$. Such a metric can be obtained
by a partition of unity argument. Denote by $I^{\prime}_{\Sigma}$
the functional on $S_I$ associated to one such metric.

Let $U$ be a Jacobi field along $f\in S_I$ and $f_{\varepsilon}$ be
a one-parameter family of elements in $S_I$ that gives rise to $U$
with $f_0=f$. One has the following first variation formula
\begin{eqnarray*}
\lefteqn{ dI_{\Sigma}(U)\;=\;dI^{\prime}_{\Sigma}(U)
  =\;\left.\frac{d}{d\varepsilon}\right|_{\varepsilon=0}\,
                          I^{\prime}_{\Sigma}(f_{\varepsilon})  }\\
 & &  =\;\int_{\Sigma}\left[  -\langle
    \mbox{tr}^{\prime}\left(\nabla^{\prime}df\right)\,,\,U
     \rangle\,\ast^{\prime}1\,+\,f^{\ast}\left(i_U dB\right) \right]\;
    +\;\int_{\partial\Sigma}\left[
  \langle f_{\ast}\mbox{\boldmath $\nu^{\prime}$},U\rangle\,
        \bar{\ast^{\prime}}1\;+\;f^{\ast}\left(i_U B\right) \right]\,,
\end{eqnarray*}
where all quantities with "$\;^{\prime}\;$" are with respect to the new
metric and {\boldmath $\nu^{\prime}$} is the outward unit normal along
$\partial\Sigma$ (with respect to the new metric). The contribution to
the formula from the metric part has been well-known in the literature
of {\it harmonic maps} ([E-L1]). That from the $B$-field results from
the following manipulation. Define
$F:\Sigma\times (-\varepsilon_0,\varepsilon_0)\rightarrow M$ by
$F(\sigma,t,\varepsilon)=f_{\varepsilon}(\sigma,t)$. Then
\begin{eqnarray*}
\lefteqn{\hspace{-4cm}
   \frac{d}{d\varepsilon}\int_{\Sigma} f^{\ast}_{\varepsilon}B\;
      =\;\int_{\Sigma\times\{\varepsilon\}}
        {\cal L}_{\frac{\partial}{\partial\varepsilon}} F^{\ast}B
      =\;\int_{\Sigma\times\{\varepsilon\}} \left[
          i_{\frac{\partial}{\partial\varepsilon}} dF^{\ast}B\,
    +\,d i_{\frac{\partial}{\partial\varepsilon}} F^{\ast}B \right]}\\
  & & \hspace{-4cm}
       =\;\int_{\Sigma} f^{\ast}(i_U dB)\,+\,\int_{\partial\Sigma}
           f^{\ast}(i_U B) \hspace{.6cm}\mbox{at $\varepsilon=0$}\,.
\end{eqnarray*}

Back to the first variation formula. Since $f\in S_I$, the integrand
of the integral over $\Sigma$ vanishes. {}From the fact that, with
respect to the new metric, the outward unit normal
{\boldmath $\nu^{\prime}$} for $\Sigma$ is $-\partial_t$ at
$S^1\times\{0\}$ and $h_{\ast}\partial_t$ at $h(S^1\times\{0\})$, the
boundary term can be expressed more explicitly as
\begin{eqnarray*}
\lefteqn{ \int_{S^1}\left[\langle U|_{h(S^1\times\{0\})},
   (f\circ h)_{\ast}(\partial_t|_{t=0})\rangle\,
       +\,B\left( U|_{h(S^1\times\{0\})},
               (f\circ h)_{\ast}\partial_{\sigma}\right)  \right]   }\\
   & &  \hspace{4cm} -\;\int_{S^1}\left[\langle U|_{S^1\times\{0\}},
                        f_{\ast}(\partial_t|_{t=0})\rangle\,
 +\,B\left(U|_{S^1\times\{0\}},f_{\ast}\partial_{\sigma}\right)\right]\,.
\end{eqnarray*}
This indicates that, if we define a 1-form $\Xi_{\cal L}$ on $S_I$ by
$$
 \left.\Xi_{\cal L}\right|_f(\,\cdot\,)\;
    =\;\int_{S^1}\left[\langle (\,\cdot\,)|_{S^1\times\{0\}},
                        f_{\ast}(\partial_t|_{t=0})\rangle\,
        +\,B\left((\,\cdot\,)|_{S^1\times\{0\}},
                          f_{\ast}(\partial_{\sigma}|_{t=0})
                                                   \right)\right]\,,
$$
then $h$ as a map on $S_I$ by pre-composition satisfies
$$
 h^{\ast}\Xi_{\cal L}\,-\,\Xi_{\cal L}\;=\;dI_{\Sigma}
$$
since $h_{\ast}U=U$. Thus the 2-form $d\Xi_{\cal L}$ on $S_I$ is
invariant under $h$.

If $S^1\times\{0\}$ and $h(S^1\times\{0\})$ intersect, they together
bound a $\Sigma=\cup_i\Omega_i$ with each $\Omega_i$ a lens domain in
$S^1\times{\Bbb R}$ bounded by two simple spacelike arcs. The first
variation formula still holds since the non-smoothness of corners of
$\Omega_i$ is insignificant under $\int$. And the rest hence follows as
well. This shows that $d\Xi_{\cal L}$ is invariant under
$\mbox{\it Conf}^{\:(+,\uparrow)}(S^1\times{\Bbb R})$.

For the other three components of $\mbox{\it Conf}\,(S^1\times{\Bbb R})$,
they are the
$\mbox{\it Conf}^{\:(+,\uparrow)}(S^1\times{\Bbb R})$-cosets of the
following simple reflections:
$$
\begin{array}{ccl}
 \mbox{Rx}_{(-,\uparrow)} &: &(\sigma,t)\mapsto (-\sigma,t)\,,\\
 \mbox{Rx}_{(-,\downarrow)} &: &(\sigma, t)\mapsto (-\sigma,- t)\,,
                                         \hspace{1cm}\mbox{and}  \\
 \mbox{Rx}_{(+,\downarrow)} &: &(\sigma,t)\mapsto (\sigma,-t)\,.
\end{array}
$$
Denote $\Xi_{\cal L}=\Xi_{\cal L}^m + \Xi_{\cal L}^B$, where
$\Xi_{\cal L}^m$ involves only the metric and $\Xi_{\cal L}^B$ only the
$B$-field. It is clear that
$$
\begin{array}{ccr}
 \mbox{Rx}_{(-,\uparrow)}^{\ast}\Xi_{\cal L}
                      &=  & \Xi_{\cal L}^m - \Xi_{\cal L}^B\,, \\
 \mbox{Rx}_{(-,\downarrow)}^{\ast}\Xi_{\cal L}
                      &=  & -\Xi_{\cal L}^m - \Xi_{\cal L}^B\,, \\
 \mbox{Rx}_{(+,\downarrow)}^{\ast}\Xi_{\cal L}
                      &=  & -\Xi_{\cal L}^m + \Xi_{\cal L}^B\,. \\
\end{array}
$$
Thus they satisfy the symplecticity properties described in the
proposition for the components of $\mbox{\it Conf}\,(S^1\times{\Bbb R})$
in which they lie. Therefore so do these components.

On the other hand, the pullback 1-form
$\mbox{\boldmath $\theta$}_{\cal L}$ on $LT_{\ast}M$ of
$\mbox{\boldmath $\theta$}$ by the Legendre transformation is exactly
$(\kappa_0^{-1})^{\ast}\Xi_{\cal L}$. Consequently, after conjugating
all back to $LT^{\ast}M$ via the Legendre tranformation, one concludes
that $\mbox{\it Conf}\,(S^1\times{\Bbb R})$ acts on $LT^{\ast}M$ as
indicated. This proves the proposition.

\noindent\hspace{14cm}$\Box$

\bigskip

\begin{flushleft}
{\bf Single-phase-space-description of interacting strings.}
\end{flushleft}
An element of $\overline{LT^{\ast}M}$ can be regarded as a parametrized
string in $M$ with an infinitesimal intent of motion. Such objects can
join or split in various ways ({\sc Figure 1\,-\,1}) (cf.\ [St1] and
[Wi1]). Formally, the joining operation "$\,\cdot\,$" induces a
{\it product} $\,\ast\,$ of string fields by convolution:
$$
 \psi_1\ast\psi_2(\gamma)\;
    =\; \int_{\gamma_1\cdot\gamma_2=\gamma}\,
                     [{\cal D}\gamma_1][{\cal D}\gamma_2]\,
                                \psi_1(\gamma_1)\psi_2(\gamma_2)\,;
$$
and the splitting operation induces a {\it coproduct} by an inverse of
convolution:
$$
 \psi\;\longrightarrow\;\{\psi_1,\psi_2\}
     \hspace{1em}\mbox{with either}\hspace{1em}
       \psi_1\ast\psi_2\,=\,\psi   \hspace{1em}\mbox{or}\hspace{1em}
       \psi_2\ast\psi_1\,=\,\psi\,.
$$
So far the coproduct as defined could be {\it multi-valued}. These
operations describe the basic interaction of strings or string fields.
It is possible to define an $A_{\infty}$-structure or a
co-$A_{\infty}$-structure associated to many-to-one joinings or
splittings ([St1-2], [Ko]).
\begin{figure}[htbp]
\setcaption{{\sc Figure 1\,-\,1.} Loops in $\overline{T^{\ast}M}$ can
            join or split in various ways. The upper ones involve
            loops in the s-boundary of $LT^{\ast}M$; while the lower
            ones loops in $LT^{\ast}M$.   }
\centerline{\psfig{figure=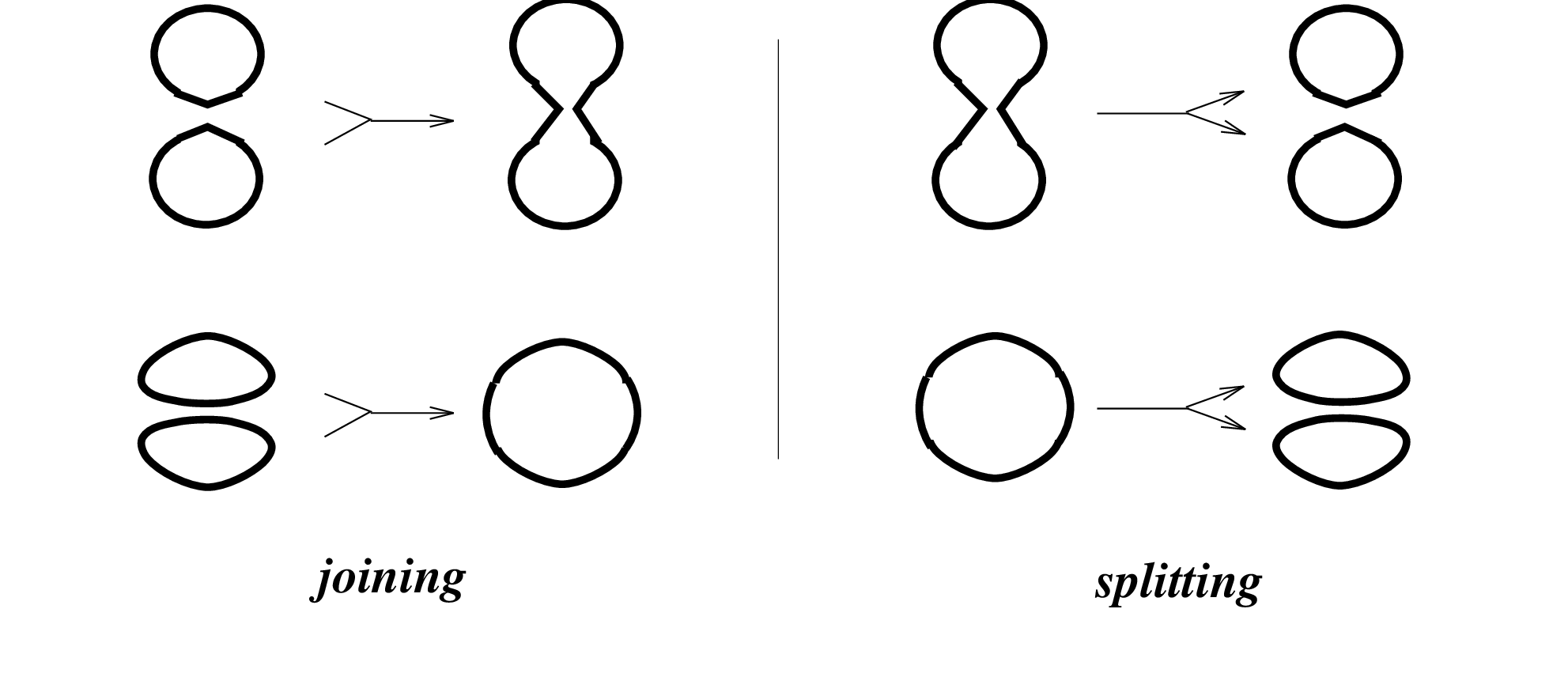,width=13cm,caption=}}
\end{figure}

The Hamiltonian system
$(L^{\ast}M, \mbox{\boldmath $\omega$}, {\cal H})$ together with these
operations are in principle enough to give a description of interacting
strings. At the classical level, a process of interactions of strings
corresponds to a {\it partially ordered} collection of
{\it integral filaments} to $X_{\cal H}$ ({\sc Figure 1\,-\,2}). At
quamtum level, it corresponds to a partially ordered collection of
one-parameter families of string fields governed by the {\it
Schr\"{o}dinger equation} (cf.\ Appendix).
\begin{figure}[htbp]
\setcaption{{\sc Figure 1\,-\,2.}  A partially ordered collection of
   integral filaments in $(LT^{\ast}M, X_{\cal H})$ represents a process
   of string interactions. The partially ordered labelling set indicates
   the events when strings join or split, i.e.\ the occurrence of loop
   operations. The dotted line $\cdots$ in $\overline{LT^{\ast}M}$
   indicates the Hamiltonian flow.    }
\centerline{
  \psfig{figure=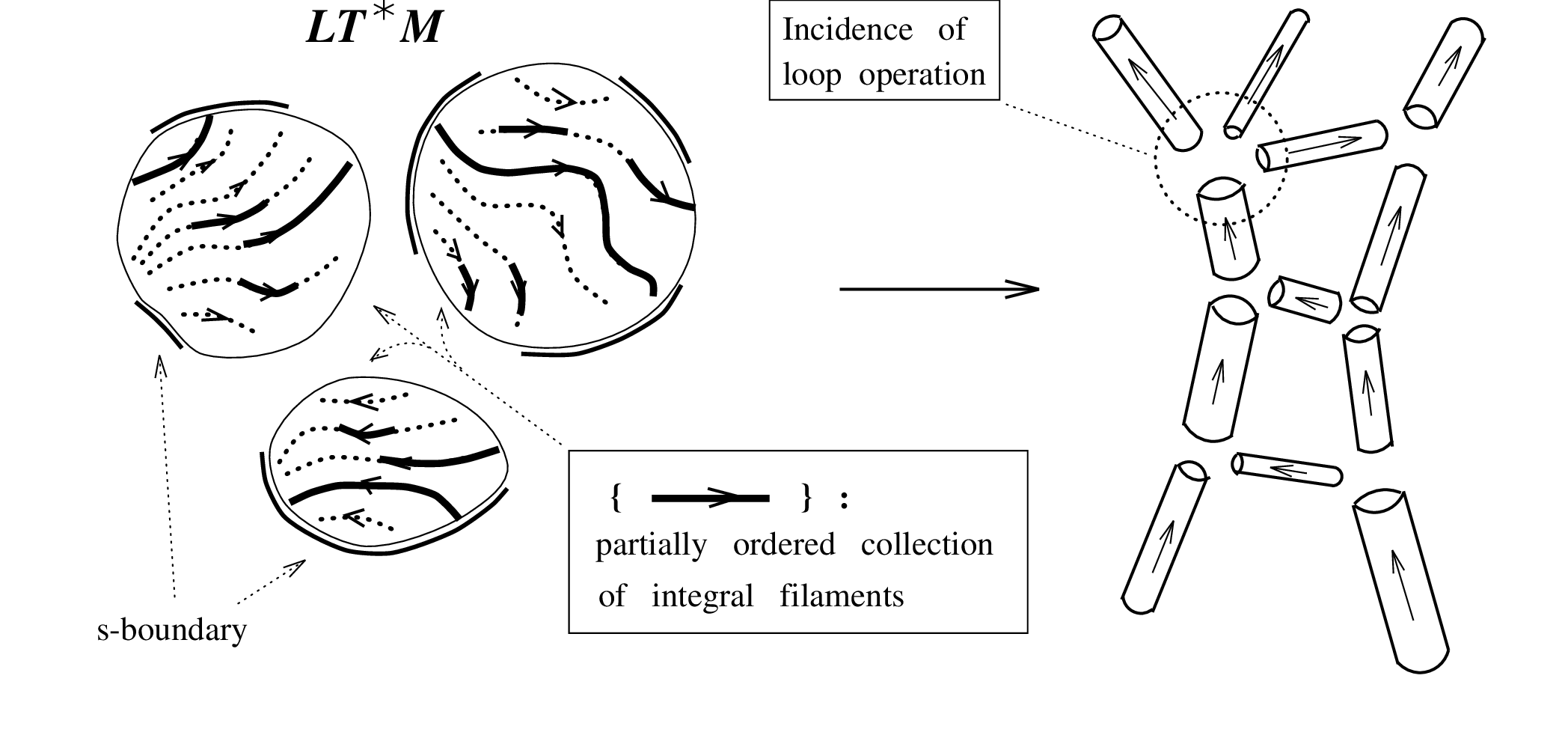,width=13cm,caption=}}
\end{figure}

For string fields ([W-Z], [Zw]), one also likes to know what action
governs their dynamics. The form of such actions for string fields are
constrained among other things by conditions due to symmetry of the
theory. In the above phase-space-picture of interacting strings, there
doesn't seem to have any limitation to the pattern of interaction of
strings at a single instant, i.e.\ the type of string vertices. However,
an action for string field should set a limitation to possible string
vertices. For example, if interacting part of the action involves only
$\psi\ast\psi\ast\psi$, then only simple trivalent string vertices
could appear. There are details to be worked out to make this picture
solid, which we are not ready. However, when turning to the Lagrangian
counterpart of this picture, one is naturally led to the study of
Lorentz surfaces.

\bigskip

\section{Lorentz surfaces and their moduli.}
A partially ordered collection of integral filaments can be realized as
a map from a Lorentzian 2-manifold $\Sigma$ to the target-space $M$.
Such $\Sigma$ has to admit metric singularities. Conformal symmetry in
the theory indicates that only the conformal structure of $\Sigma$
matters. The whole setting thus leads to a study of to-be-defined
Lorentz surfaces. This is a fundamental ingredient toward an
{\it un-Wick-rotated string theory}. We like to mention the papers
[De1-2] by T.\ Deck, which are likely a pioneering work in this less
popular direction. Readers are referred to [Ar1], [A-A], [C-L-W],
[D-R-R] and [Ta1] for facts in smooth dynamical systems. The surfaces
$S^1\times{\Bbb R}$, $S^1\times (-\infty,0]$, $S^1\times [0,\infty)$,
and $S^1\times[0,T]$ in the discussion are all equipped with the
standard Lorentzian metric unless otherwise noted.

\bigskip

\subsection{Lorentz surfaces and Mandelstam diagrams.}
Analogous to a Riemann surface being a conformal class of Riemannian
2-manifolds, a {\it Lorentz surface} is meant to be a conformal class
of Lorentzian 2-manifolds. However, while a Riemannian structure resides
on every paracompact smooth manifold, a Lorentzian structure on a
smooth manifold in general has to accommodate singularities where the
metric is degenerate due to topological reasons.

\bigskip

\noindent {\bf Definition 2.1.1 [s-d-l Lorentzian 2-manifold].}
 A {\it string-diagram-like (s-d-l) Lorentzian 2-manifold} is a smooth
 2-manifold of finite type with a smooth Lorentzian structure that
 satisfies the following conditions: (1) All of its singularities are
 isolated points and they lie in the interior of the manifold. (2) There
 is no trapped set in the interior of the manifold. (3) The complement
 of singularities is time-orientable and contains no causal loops. (4)
 Every boundary component is spacelike and every end is conformal to
 either $S^1\times(-\infty,0]$ or $S^1\times[0,\infty)$.

\bigskip

\noindent
The conditions listed in the definition reflect the assumptions for
interacting strings: (1) Points on a string are causally independent.
(2) Interacting strings are both past and future asymptotically
non-interacting. (3) Each interaction takes place only at an instant.

For simplicity of phrasing, in this article we shall mainly consider
s-d-l Lorentzian 2-manifolds with ends unless otherwise noted. We shall
call a such time-oriented $\Sigma$ {\it of type} $(+,\hat{\chi},m,n)$
(resp.\ $(-,\hat{\chi},m,n)$) if $\Sigma$ is orientable (resp.\
non-orientable) with $m$ past- and $n$ future-ends such that the
Euler characteristic of the surface after all the ends are capped by a
disk is $\hat{\chi}$.

\bigskip

\begin{flushleft}
{\bf Singularities.}
\end{flushleft}
Let $\Sigma$ be a time-oriented s-d-l Lorentzian 2-manifold. One can
assign an index $\mbox{ind}_p(C)$ to the light-cone field $C$ at any
$p$ in $\Sigma$ to be the index of any causal line field in a
neighborhood of $p$. The Poincar\'{e}-Bendixson theorem implies that a
generic singularity $p$ could only have index $-1$ in order not to
violate either the time-orientability or no-trapped-set condition. A
non-generic one arises from fusion of some $s$ generic ones and is of
index $-s$. The light-cone structure around such is illustrated in
{\sc Figure 2\,-\,1}.
\begin{figure}[htbp]
\setcaption{{\sc Figure 2\,-\,1}  A light-cone structure around a
    singularity $p$ of index $-s$. The shaded cones in (a) are the
    future light-cones. In (b) the $J_i^+$'s (resp.\ $J_i^-$'s) form the
    causal future (resp.\ past) of $p$.              }
\centerline{\psfig{figure=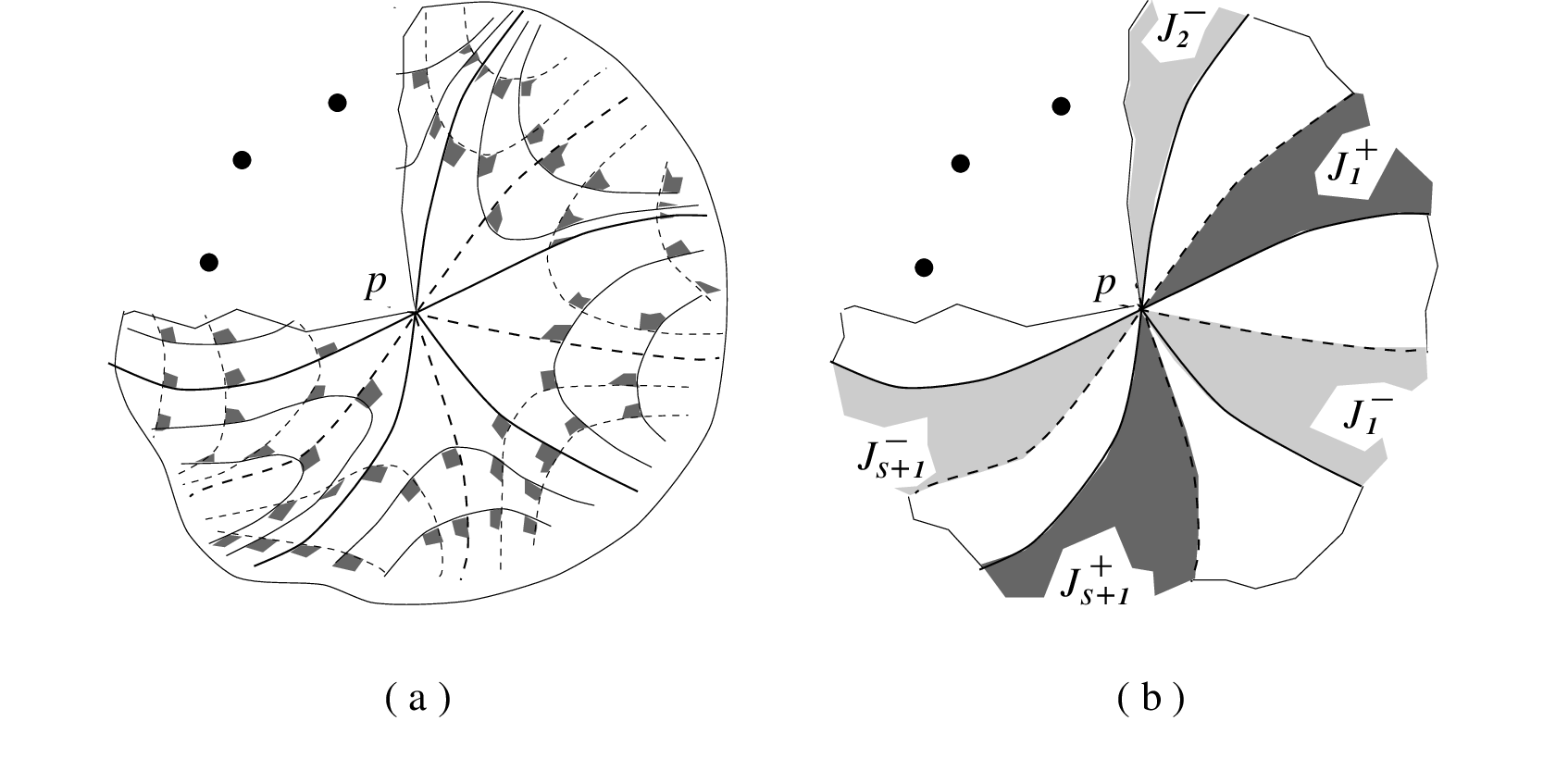,width=13cm,caption=}}
\end{figure}
In a neighborhood of such $p$, each of the causal future $J^+(p)$ and
causal past $J^-(p)$ of $p$ has $s+1$ components. Together they form a
{\it causal flower} at $p$ with the petals alternating between future
and past. Such singularity can be perturbed and disintegrated back into
a collection of generic ones.

\bigskip

\begin{flushleft}
{\bf The coarse and fine conformal equivalences.}
\end{flushleft}
Unlike the case of Riemannian manifolds, singularities distinguish
themselves in a Lorentzian manifold. In defining a conformal equivalence
$f$ between two generic Lorentzian manifolds, one has the following two
choices: (1) {\it The coarse sense}: up to some mild extra demand,
$f$ is required to be a homeomorphism but is smooth only in the
complement of singularities. (2) {\it The fine sense}: $f$ is required
to be a diffeomorphism.

For the coarse conformal Lorentzian geometry, one may as well enlarge
the scope of the metric tensors under consideration to those that are
smooth in the complement of their singularities and satisfy only some
weaker conditions around the singularities - basically that an
appropriate metric completion of the complement around a singularity
should give back exactly that original singularity. Due to this
looseness, the coarse conformal Lorentzian geometry is conceivably much
less rigid than the fine conformal Lorentzian geometry. In the
2-dimensional case, associated to each singularity of the metric, there
is a fine conformal invariant, i.e.\ {\it the modulus of the smooth
orbital equivalence class} of the transverse pair of null line fields
around that singularity. This may not be preserved under a coarse
conformal deformation. Though both are interesting, details concerning
the above moduli of singularities does not seem available at the moment.
Fortunately, both our string action and its conformal invariance can be
readily extended to Lorentzian 2-manifolds in the coarse category.

\bigskip

\noindent {\bf Definition 2.1.2 [Lorentz surface].}
 A {\it coarse} (resp.\ {\it fine}) {\it Lorentz surface} is a coarse
 (resp.\ fine) conformal equivalence class of coarse (resp.\
 fine) Lorentzian 2-manifolds.

\bigskip

In this article we shall discuss only coarse s-d-l Lorentz surfaces
unless otherwise noted.

\bigskip

\begin{flushleft}
{\bf Mandelstam diagrams.}
\end{flushleft}
Let ${\Bbb R}^{1+1}$ be the standard 2-dimensional Minkowski space-time
with metric $d\sigma^+\cdot d\sigma^-$, where $(\sigma^+,\sigma^-)$ are
the light-cone coordinates of ${\Bbb R}^{1+1}$. The global time function
$t$ is then given by $\frac{1}{2}(\sigma^++\sigma^-)$. Notice that all
infinite Minkowskian cylinders
${\Bbb R}^{1+1}/$\raisebox{-.6ex}{${\Bbb Z}v$} with $v$
spacelike are homothetic to each other. We shall fix $v_0=(2\pi,-2\pi)$
throughout the paper. The time function $t$ on ${\Bbb R}^{1+1}$
then descends to the standard cylinder
$S^1\times{\Bbb R}={\Bbb R}^{1+1}/$\raisebox{-.6ex}{${\Bbb Z}v_0$}.
So do the two 1-forms $d\sigma^+$ and $d\sigma^-$. A {\it Mandelstam
diagram} is a coarse s-d-l Lorentzian 2-manifold $\Xi$ that satisfies
the following conditions: (1) $\Xi$ admits an annuli decomposition
$\{A_{\alpha}\}$ with each annulus $A_{\alpha}$ homothetic via an
$f_{\alpha}$ to one of the standard annuli: $S^1\times{\Bbb R}$,
$S^1\times (-\infty,0]$, $S^1\times [0,\infty)$, and $S^1\times [0,T]$
for some $T$. (2) When boundaries of these annuli are pasted in $\Xi$,
the pulled-back local 1-forms $\{f_{\alpha}^{\ast}d\sigma^+\}$ and
$\{f_{\alpha}^{\ast}d\sigma^-\}$ on $\Xi$ can also be pasted. The result
is a bi-valued 1-form $\mu$ on $\Xi$ singular only at the singularities
of $\Xi$ (or better a bi-valued transverse measure to the null
foliations of $\Xi$, cf.\ Sec.\ 2.2). For $\Xi$ orientable, $\mu$
splits into two single-valued 1-forms $\mu_L$, $\mu_R$ on $\Xi$. For
$\Xi$ non-orientable, $\mu$ can be lifted to the orientation covering
space $\Xi^{\rm ornt}$ and becomes single-valued. We shall call these
1-forms {\it characteristic 1-forms} on $\Xi$. (3) Up to constant
shifts, one for each $A_{\alpha}$, the collection of local time
functions $\{f_{\alpha}^{\ast}t\}$ on $\Xi$ form a globally
well-defined time function (still denoted by $t$) on $\Xi$.

\bigskip

\subsection{Basic structures and coarse conformal groups.}
The light-cone structure determines the conformal structure of a
regular Lorentzian manifold. In two-dimensions it determines the
coarse conformal structure of an s-d-l Lorentzian manifold and hence
a coarse s-d-l Lorentz surface $\Sigma$. There are some basic
structures on $\Sigma$ that follow from its light-cone structure.
They play important roles in our study.

\bigskip

\begin{flushleft}
{\bf Basic structures for $\Sigma$ orientable.}
\end{flushleft}
Let $\Sigma$ be both oriented and time-oriented. One can then define a
{\it left} (resp.\ {\it right}) {\it null line element} on $\Sigma$ as
one whose future direction to the time-orientation is opposite to
(resp.\ accordant with) the given orientation of the surface. They form
the {\it left null line field} on $\Sigma$. Its integral trajectories
give the {\it left null foliations} ${\cal F}_L$ of $\Sigma$. It is a
foliation with singularities the same as $\Sigma$. The time-orientation
of $\Sigma$ gives a direction for leaves of ${\cal F}_L$. Generically,
they emit from past-ends of $\Sigma$ and go into future-ends. There are
finitely many directed leaves that either emit from or land on some
singularity. We shall call them {\it characteristic leaves}. The
complement of these leaves and singularities is a collection of strips,
which we shall call the {\it left characteristic strips}.

Let $\Sigma_0$ be the complement of
singularities. Then the {\it left leaf-space}
$\Sigma_0/$\raisebox{-.6ex}{${\cal F}_L$} with the quotient topology is
a smooth non-Hausdorff 1-manifold whose non-Hausdorff points correspond
exactly to the characteristic leaves of ${\cal F}_L$. A smooth 1-form
on $\Sigma_0/$\raisebox{-.6ex}{${\cal F}_L$} can be pulled back to a
{\it directed transverse measures} on $\Sigma_0$ with respect to
${\cal F}_L$. It has the property that the total measure along any
small loop encircling a singularity is 0. Consequently, homologous
1-cycles on $\Sigma$ have the same transverse measures.

Analogously, one has the {\it right null foliation} ${\cal F}_R$, {\it
right characteristic strips}, and the {\it right leaf-space}
$\Sigma_0/$\raisebox{-.6ex}{${\cal F}_R$} for $\Sigma$. They share
similar properties as their corresponding left partners. The two
foliations ${\cal F}_L$ and ${\cal F}_R$ are transverse to each other.

For generic $\Sigma$, all its singularities are of index $-1$ and there
are no characteristic leaves that connect any two of them. One can then
identify to a point each collection of non-Hausdorff points in
$\Sigma_0/$\raisebox{-.6ex}{${\cal F}_L$} associated to a singularity.
The result is a regular $4$-valence graph. It can be embedded as a graph
$\Gamma_L$ in $\Sigma$ such that its vertex set coincides the set of
singularities of $\Sigma$ and that its edges are transverse to
${\cal F}_L$. Any two different such embeddings are isotopic in
$\Sigma$ relative to the vertex set and $\Gamma_L$ is a deformation
retract of $\Sigma$ along the leaves of ${\cal F}_L$. For non-generic
$\Sigma$, $\Sigma$ can be obtained after {\it squashing} a finite
collection of characteristic strips of some generic $\Sigma^{\|}$
{\it along their partner foliation}. The $\Gamma_L^{\|}$ in
$\Sigma^{\|}$ then descends to a $\Gamma_L$ in $\Sigma$ following a
corresponding sequence of either null-operation (which preserves
topology of graphs) if squashing a left strip or pinching of some part
of an edge if squashing a right strip. Different $\Sigma^{\|}$'s that
lead to same $\Sigma$ could however give different $\Gamma_L$. In both
cases, the complement of $\Gamma_L$ consists of topological cylinders,
one for each end of $\Sigma$. Similarly, one has also $\Gamma_R$ with
the same properties. ({\sc Figure 2\,-\,2.})
\begin{figure}[htbp]
\setcaption{{\sc Figure 2\,-\,2.} Non-generic $\Sigma$ can be obtained
  from a generic one $\Sigma^{\|}$ by a finite sequence of squashing.
  For clarity, the strip to be squashed at each step is shadowed. }
\centerline{\psfig{figure=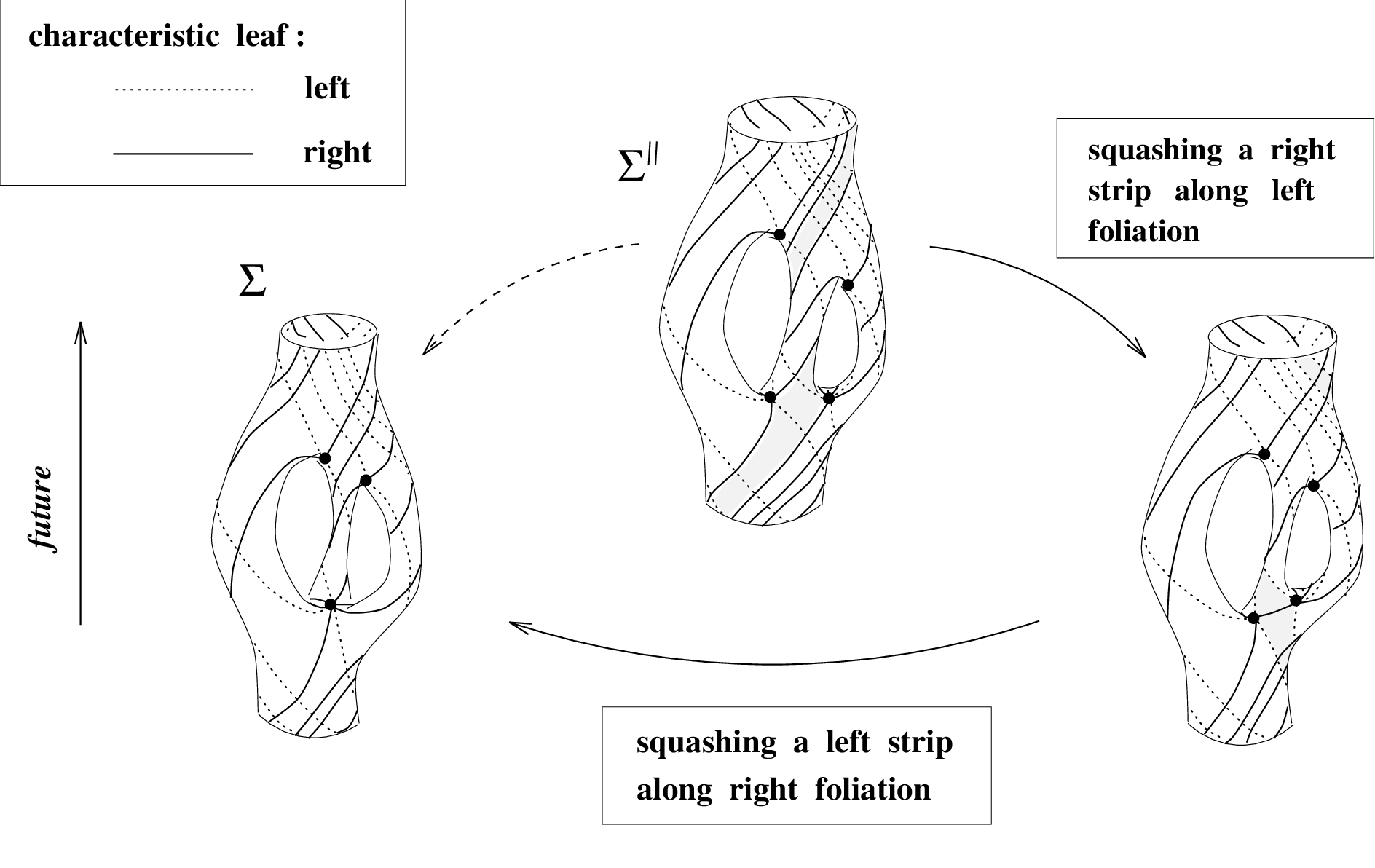,width=13cm,caption=}}
\end{figure}

The time-orientation on $\Sigma$ as a timelike vector field on
$\Sigma_0$ is transverse to both ${\cal F}_L$ and ${\cal F}_R$. Hence
it induces an orientation for
$\Sigma_0/$\raisebox{-.6ex}{${\cal F}_L$} and
$\Sigma_0/$\raisebox{-.6ex}{${\cal F}_R$}. This in turn leads to an
orientation for $\Gamma_L$ and $\Gamma_R$. One can thus define
{\it positive 1-forms} on $\Sigma_0/$\raisebox{-.6ex}{${\cal F}_L$} and
$\Sigma_0/$\raisebox{-.6ex}{${\cal F}_R$} to be those whose evaluation
along the orientation is positive.

The union of the left and right set of characteristic leaves together
bind $\Sigma$. Its complement is a disjoint union of
{\it light-cone-diamonds}, each of the form $I^+(q_1)\cap I^-(q_2)$ for
some chronological pair of points $(q_1,q_2)$. This gives the
{\it (characteristic) light-cone-diamond} ({\it l-c-d})
{\it tessellation} of $\Sigma$ ({\sc Figures} 2\,-\,2, 2\,-\,3, 3\,-\,4,
 and 4\,-\,3).

The union of all left characteristic leaves and strips that come from
the same past-end will be called a {\it past-left crown} of $\Sigma$
({\sc Figure 2\,-\,3}). One can border it by the left characteristic
leaves that lie in the closure of the strips involved. Similarly, one
has {\it past-right}, {\it future-left}, and {\it future-right crowns}.
\begin{figure}[htbp]
\setcaption{{\sc Figure 2\,-\,3.} The left characteristic leaves and
    strips of $\Sigma$ and the past-left crown associated to one of its
    past-ends. }
\centerline{\psfig{figure=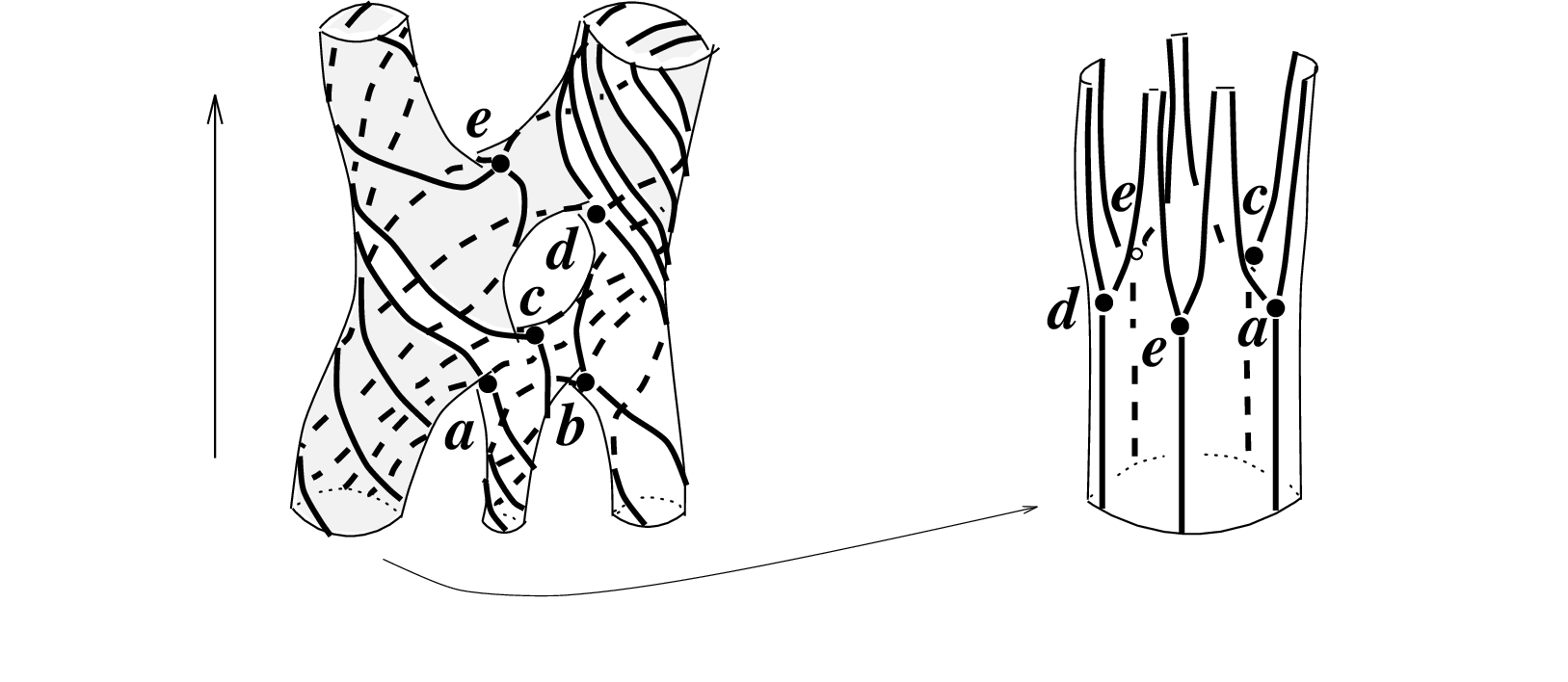,width=13cm,caption=}}
\end{figure}

\bigskip

\begin{flushleft}
{\bf Basic structures for $\Sigma$ non-orientable.}
\end{flushleft}
When $\Sigma$ is non-orientable, the above pairs of structures can be
associated to its orientation covering space $\Sigma^{\rm ornt}$ with
the lifted Lorentzian conformal structure. The latter is also s-d-l.
The non-trivial deck transformation on $\Sigma^{\rm ornt}$ is an
involution on $\Sigma$ that preserves time-orientation while exchanging
left and right. Any left structure for $\Sigma^{\rm ornt}$ and its right
partner are isomorphic to each other under this involution. A
{\it characteristic leaf} to the light-cone field on $\Sigma$ is defined
to be the projection of one on $\Sigma^{\rm ornt}$. Similarly for a
{\it characteristic strip}. They are embedded in $\Sigma$. Either
foliation, ${\cal F}_L$ or ${\cal F}_R$, on $\Sigma^{\rm ornt}$ projects
to the locally transverse bi-foliation on $\Sigma$ associated to the
light-cone field. We shall take the leaf space
$\Sigma^{\rm ornt}_0/$\raisebox{-.6ex}{${\cal F}$}, where $\cal F$ is
either ${\cal F}_L$ or ${\cal F}_R$, as the leaf space associated to
$\Sigma$. Recall that it is oriented. As in the orientable case,
non-generic $\Sigma$ can be obtained from generic one by
{\it squashing}. The {\it graph} $\Gamma(\Sigma)$ is taken to be
either $\Gamma_L(\Sigma^{\rm ornt})$ or $\Gamma_R(\Sigma^{\rm ornt})$.
When $\Sigma$ is non-generic, it is understood that one considers
squashings invariant under the deck transformation. The collection of
characteristic leaves on $\Sigma$ give likewise the
{\it l-c-d tessellation} of $\Sigma$.

\bigskip

\noindent{\it Remark 2.2.1.} The l-c-d tessellation resembles a
 {\it circle packing} on a Riemann surface ([R-S], [Thu]). While the
 latter is only an approximate conformal structure to a Riemann surface,
 the former is intrinsic to and completely determines the coarse
 conformal structure of a Lorentz surface.

\bigskip

\begin{flushleft}
{\bf Coarse conformal groups.}
\end{flushleft}
The following discussion refines and recasts Theorems 2.2 and 2.3 in
[De] to the current setting.

Let $\Sigma$ be a time-oriented s-d-l Lorentz surface and
$\mbox{\it Conf}\,^{(c)}(\Sigma)$ be the group of coarse conformal
automorphisms of $\Sigma$. Assume first that $\Sigma$ is oriented. Let
$$
 \mbox{\it Diff}^{\,\pm}\left(
   \Sigma_0/\mbox{\raisebox{-.6ex}{${\cal F}_L$}}
    \amalg \Sigma_0/\mbox{\raisebox{-.6ex}{${\cal F}_R$}}\right)
$$
be the group of diffeomorphisms of the disjoint union
$\Sigma_0/\mbox{\raisebox{-.6ex}{${\cal F}_L$}}
    \amalg \Sigma_0/\mbox{\raisebox{-.6ex}{${\cal F}_R$}}$
that are either orientation-preserving or orientation-reversing and
$$
 \mbox{\it Diff}_{\,0}\left(
   \Sigma_0/\mbox{\raisebox{-.6ex}{${\cal F}_L$}}
    \amalg \Sigma_0/\mbox{\raisebox{-.6ex}{${\cal F}_R$}}\right)\;
  =\;\mbox{\it Diff}_{\,0}
  (\Sigma_0/\mbox{\raisebox{-.6ex}{${\cal F}_L$}})
    \times\mbox{\it Diff}_{\,0}
       (\Sigma_0/\mbox{\raisebox{-.6ex}{${\cal F}_R$}})
$$
be its identity component. Let
$\mbox{\it Aut}\,(\mbox{\rm Tessln}_{\Sigma})$ be the group of isotopy
classes of (topological) automorphisms of $\Sigma$ that preserve the
l-c-d tessellation by sending tiles only to tiles. Since an
automorphism of surface $\Sigma$ is coarse conformal if and only if it
preserves $\{{\cal F}_L,{\cal F}_R\}$, hence their characteristic
leaves, and is smooth outside singularities, there are two group
homomorphisms
$$
 \varphi_1 \;:\; \mbox{\it Conf}\,^{(c)}(\Sigma)\; \longrightarrow\;
  \mbox{\it Diff}^{\,\pm}\left(
   \Sigma_0/\mbox{\raisebox{-.6ex}{${\cal F}_L$}}
    \amalg \Sigma_0/\mbox{\raisebox{-.6ex}{${\cal F}_R$}}\right)
$$
and
$$
 \varphi_2 \;:\; \mbox{\it Conf}\,^{(c)}(\Sigma)\; \longrightarrow\;
             \mbox{\it Aut}\,(\mbox{\rm Tessln}_{\Sigma})\,.
$$

When $\Sigma$ is topologically an infinite cylinder, it is represented
by the standard $S^1\times{\Bbb R}$ and hence
$\mbox{\it Conf}\,^{(c)}(\Sigma)$ is
$\mbox{\it Conf}\,(S^1\times{\Bbb R})$. In this case,
$\mbox{\it Diff}^{\,\pm}\left(
   \Sigma_0/\mbox{\raisebox{-.6ex}{${\cal F}_L$}}
    \amalg \Sigma_0/\mbox{\raisebox{-.6ex}{${\cal F}_R$}}\right)
  =\mbox{\it Diff}^{\,\pm}\left(S^1\amalg S^1\right)$
has four components:
$\mbox{\it Diff}^{\,+}(S^1)\times \mbox{\it Diff}^{\,+}(S^1)$
(resp.\ $\mbox{\it Diff}^{\,-}(S^1)\times \mbox{\it Diff}^{\,-}(S^1)$)
for orientation-preserving (resp.\ -reversing) diffeomorphisms with each
$S^1$ mapped to itself; and
$\mbox{\it Diff}^{\,+\leftrightarrow}\left( S^1 \amalg S^1 \right)$
(resp.\
$\mbox{\it Diff}^{\,-\leftrightarrow}\left(S^1 \amalg S^1 \right)$)
for orientation-preserving (resp.\ -reversing) diffeomorphisms with the
two  $S^1$ exchanged. And
$$
\varphi_1\;:\; \left.
 \begin{array}{ccl}
  \mbox{\it Conf}^{\:(+,\uparrow)}(S^1\times{\Bbb R}) & \longrightarrow
     & \mbox{\it Diff}^{\,+}(S^1)\times \mbox{\it Diff}^{\,+}(S^1)  \\
  \mbox{\it Conf}^{\:(-,\uparrow)}(S^1\times{\Bbb R}) & \longrightarrow
     & \mbox{\it Diff}^{\,+\leftrightarrow}\left(S^1\amalg S^1\right) \\
  \mbox{\it Conf}^{\:(-,\downarrow)}(S^1\times{\Bbb R}) &\longrightarrow
     & \mbox{\it Diff}^{\,-}(S^1)\times \mbox{\it Diff}^{\,-}(S^1)  \\
  \mbox{\it Conf}^{\:(+,\downarrow)}(S^1\times{\Bbb R}) &\longrightarrow
     & \mbox{\it Diff}^{\,-\leftrightarrow}\left(S^1\amalg S^1\right) \\
\end{array} \right.
$$
is a non-trivial $\Bbb Z$-covering. On the other hand, no canonical
l-c-d tessellations exist for $S^1\times{\Bbb R}$ and $\varphi_2$ is
vacuous.

For all other topologies, while $\varphi_1$ may not be surjective (this
can be seen particularly using the concept of "grafting" (cf.\ Sec.\
2.3)), $\varphi_2$ is always surjective since any surface automorphism
of $\Sigma$ that preserves the l-c-d tesselllation can be isotoped into
a coarse conformal one. Its kernel is isomorphic to
$\mbox{\it Diff}_{\,0}
  (\Sigma_0/\mbox{\raisebox{-.6ex}{${\cal F}_L$}})
    \times\mbox{\it Diff}_{\,0}
       (\Sigma_0/\mbox{\raisebox{-.6ex}{${\cal F}_R$}})$ via $\varphi_1$.
It is contractable and hence is the identity component of
$\mbox{\it Conf}\,^{(c)}(\Sigma)$.

For $\Sigma$ non-orientable, one has likewise the surjective group
homomorphism $\varphi_2$
$$
 \varphi_2 \;:\; \mbox{\it Conf}\,^{(c)}(\Sigma)\; \longrightarrow\;
             \mbox{\it Aut}\,(\mbox{\rm Tessln}_{\Sigma})\,.
$$
Its kernel is isomorphic to
$\mbox{\it Diff}_{\,0}\left(
   \Sigma^{\rm ornt}_0/\mbox{\raisebox{-.6ex}{${\cal F}$}}\right)$
and is the identity component of $\mbox{\it Conf}\,^{(c)}(\Sigma)$.

\bigskip

\noindent{\it Remark 2.2.2.} For $\Sigma$ not a cylinder,
$\mbox{\it Aut}\,(\mbox{\rm Tessln}_{\Sigma})$ is finite. (In fact,
since for $[f]$ in $\mbox{\it Aut}\,(\mbox{\rm Tessln}_{\Sigma})$,
$f$ takes a singularity to a singularity and its restriction to a tile
determines the whole isotopy class $[f]$ by tilewise continuation, the
order of $\mbox{\it Aut}\,(\mbox{\rm Tessln}_{\Sigma})$ is bounded by
$2\cdot(-8\chi(\Sigma))$, where $-8\chi(\Sigma)$ is the bound counted
with multiplicity for the number of tiles that are adjacent to some
singularity and factor $2$ is due to a possible flip under $f$.)
Comparing the theory of Riemann surfaces, it is instructive to
regard $\mbox{\it Aut}\,(\mbox{\rm Tessln}_{\Sigma})$ as the {\it true}
symmetry group of a Lorentz surface. The infinite dimensional identity
component
$\mbox{\it Diff}_{\,0}
       (\Sigma_0/\mbox{\raisebox{-.6ex}{${\cal F}_L$}})
    \times\mbox{\it Diff}_{\,0}
          (\Sigma_0/\mbox{\raisebox{-.6ex}{${\cal F}_R$}})$ or
$\mbox{\it Diff}\left(\Sigma^{\rm ornt}_0/
                  \mbox{\raisebox{-.6ex}{${\cal F}$}}\right)$
of $\mbox{\it Conf}\,^{(c)}(\Sigma)$ reflects the {\it local
non-rigidity} of a Lorentzian conformal map in two dimensions. This is
contrasted by the local rigidity of a holomorphic map in the Riemannian
case. Since $\mbox{\it Aut}\,(\mbox{\rm Tessln}_{\Sigma})$ is finite,
elements in $\mbox{\it Conf}\,^{(c)}(\Sigma)$ are {\it of periodic
type} in the Nielsen-Thurston's classification of surface automorphisms.
This is a parallel to the fact that automorphisms of Riemann surfaces of
negative Euler characteristic are also of periodic type and they form a
finite group.

\bigskip

\noindent
{\it Remark 2.2.3.} The algebra
$\mbox{\rm Vect}_{\,\supbscriptsizeBbb C}(S^1)$
of complex-valued smooth vector fields on the circle and its central
extensions, {\it Virasoro algebras}, have been of importance to string
theory. Their generalizations to a Riemann surface have been studied and
led to some {\it algebras of Virasoro type} ([K-N1 - 3], [J-K-L],
[M-N-Z]). In the current Lorentzian setting, one may regard
$\mbox{\it Conf}\,^{(c)}(\Sigma)$ as a generalization of
$\mbox{\it Diff}\,(S^1)$ and central extensions of its complexified Lie
algebra as other algebras of Virasoro type. Assume $\Sigma$ has $m$
past- and $n$ future-ends. By restricting automorphisms to the ends of
$\Sigma$, one has the following double inclusion:
$$
\begin{array}{cccl}
    & \hspace{-1em}\mbox{\it Conf}\,^{(c)}(\Sigma) & & \\
  \hspace{8em}\swarrow  & & \hspace{-8em}\searrow  & \\
  \prod_m \mbox{\it Conf}\,(S^1\times{\Bbb R})
          &  & \prod_n \mbox{\it Conf}\,(S^1\times{\Bbb R})  &.
\end{array}
$$
A representation of such diagrams or their central extensions into the
category of Hilbert spaces should give a picture of how a $\Sigma$
selects incoming past states, how it transmutes them, and then produces
outgoing future states. Unfortunately, we are still far from realizing
this goal.

\bigskip

\subsection{Rompers decompositions and coarse moduli spaces.}
Pants decompositions have played important roles in understanding
Riemann surfaces. We shall now discuss their Lorentzian analogue and
use it to understand the coarse moduli spaces of Lorentz surfaces.

\bigskip

\begin{flushleft}
{\bf Rompers decompositions of an s-d-l Lorentz surface.}
\end{flushleft}
A loop $C$ in an s-d-l Lorentz surface $\Sigma$ is called {\it achronal}
if $I^+(C)\cap C$ is empty, in other words if there are no two points of
$C$ with timelike separation ([H-E]) ({\sc Figure 2\,-\,4}). A loop in
$\Sigma$ is called {\it peripheral} if it can be homotoped into either
a boundary component or an end of $\Sigma$; otherwise it is called
{\it non-peripheral}.
\begin{figure}[htbp]
\setcaption{{\sc Figure 2\,-\,4.}  Achronal and non-achronal spacelike
     simple loops in an s-d-l Lorentz surface $\Sigma$. Notice that a
     non-achronal one could travel in a complicated way in $\Sigma$. }
\centerline{\psfig{figure=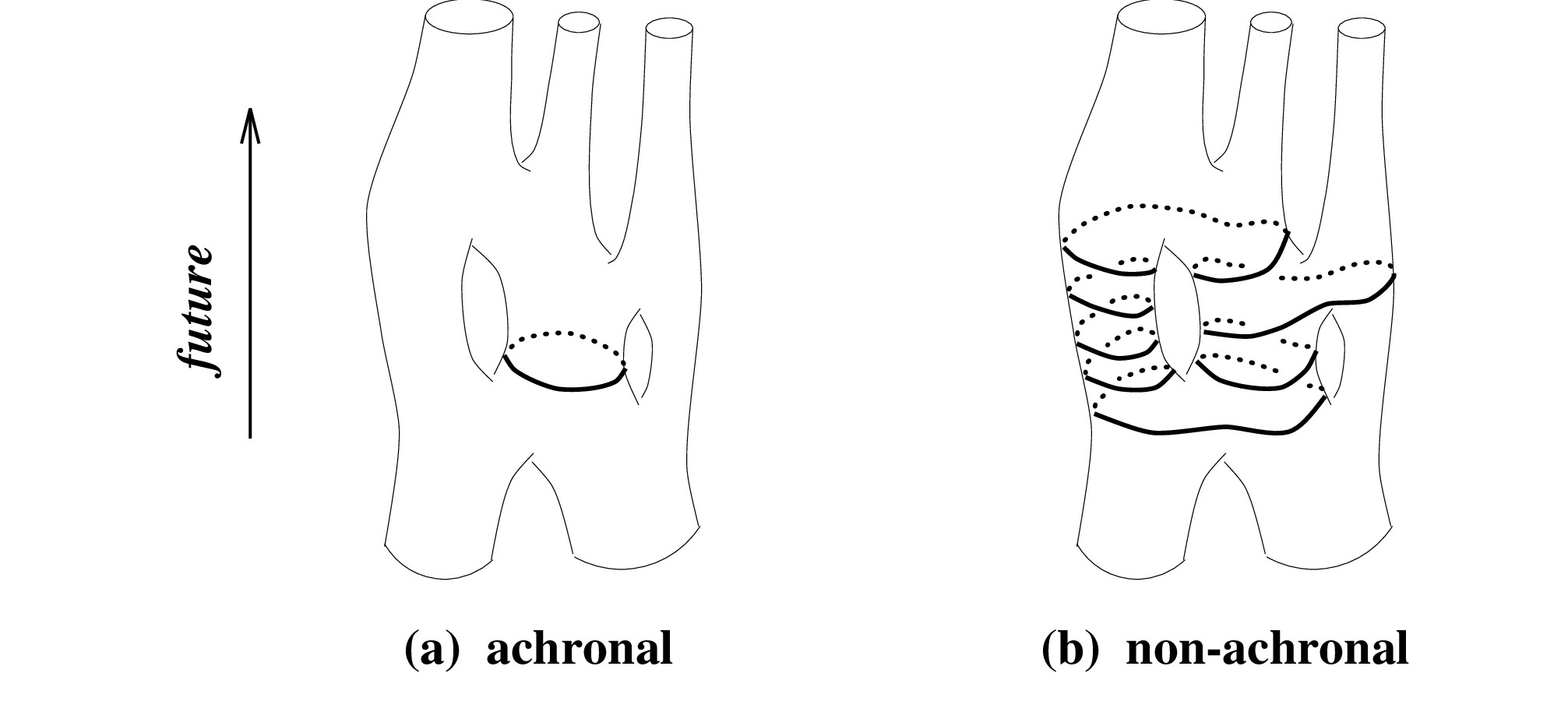,width=13cm,caption=}}
\end{figure}

\bigskip

\noindent{\bf Lemma 2.3.1 [simplest loop].} {\it
 Let $\Sigma$ be an s-d-l Lorentz surface that has more than one
 singularity. Then there exists an achronal spacelike simple loop $C$
 in $\Sigma$ that is non-peripheral. Furthermore, there are only
 finitely many free homotopy classes of such loops.  }

\bigskip

\noindent We shall call a non-peripheral achronal spacelike simple loop
in $\Sigma$ a {\it simplest} loop.

\bigskip

\noindent{\it Proof.} Since there are only finitely many singularities
in $\Sigma$, the no-causal-loop condition implies that there exists a
singularity $p_0$ whose causal past $J^-(p_0)$ contains no
singularities. Let $\cal N$ be a closed neighborhood of $J^-(p_0)$ in
the form of a submanifold-with-boundary in $\Sigma$ that contains only
the singularity $p_0$ and is deformation retractable to $J^-(p_0)$.

Since both ${\cal N}$ and $\Sigma-{\cal N}$ contain a singularity, the
future boundary $\partial^+{\cal  N}$ of $\cal N$ contains at least a
simple loop $C_0$ that is non-peripheral. As a boundary of a submanifold
in $\Sigma$, $C_0$ has an orientable neighborhood. Since $\cal N$ is
deformation retractable to $J^-(p_0)$, the complement
${\cal N}-J^-(p_0)$ is a collection of annuli $A_i$. The completion
$\overline{A_i}$ of $A_i$ (with respect to the topology of $\Sigma$)
contains as the past boundary component a broken null loop alternating
between future and past directions. This is the boundary shared with
$J^-(p_0)$. Such a boundary loop can be isotoped into a spacelike
loop in $A_i$. The latter in turn is isotopic to the future boundary
component of $A_i$ that serves also as a future boundary component of
$\cal N$. Applying this to the $A_i$ that has $C_0$ as the future
boundary, one then obtains a simple spacelike loop $C$. Since $C$ comes
from perturbing a broken null loop in $\partial J^-(p_0)$, which has to
be achronal, $C$ itself has empty $I^-(C)\cap C$. This shows that $C$ is
achronal.

To see that there are only finitely many free homotopy classes of such
$C$, first notice that any achronal spacelike loop that crosses a tile
in the l-c-d tessellation of $\Sigma$ that lies far enough in an end has
to be peripheral. Consequently, there are only finitely many tiles that
can admit some non-peripheral achronal spacelike simple loops to cross
them. Up to free homotopy, we may assume that all the loops considered
do not hit the corners of any tiles. If $C$ is achronal, then it can
pass through a tile at most once. Any two such loops with the same
pattern of crossing the edges of tiles are free homotopic; and there
are only finitely many such patterns since there are only finitely
admissible tiles.

This completes the proof.

\noindent\hspace{14cm}$\Box$

\bigskip

Notice that a spacelike simple loop $C$ in an s-d-l $\Sigma$ cannot be
null-homotopic. Its tubular neighborhood is always a cylinder; and the
complement $\Sigma-C$ remains s-d-l (after appropriately bordered).
By applying the above lemma finitely many times with slight modification
in the proof to take $\Sigma$ with spacelike boundary also into account,
one then has

\bigskip

\noindent{\bf Corollary 2.3.2 [simple cut system].} {\it
 Let $\Sigma$ be an s-d-l Lorentz surface that has more than one
 singularity. Then there exists a system of achronal spacelike simple
 loops $C_{\alpha}$ such that the complement
 $\Sigma-\{C_{\alpha}\}_{\alpha}$ is a collection of s-d-l Lorentz
 surfaces that have exactly one singularity.}

\bigskip

\noindent We shall call such $\{C_{\alpha}\}_{\alpha}$ a {\it simple
cut system} of $\Sigma$. Two such systems are {\it equivalent} if they
differ by a homotopy of loops. The following lemma characterizes the
component of its complement.

\bigskip

\noindent{\bf Lemma 2.3.3 [s-d-l with one singularity].} {\it
 Let $\Sigma$ be an s-d-l Lorentz surface (without boundary) that has
 only one singularity. Then $\Sigma$ is topologically a sphere with some
 $m+n\geq 3$ punctures, where $m$ and $n$ are the number of past- and
 future-ends respectively. All positive integer pairs $(m,n)$ with
 $m+n\geq 3$ are admissible; and, up to coarse conformal equivalences,
 there are only finitely many such $\Sigma$ for each admissible
 $(m,n)$.    }

\bigskip

\noindent{\it Proof.} Let $p$ be the singularity. Let us first show
that $\Sigma$ must be orientable. Since there is only one singularity,
no-causal-loop condition implies that every characteristic strip
$\Omega$ of $\Sigma$ must be of the either form in
{\sc Figure 2\,-\,5\,}(a). The singularity $p$ as appears exactly once
on each side of $\Omega$ must be of spacelike separation in
$\Omega$. On the other hand, every loop at $p$ can be deformed into a
product of loops at $p$ that lie completely in some characterristic
strips. These two observations imply that one can choose a set of
generators for $\pi_1(\Sigma,p)$ that consists only of spacelike loops.
Such loops must have orientable tubular neighborhood. Consequently,
$\Sigma$ is orientable.
\begin{figure}[htbp]
\setcaption{{\sc Figure 2\,-\,5.} Admissible characteristic strips on a
      Lorentz surface with only one singularity.   }
\centerline{\psfig{figure=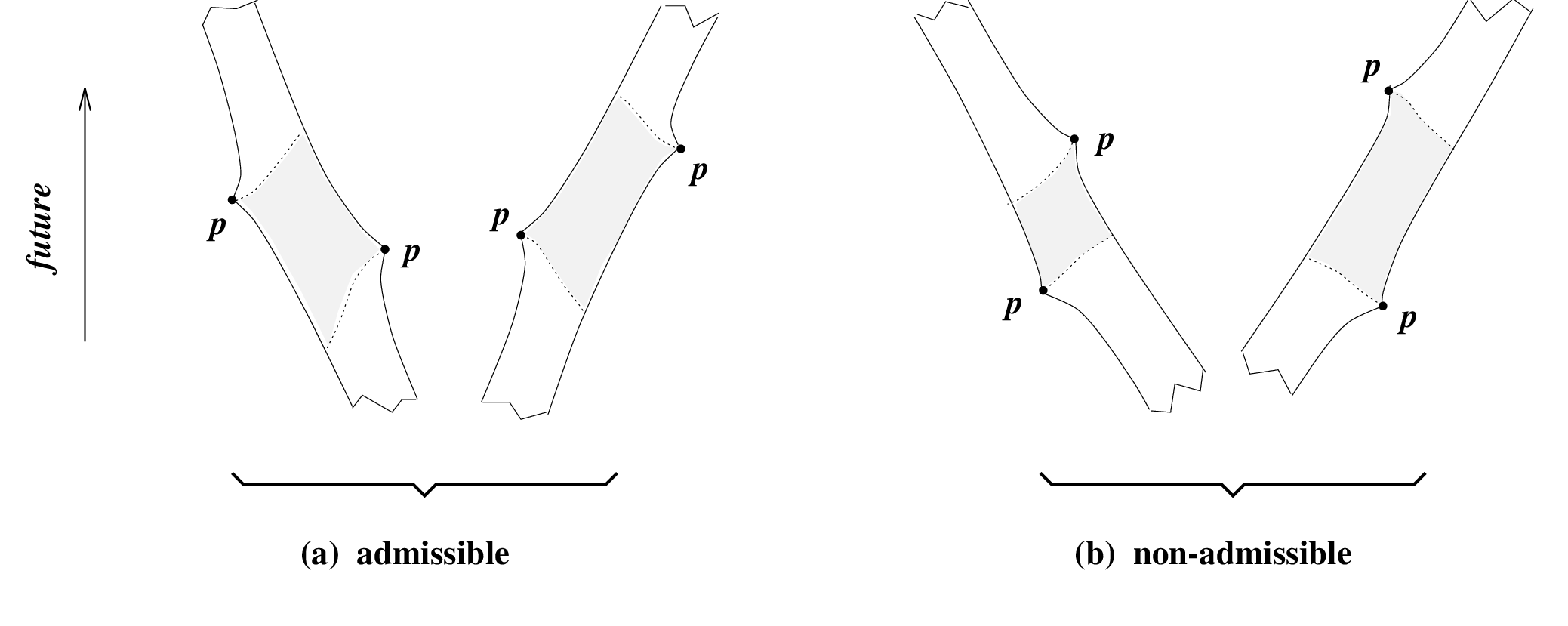,width=13cm,caption=}}
\end{figure}

Let $\Sigma$ now be oriented. Then one may reconstruct it from, say, its
past-left crowns. Since $\Sigma$ has only one singularity, each of the
crowns with their own characteristic leaves of the restricted Lorentz
structure is as indicated in {\sc Figure 2\,-\,6}. To get back $\Sigma$,
one has to identify some future bordering characteristic leaf labelled
by $A$ to another labelled by $B$. There are only finitely many such
pairings. Each pairing determines the topology of the resulting
2-complex with a tessellation by extending the characteristic leaves
following the causal structure of the crowns.
For a pairing that gives a connected manifold structure, the
resulting topology has to be a punctured sphere. The existence of
singularity $p$ implies that the total number of ends has to be at
least three. The coarse conformal structure is also determined by the
pairing since different ways of pasting along the same paired
characteristic leaves will give the same l-c-d tessellation pattern.
{\sc Figure 2\,-\,7} indicates how one can construct such Lorentz
surface with arbitrary $m$ past- and $n$ future-ends as long as
$m+n\geq 3$.

This completes the proof.

\begin{figure}[htbp]
\setcaption{{\sc Figure 2\,-\,6.} A bordered past-left crown that
       appears in an s-d-l Lorentz surface with one singularity. For
       clarity, it is drawn as a planar domain.   }
\centerline{
   \psfig{figure=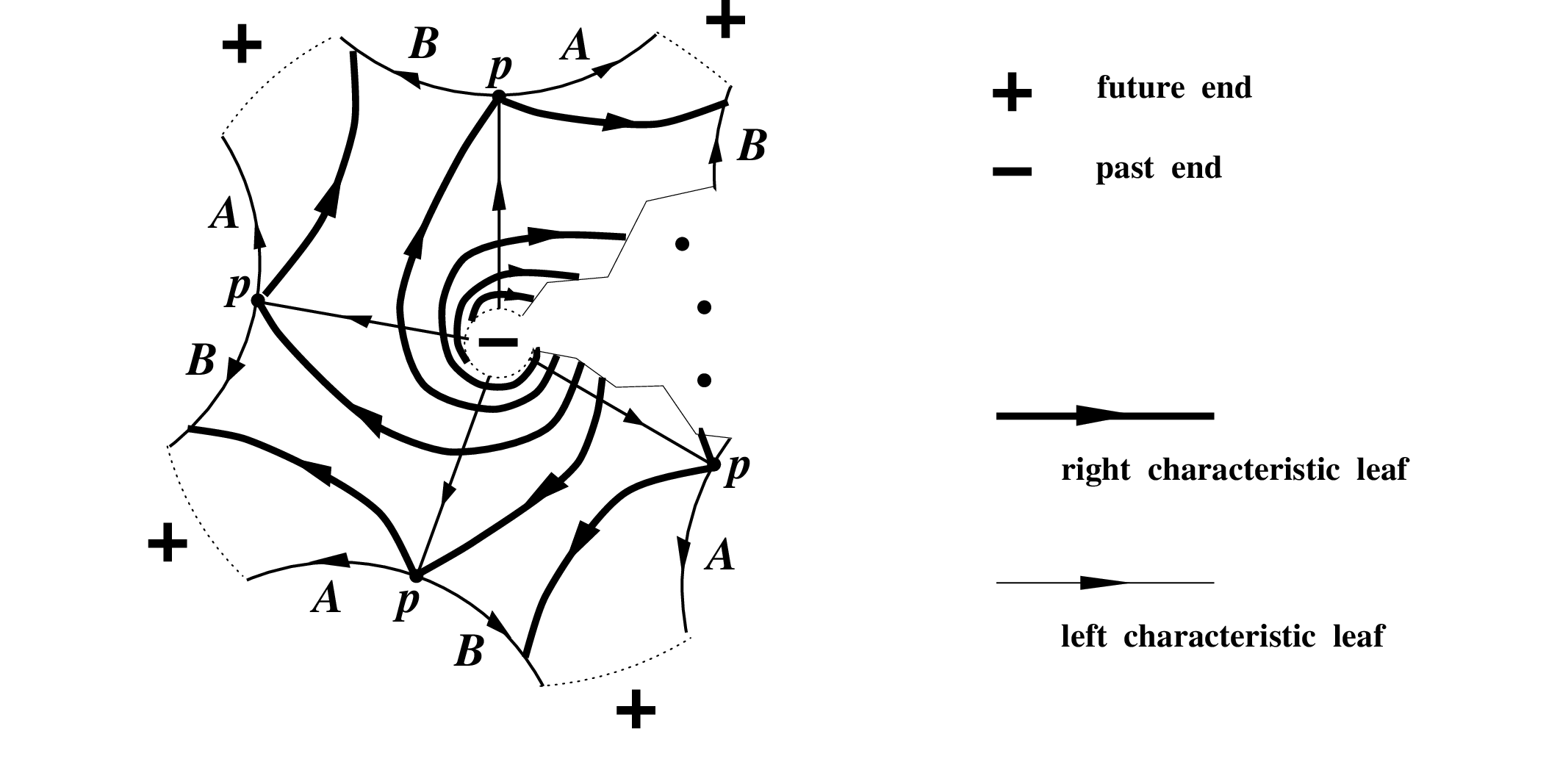,width=13cm,caption=}}
\end{figure}
\begin{figure}[htbp]
\setcaption{{\sc Figure 2\,-\,7.} By cut-and-paste with
       $S^1\times{\Bbb R}$ along a timelike ray at the singularity, one
       can obtain an s-d-l Lorentz surface with one singularity that has
       any $m$ past- and $n$ future-ends for $m+n\geq 3$. These surfaces
       are called {\it rompers}.  }
\centerline{
   \psfig{figure=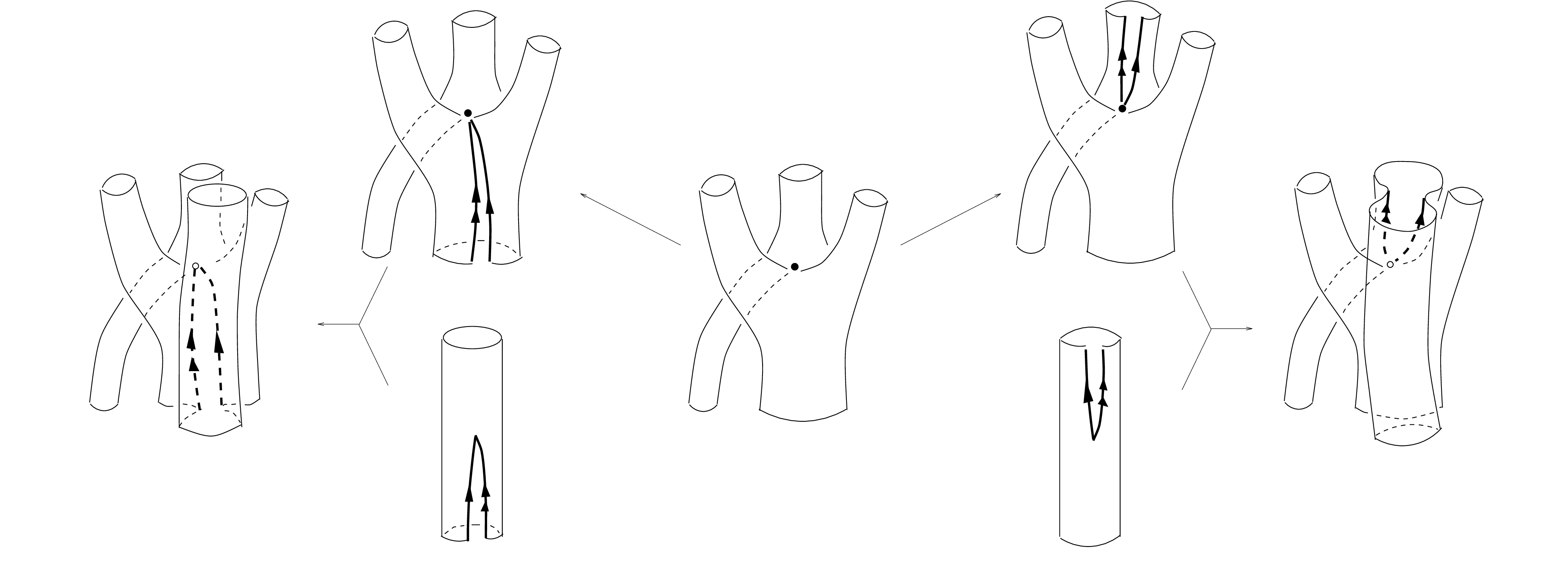,width=13cm,caption=}}
\end{figure}

\noindent\hspace{14cm}$\Box$

\bigskip

Due to its topology, we shall call an s-d-l Lorentz surface with
one singularity a set of {\it rompers} and $(m,n)$ its {\it type}. Each
component of the complement of a simple cut system of $\Sigma$ is a set
of rompers with ends truncated. And we shall call such decomposition a
{\it rompers decomposition}. Notice that since there are only finitely
many achronal spacelike simple loops up to homotopy, there are only
finitely many non-equivalent simple cut systems for $\Sigma$.

Rompers, and hence all Lorentz surfaces, admit foliations whose generic
leaves are simplest loops. By pinching these leaves, one then obtains a
{\it directed network}
$\mbox{\it Net}\,(\Sigma,\{C_{\alpha}\}_{\alpha})$ associated to a
rompers decomposition of $\Sigma$. We shall also called it a
{\it sewing-diagram} ({\sc Figure 2\,-\,9}).

\bigskip

\begin{flushleft}
{\bf Graftings and coarse moduli spaces.}
\end{flushleft}
%
\noindent{\bf Definition 2.3.4 [grafting].}
 Let $\Sigma$ be an s-d-l Lorentz surface and $C$ be a simplest loop
 in $\Sigma$. Recall that its tubular neighborhood is a cylinder.
 Define a {\it grafting} of $\Sigma$ along $C$ of {\it step} $k$,
 $k\in{\Bbb N}$, by cutting $\Sigma$ along $C$ and then pasting to it
 conformally the standard Minkowskian cylinder $S^1\times [0,2\pi k]$,
 as illustrated in {\sc Figure 2\,-\,8}. This leads to a new s-d-l
 Lorentz surface, denoted by $(\Sigma,C;k)$. The reverse of this shall
 be called an {\it excision} of step $k$.
\begin{figure}[htbp]
\setcaption{{\sc Figure 2\,-\,8.} Grafting of an s-d-l Lorentz surface
       along a simplest loop $C$. }
\centerline{\psfig{figure=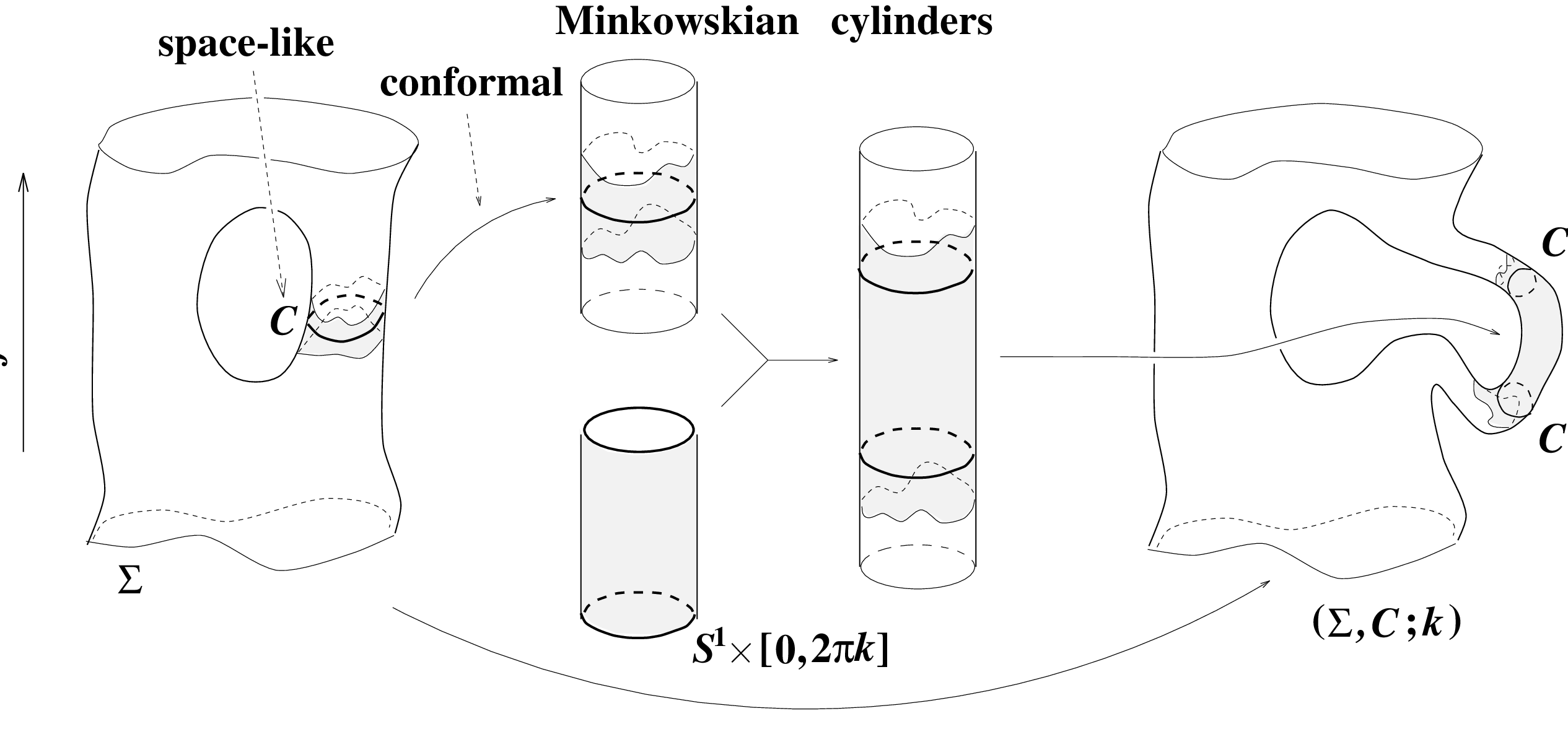,width=13cm,caption=}}
\end{figure}
%

\bigskip

Notice that a grafting on $\Sigma$ determines uniquely a conformal
structure on the new surface $\Sigma^{\prime}$. The effect of grafting
along $C$ on $\Gamma_L$ and $\Gamma_R$ (or, with suitable modification,
$\Gamma(\Sigma)$ if $\Sigma$ non-orientable) is to make their edges
that cross $C$ wind more; but it leaves their topologies unchanged. The
identity component of $\mbox{\it Conf}\,^{(c)}(\Sigma^{\prime})$ is
canonically isomorphic to that of $\mbox{\it Conf}\,^{(c)}(\Sigma)$ by
the restriction map from $\Sigma^{\prime}$ to $\Sigma$.

\bigskip

\noindent{\it Remark 2.3.5.} We define graftings only for $k\in{\Bbb N}$
 so that the l-c-d tessellation remains the same on the original part of
 the surface. It should be noted that a grafting for $k\in{\Bbb R}_+$
 is also well-defined in exactly the same way.

\bigskip

Let ${\cal M}_{\rm Lorz}^{(c)}(\pm,\hat{\chi},m,n)$ be the moduli space
of s-d-l (coarse) Lorentz surfaces of type $(\pm,\hat{\chi},m,n)$. It is
a discrete set. Using grafting, one can define a relation $\prec$ on
${\cal M}_{\rm Lorz}^{(c)}(\pm,\hat{\chi},m,n)$ by setting
$\Sigma\prec\Sigma^{\prime}$ if $\Sigma^{\prime}$ can be obtained by a
finite sequence of graftings along {\it simplest} loops beginning with
$\Sigma$. (We shall say that "{\it $\Sigma$ precedes $\Sigma^{\prime}$}"
or that "{\it $\Sigma^{\prime}$ follows $\Sigma$}".) This defines a
partial ordering on ${\cal M}_{\rm Lorz}^{(c)}(\pm,\hat{\chi},m,n)$. We
shall call a minimal element relative to $\prec$ a {\it primitive}
Lorentz surface.

\bigskip

\noindent{\bf Proposition 2.3.6 [primitive finite].} {\it
 There are only finitely many minimal elements in
 $\left({\cal M}_{\rm Lorz}^{(c)}(\pm,\hat{\chi},m,n),\prec\right)$. }

\bigskip

\noindent{\it Proof.} Let $\Sigma$ be a Lorentz surface in
${\cal M}_{\rm Lorz}^{(c)}(\pm,\hat{\chi},m,n)$. Then any of the
components of its rompers decompositions must have Euler characteristic
negative but greater than $\hat{\chi}-(m+n)$. There are only finitely
many of them. Up to grafting and excision, their different ways of
pasting will lead only to finitely many different l-c-d tessellation
patterns on Lorentz surfaces of type $(\pm,\hat{\chi},m,n)$. This proves
the proposition.

\noindent\hspace{14cm}$\Box$

\bigskip

To capture its geometry, it is instructive to define a directed graph
structure on ${\cal M}_{\rm Lorz}^{(c)}(\pm,\hat{\chi},m,n)$. The vertex
set is the moduli space itself. $\Sigma_1$ and $\Sigma_2$ is connected
by a directed edge $(\Sigma_1,\Sigma_2)$ if $\Sigma_2$ is obtained from
$\Sigma_1$ by a grafting of step $1$ along a simplest loop. This is an
infinite graph with finitely many source but no sink vertices. Its
valence at a vertex $\Sigma$ is bounded by twice the number of homotopy
classes of simplest loops in $\Sigma$. In general, points in
${\cal M}_{\rm Lorz}^{(c)}(\pm,\hat{\chi},m,n)$ are not related just by
graftings and excisions; and hence this graph can have several
components. Roughly, the set of rays in the directed graph gives a
compactification of the moduli space. The geometry associated to an
added point is an s-d-l Lorentz surface obtained from a $\Sigma$ in
${\cal M}_{\rm Lorz}^{(c)}(\pm,\hat{\chi},m,n)$ by cutting along some
collection of simplest loops and then attach an $S^1\times [0,\infty)$
or $S^1\times (-\infty,0]$ to each pair of the newly created boundary
components ({\sc Figure 2\,-\,10}).

\bigskip

\noindent{\it Remark 2.3.7.} Technically, a ray can have more than one
 limit geometry; hence the above prescription of compactification
 needs to be refined. The fine moduli space
 ${\cal M}_{\rm Lorz}^{(f)}(\pm,\hat{\chi},m,n)$ is stratified by the
 l-c-d tessellation patterns of Lorentz surfaces of type
 $(\pm,\hat{\chi},m,n)$; since in addition graftings can be defined for
 all ${\Bbb R}_+$ (cf.\ Remark 2.3.5), one conceives that the directed
 graph for ${\cal M}_{\rm Lorz}^{(c)}(\pm,\hat{\chi},m,n)$ can be
 embedded in ${\cal M}_{\rm Lorz}^{(f)}(\pm,\hat{\chi},m,n)$
 uniformally, picking one point for the image of a vertex in each
 stratum.

\bigskip

\noindent{\it Remark 2.3.8.} When compared with the Fenchel-Nielsen
 coordinates for the Teichm\"{u}ller and moduli spaces of Riemann
 surfaces, the set of primitive elements in
 ${\cal M}_{\rm Lorz}^{(c)}(\pm,\hat{\chi},m,n)$ is a parallel to a
 coordinate-plane for the parameters of twisting angles, while the step
 of graftings is a parallel to the inverse of the length parameters.

\bigskip

\begin{flushleft}
{\bf Sketch of examples.}
\end{flushleft}
To illustrate the ideas, let us give a brief sketch of how to build the
coarse moduli spaces ${\cal M}_{\rm Lorz}^{(c)}(+,-2,1,1)$ and
${\cal M}_{\rm Lorz}^{(c)}(-,-2,1,1)$. For $\Sigma$ of type
$(\pm,-2,1,1)$, by considering first pants decompositions and then their
degenerates, one finds that the rompers that build $\Sigma$ can only be
of types $(1,3)$, $(1,2)$, $(2,2)$, $(2,1)$, and $(3,1)$. They together
create seven admissible sewing-diagrams ({\sc Figure 2\,-\,9}).
\begin{figure}[htbp]
\setcaption{{\sc Figure 2\,-\,9} Admissible rompers and sewing-diagrams
    for an s-d-l Lorentz surface of type $(\pm,-2,1,1)$. Characteristic
    leaves in each rompers and degeneracy relations between diagrams are
    also indicated.     }
\centerline{\psfig{figure=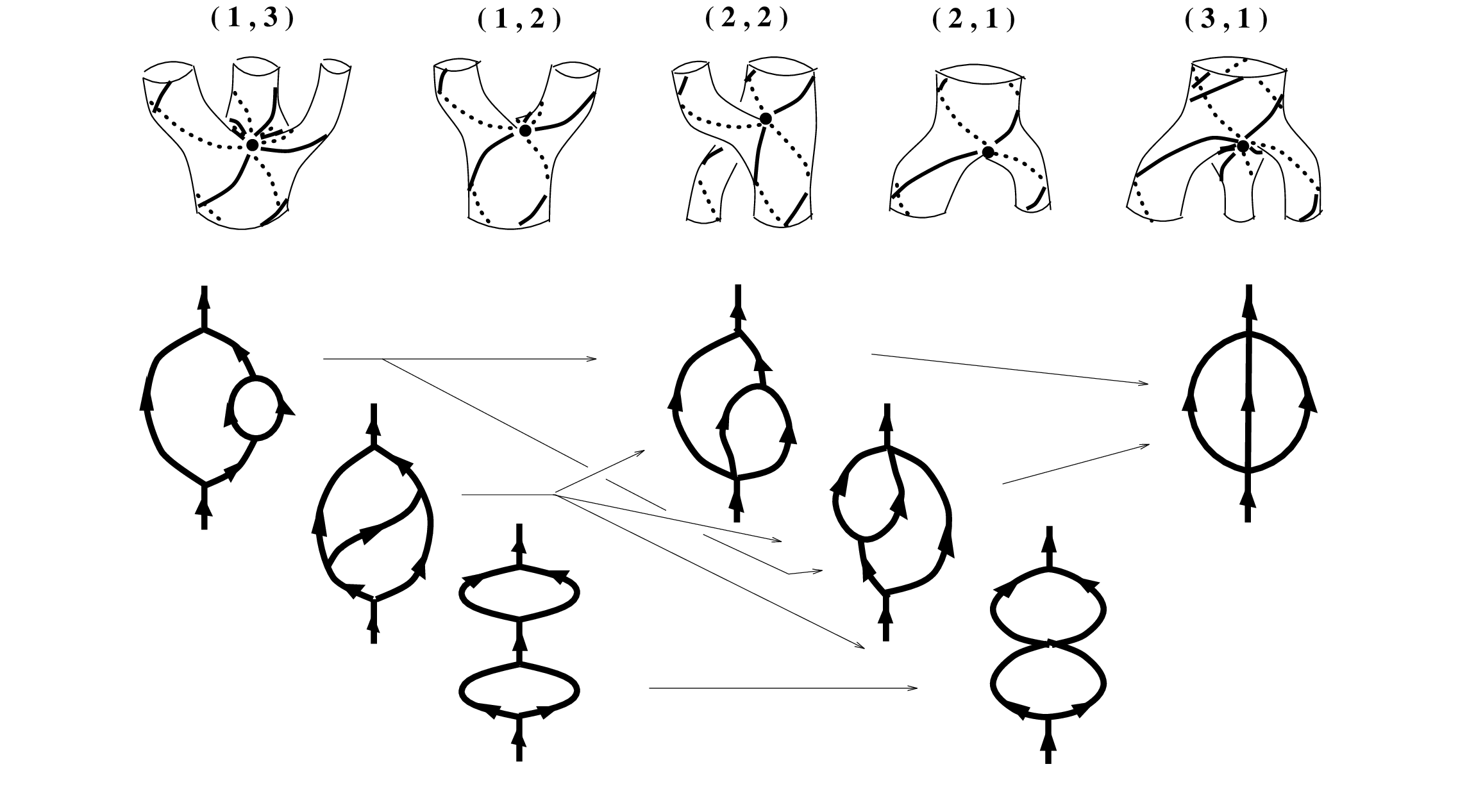,width=13cm,caption=}}
\end{figure}

For these simpler types, each $(m,n)$ gives only one set of rompers. We
will think of them as Mandelstam diagrams (cf.\ Sec.\ 3.1 and
Remark 3.1.3). To build a primitive elememt, we truncate these
Mandelstam diagrams by equal-time curves so that, after truncation, any
collar of the boundary that is singularity-free and has spacelike
boundary contains no larger cylinders than $S^1\times [0,2\pi)$. Twists
(and also flips for ${\cal M}_{\rm Lorz}^{(c)}(-,-2,1,1)$) when sewing
truncated rompers following an admissible sewing-diagram will create a
collection of Lorentz surfaces of type $(\pm,-2,1,1)$. After sewn, the
characteristic leaves on each set of rompers extend to those on the
whole surface by following null leaves in other rompers. After excisions
of step at most $1$ along sewing loops, one obtains all the primitive
elements. Together with the homotopy classes of simplest loops thereon,
one can then build the whole moduli space. Each simple cut system (or,
equivalently, sewing-pattern) gives a Fenchel-Nielsen type coordinate
chart on the moduli space by the step number of graftings. These charts
look like a collection of lattice cones and they overlap at most
marginally at where some grafting step remains small.
({\sc Figure 2\,-\,10}.)
\begin{figure}[htbp]
\setcaption{{\sc Figure 2\,-\,10.} Directed graph gives a way to
     describe the geometry of
     ${\cal M}_{\rm Lorz}^{(c)}(\pm,\hat{\chi},m,n)$ and its
     compactification. Only some components of
     ${\cal M}_{\rm Lorz}^{(c)}(+,-2,1,1)$ are shown sketchily with
     the Fenchel-Nielsen type of coordinate charts. Due to symmetry,
     there can be more redundancy of the coordinates, e.g.\
     $01010=10001$, $01110=10101$, etc.. Two examples of the changes of
     the geometry along a ray and their limit are also indicated. }
\centerline{\psfig{figure=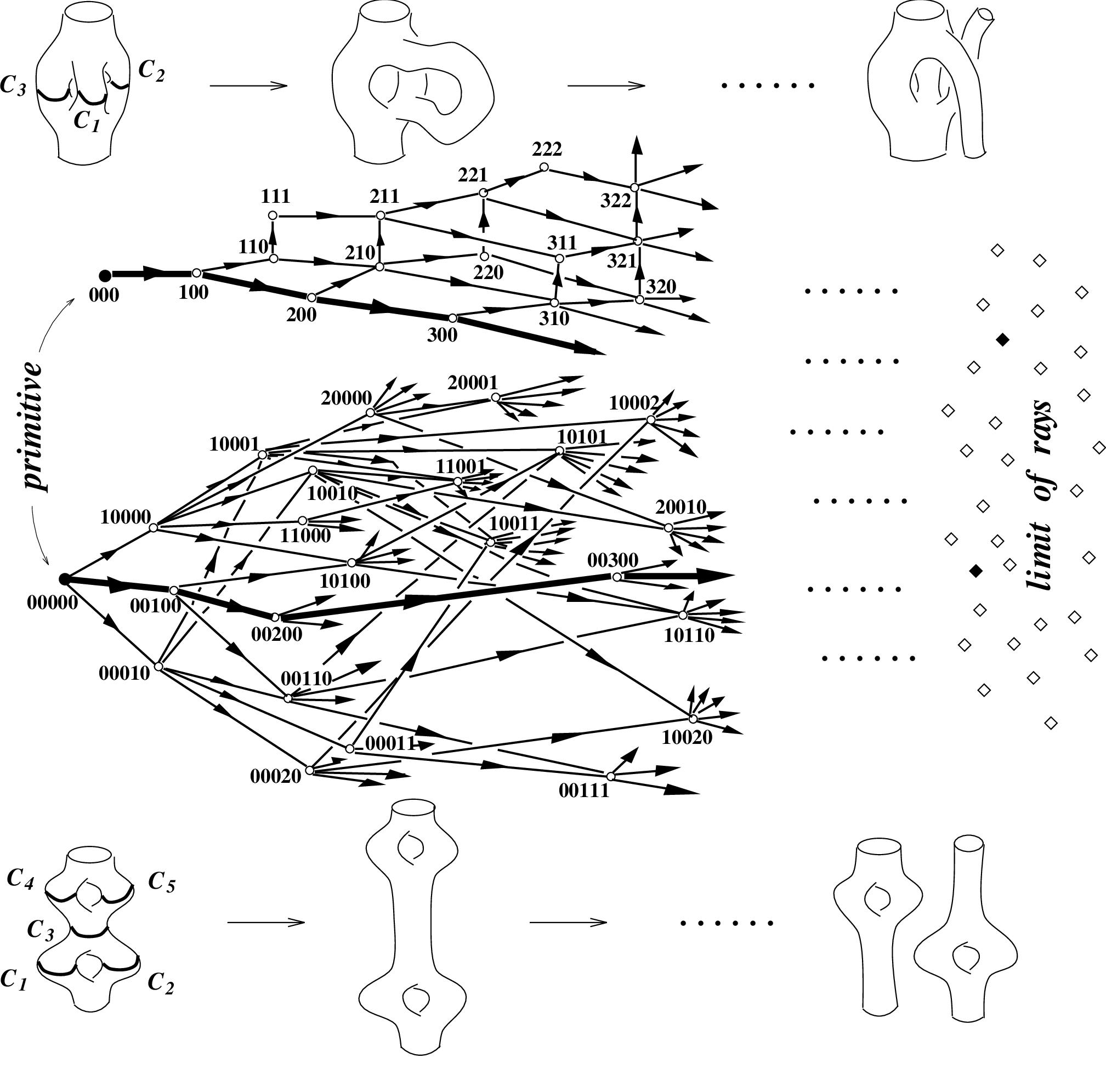,width=13cm,caption=}}
\end{figure}

We omit the details here since it is tedious. We would like to know
if there are more efficient ways to understand these moduli spaces.

\bigskip

Though the discussion here on the moduli spaces of Lorentz surfaces is
far from enough nor complete, we shall be contented to stop here. These
moduli spaces and their asymptotic behaviors when $(\hat{\chi},m,n)$ get
large should be important for string theory.

\bigskip

\section{Rectifiability into Mandelstam diagrams.}
Mandelstam diagrams are related to both Hamiltonian and light-cone
string theory. In the Riemannian case, it is known that a Riemannian
2-manifold with at least two punctures is coarse conformal to a
Euclidean Mandelstam diagram [G-W]. In contrast to this, we shall show
in this section that the analogue does not hold for the Lorentzian case.

\bigskip

\subsection{\bf Branched coverings, positive cones, and rectifiability.}
An s-d-l Lorentz surface $\Sigma$ is called {\it rectifiable} if it has
a Mandelstam diagram $\Xi$ as a representative. The following
proposition characterizes rectifiability.

\bigskip

\noindent{\bf Proposition 3.1.1 [rectifiability].} {\it Let $\Sigma$ be
 an s-d-l Lorentz surface. Then (1), for $\Sigma$ orientable,
 $\Sigma$ is rectifiable if and only if $\Sigma$ is a coarse-conformal
 branched covering over a Minkowskian cylinder; (2), for $\Sigma$
 non-orientable, $\Sigma$ is rectifiable if and only if its
 orientation covering $\Sigma^{\rm ornt}$ with the lifted coarse
 conformal structure is rectifiable.        }

\bigskip

\noindent
{\it Remark 3.1.2.} When $\Sigma$ is orientable and rectifiable, coarse
conformalness implies that the branched points over the cylinder are
exactly the singularities of $\Sigma$.

\bigskip

\noindent{\it Proof of Proposition 3.1.1.} We shall assume that
$\chi(\Sigma)<0$ since otherwise $\Sigma$ is an s-d-l cylinder and the
proposition is readily true. Recall that the standard cylinder
$S^1\times{\Bbb R}$ is given by
${\Bbb R}^{1+1}/$\raisebox{-.6ex}{${\Bbb Z}v_0$} with $v_0=(2\pi,-2\pi)$
in the light-cone coordinates.

Assume first that $\Sigma$ is generic and oriented. Recall that a
positive 1-form $\mu_L$ on
$\Sigma_0/\mbox{\raisebox{-.6ex}{${\cal F}_L$}}$ and a positive 1-form
$\mu_R$ on $\Sigma_0/\mbox{\raisebox{-.6ex}{${\cal F}_R$}}$ together
give a pair $(\mu_L,\mu_R)$ of measures on $\Sigma$ that are transverse
to each other. It defines a map $\mbox{Hol}_0$ from the based path
space $\mbox{\it Path}\,(\Sigma, q_0)$ to ${\Bbb R}^{1+1}$ by
$$
 \mbox{Hol}_0\,(\gamma)\;
       =\;(\int_{\gamma}\mu_L\,,\,\int_{\gamma}\mu_R)\,.
$$
Due to homotopy invariance, $\mbox{Hol}_0$ descends to a map
$\mbox{Hol}_1$ from the universal covering $\widetilde{\Sigma}$ of
$\Sigma$ to ${\Bbb R}^{1+1}$. It descends further to a branched
covering map from $\Sigma$ to the cylinder
${\Bbb R}^{1+1}/$\raisebox{-.6ex}{${\Bbb Z}v_0$} if and only if
$$
 \hspace{-8em}(\ast)\mbox{ [covering] }\hspace{3em}
  \mbox{Hol}_0(\gamma)\,\in\,{\Bbb Z}\cdot v_0\,,\hspace{.5em}
    \mbox{for all $\gamma\in\pi_1(\Sigma,q_0)$}\,.
$$
This condition contains two parts:
$$
 \hspace{-8.2em}(\ast_1)\mbox{ [slope] }\hspace{5em}
  \frac{\int_{\gamma}\mu_R}{\int_{\gamma}\mu_L}\; =\; -1\,,
   \hspace{.5em}\mbox{for all $\gamma\in\pi_1(\Sigma,q_0)$}\,;
$$

\vspace{-2ex}
\noindent and
\medskip
$$
 \hspace{-8em}(\ast_2)\mbox{ [integral] } \hspace{4em}
     \int_{\gamma}\mu_L\,\in\,2\pi{\Bbb Z} \,,\hspace{.5em}
        \mbox{for all $\gamma\in\pi_1(\Sigma,q_0)$}\,.
$$

Observe that $(\Sigma,\mu_L\cdot\mu_R)$, where "$\cdot$" is the
symmetric product, is a representative of $\Sigma$. The slope condition
$(\ast_1)$ means that the 1-form $\mu_L+\mu_R$ is exact. Its integral
$t$ is then a global time function on $\Sigma$ whose level curves are
transverse to both leaves of ${\cal F}_L$ and ${\cal F}_R$. The equal
time trajectories through singularities thus give an annuli
decomposition of $\Sigma$ and the $\mbox{Hol}_0$-type maps on these
annuli send them to standard cylinders in a way that meets the
requirements for a Mandelstam diagram. This shows that $\Sigma$ is
rectifiable if and only if there exists a positive $(\mu_L,\mu_R)$ that
satisfies Cond.\,$(\ast_1)$.

Let $\{E^L_i\}_{i=1}^k$, $\{E^R_i\}_{i=1}^k$ be the set of directed
edges of $\Gamma_L$, $\Gamma_R$ respectively. Let
$\{\gamma\}_{i=1}^{k_0}$ be a set of generators of $\pi_1(\Sigma,q_0)$.
One has $k=-2\chi(\Sigma)$ and $k_0=-\chi(\Sigma)+1$. Cond.\,$(\ast_1)$
for a positive pair $(\mu_L,\mu_R)$ is equivalent to the existence of
a positive solution to the following system of linear equations
$$
 \sum_{j=1}^k\, a^R_{ij}x^R_j\;
     =\; -\sum_{j=1}^k\, a^L_{ij}x^L_j\;
           \hspace{1em}\mbox{for}\hspace{1em} i=1,\,\ldots,\,k_0\;,
$$
where $a^R_{ij}$ (resp.\ $a^L_{ij}$) is the {\it multiplicity} of
$\gamma_i$ with respect to $E^R_j$ (resp.\ $E^L_j$) defined as the
{\it signed number of times} that $\gamma_i$ would go over $E^R_j$
(resp.\ $E^L_j$) under the deformation retract that takes $\Sigma$ to
$\Gamma_R$ (resp.\ $\Gamma_L$) and the unknowns $x^R_j$ (resp.\
$x^L_j$) are the prospect values for $\int_{E^R_j}\mu_R$ (resp.\
$\int_{E^L_j}\mu_L$). Due to homogeniety of the system and integralness
of its coefficients, the set $\mbox{\it Sol}^{\,+}$ of positive
solutions to the system, if not empty, is a cone in the $2k-k_0$
dimensional solution space of the linear system in
${\Bbb R}^k\oplus{\Bbb R}^k$ with non-empty
$\mbox{\it Sol}^{\,+}
      \cap\left((2\pi{\Bbb Z})^k\oplus(2\pi{\Bbb Z})^k\right)$.
{}From this intersection, one obtains $(\mu_L,\mu_R)$ that satisfies
both Cond.\,$(\ast_1)$ and Cond.\,$(\ast_2)$; and hence
Cond.\,$(\ast)$. This shows that the existence of a positive
$(\mu_L,\mu_R)$ that satisfies the covering condition $(\ast)$ is
equivalent to the existence of a positive $(\mu_L,\mu_R)$ that
satisfies the slope condition $(\ast_1)$. This concludes the proof
of Part (1) for $\Sigma$ generic.

When $\Sigma$ is non-generic, recall that it can be obtained from a
generic $\Sigma^{\|}$ by squashing strips along transverse foliation.
Applying the above argument to $\Sigma^{\|}$ with the modification
that allows non-negative solutions to the linear system, whose zero
components correspond to the squashed strips, one obtains the same
result for $\Sigma$. This completes the proof of Part (1).

For $\Sigma$ non-orientable, since the orientation covering
$\Xi^{\rm ornt}$ of a Mandelstam diagram $\Xi$ with the lifted structure
is also a Mandelstam diagram, that $\Sigma$ is rectifiable implies that
$\Sigma^{\rm ornt}$ is also rectifiable. Assume now the converse that
$\Sigma^{\rm ornt}$ is rectifiable to a Mandelstam diagram
$\Xi^{\prime}$. Let $f$ from $\Xi^{\prime}$ to itself be the involution
that comes from the non-trivial deck transformation on
$\Sigma^{\rm ornt}$. Since the Lorentz structure on $\Sigma$ is simply
the average of that in $\Sigma^{\rm ornt}$ with respect to the
covering projection, we only need to show that the quotient
$\Xi=\Xi^{\prime}/$\raisebox{-.6ex}{$f$} with the averaged Lorentzian
structure from that of $\Xi^{\prime}$ under the locally conformal
quotient map is also a Mandelstam diagram.

Let $\Xi^{\prime}$ be oriented and $\mu_L^{\prime}$, $\mu_R^{\prime}$
be the left and right characteristic 1-forms. Recall that the
Lorentzian metric on $\Xi^{\prime}$ is then
$\mu_L^{\prime}\cdot\mu_R^{\prime}$. For a simply-connected region
$\Delta$ in $\Xi$, let $\Delta_1$, $\Delta_2$ be the two components
of its corresponding region in $\Xi^{\prime}$. The transverse pair of
local 1-forms on $\Sigma$
$$
 \left.\mu_1\right|_{\Delta}\;
     =\;\frac{1}{2}\left( \left.\mu_L^{\prime}\right|_{\Delta_1}
           + \left.\mu_R^{\prime}\right|_{\Delta_2}   \right)
        \hspace{.6cm}\mbox{and}\hspace{.6cm}
 \left.\mu_2\right|_{\Delta}\;
     =\;\frac{1}{2}\left( \left.\mu_R^{\prime}\right|_{\Delta_1}
           + \left.\mu_L^{\prime}\right|_{\Delta_2}   \right)
$$
cannot be globally well-defined since extending the local $\mu_1$,
$\mu_2$ along a loop with non-orientable neighborhood will turn
$\mu_1$, $\mu_2$ into each other. Nevertheless, the two mixed locally
defined objects,
$$
 \mu|_{\Delta}\;=\;\frac{1}{2}\left(\left.\mu_1\right|_{\Delta}
                  + \left.\mu_2\right|_{\Delta}\right)
   \hspace{.6cm}\mbox{and}\hspace{.6cm}
 dh^2|_{\Delta}\;=\;\left.\mu_1\right|_{\Delta}\,\cdot\,
                                   \left.\mu_2\right|_{\Delta}\,,
$$
can always be globally extended to a $\mu$ and $dh^2$. Exactness of
$\mu_L^{\prime}+\mu_R^{\prime}$ implies exactness of $\mu$; and its
integral gives a time function $t$ on $\Xi$ that can be realized as the
average of one on $\Xi^{\prime}$. Its level curves have to be transverse
to the bi-foliation on $\Xi$ coming from the projection of, say,
${\cal F}_L$ on $\Xi^{\prime}$. Hence, as in showing the relation of
Cond.\ $(\ast_1)$ to rectifiability, time level trajectories through
singularities give an annuli decomposition of $\Xi$ and the
$\mbox{\rm Hol}_0$-type maps using the local $(\mu_1,\mu_2)$, now
well-defined on each annulus, provide the required homotheties to
standard cylinders. This shows that $(\Xi, dh^2)$ is indeed a Mandelstam
diagram.

This concludes the proof of Part (2); and hence the proposition.

\noindent\hspace{14cm}$\Box$

\bigskip

\noindent{\it Remark 3.1.3.} Since an s-d-l Lorentz surface
 with only one singularity has spacelike generators for its fundamental
 group, the slope condition can be satisfied by some positive
 $(\mu_L,\mu_R)$; and hence all such Lorentz surfaces are rectifiable.

\bigskip

For $\Sigma$ generic and oriented, there is some geometry related to
the linear system that appears in the above proof. We shall now take a
look at this. It will be used in the next subsection for checking
rectifiability of $\Sigma$.

Let $\Omega^L_i$ (resp.\ $\Omega^R_j$) be the left (resp.\ right)
characteristic strip associated to $E^L_i$ (resp.\ $E^R_j$). One can
define a {\it pairing} between $\{E^L_i\}$ and $\{E^R_j\}$ as
illustrated in {\sc Figure 3\,-\,1}. Each characteristic strip $\Omega$
contains a unique light-cone-diamond $D$ that have the two singularities
at the border of $\Omega$ as two of its four corners. The pairing
$(E_i,E_j)$ of two edges $E_i$ and $E_j$, one left and one right, takes
an integer value that counts how many times $\Omega_j$ crosses $D_i$
with the orientation of $E_i$ taken into consideration.
\begin{figure}[htbp]
\setcaption{{\sc Figure 3\,-\,1.} A pairing between $\{E^L_i\}$ and
            $\{E^R_i\}$, where $k\in {\Bbb N}\cup\{0\}$. }
\centerline{\psfig{figure=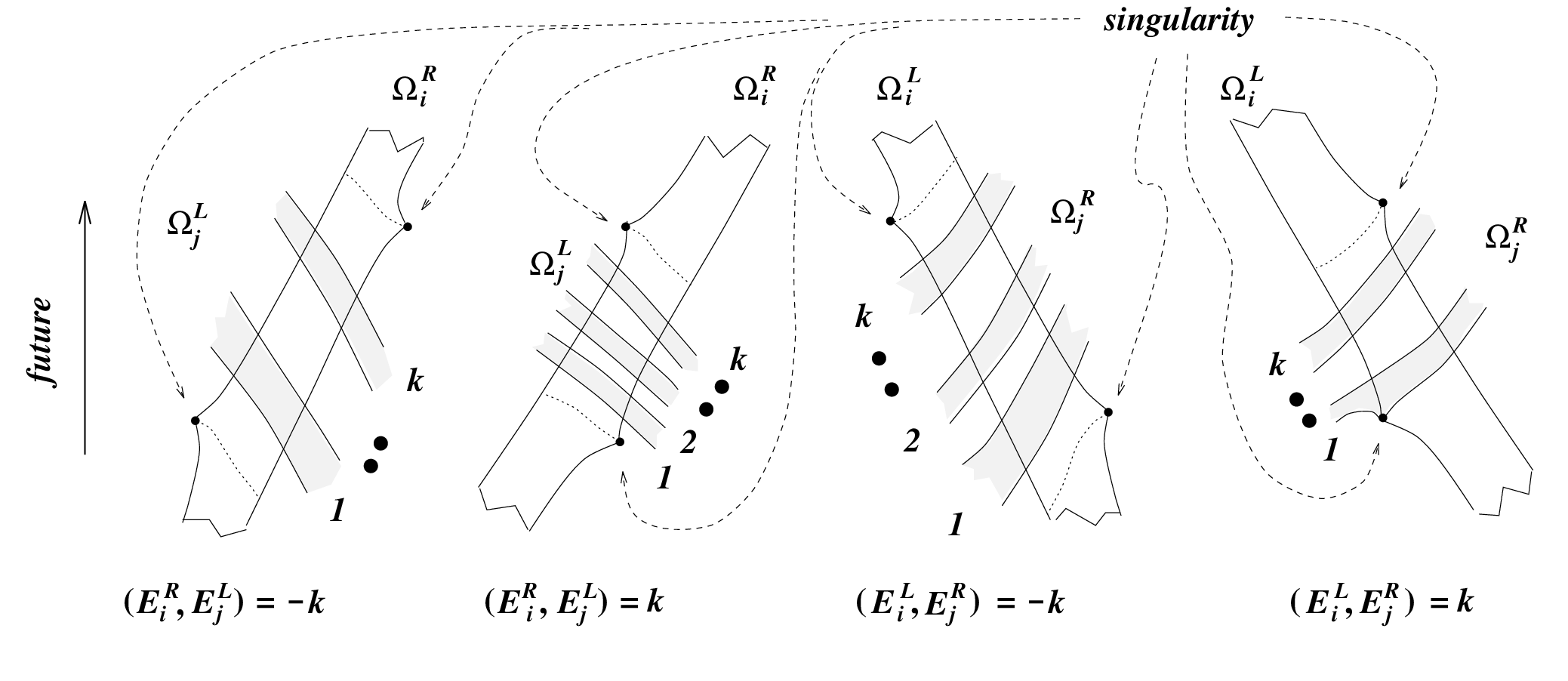,width=13cm,caption=}}
\end{figure}
An equivalent definition is given by assigning to each $E^L_i$, $E^R_j$
a directed positive measure $\mu^L_i$, $\mu^R_j$ of total mass 1
compatible with the orientation of edges. Then
$$
(E^R_i,E^L_j)\;=\;\int_{E^R_i}\,\mu^L_j
          \hspace{1cm}\mbox{and}\hspace{1cm}
                       (E^L_j,E^R_i)\;=\;\int_{E^L_j}\,\mu^R_i\,,
$$
where $\mu^L_j$, $\mu^R_i$ in the integrand are now regarded as the
induced directed transverse measures with respect to ${\cal F}_L$,
${\cal F}_R$.

Let ${\cal V}_L$ be the real {\it left-edge space}
$\mbox{\it Span}_{\supbscriptsizeBbb R}\{E^L_1,\ldots,E^L_k\}$ with the
positive definite inner product that takes $\{E^L_1,\ldots,E^L_k\}$ as
an orthonormal basis. Let ${\cal Z}_L$ and ${\cal U}_L$ be respectively
the {\it cycle space} and the {\it cut space} of $\Gamma_L$. (Recall
that the cut space of a graph $\Gamma$ can be regarded as the space of
functions on the vertex set of $\Gamma$ modulo $\Bbb R$; hence as the
space of exact 1-cocycles of $\Gamma$ as a simplicial 1-complex.)
Let ${\cal V}_R$, ${\cal Z}_R$, and ${\cal U}_R$ be defined similarly.
Recall the orthogonal decomposition [Bo]:
$$
{\cal V}_L\;=\;{\cal Z}_L\oplus {\cal U}_L\,,\hspace{1cm}
 {\cal V}_R\;=\;{\cal Z}_R\oplus {\cal U}_R\,.
$$
Define $T_{R\rightarrow L}$ from ${\cal V}_R$ to ${\cal V}_L$ by
linearly extending
$$
T_{R\rightarrow L}(E^R_i)\;=\;\sum_j (E^R_i,E^L_j)\,E^L_j\,;
$$
and, similarly, $T_{L\rightarrow R}$ from ${\cal V}_L$ to ${\cal V}_R$
by linearly extending
$$
T_{L\rightarrow R}(E^L_i)\;=\;\sum_j (E^L_i,E^R_j)\,E^R_j\,.
$$

Notice that $T_{R\rightarrow L}(E^R_i)$ is the edge-path in $\Gamma_L$
obtained by homotoping $E^R_i$ into $\Gamma_L$ relative to its
end-points; and similarly for $T_{L\rightarrow R}(E^L_i)$. Hence,
$T_{R\rightarrow L}\circ T_{L\rightarrow R}(E^R_i)$ and $E^R_i$ are
homotopic relative to the end-points; and so are
$T_{L\rightarrow R}\circ T_{R\rightarrow L}(E^L_i)$ and $E^L_i$. This
implies that
$$
T_{R\rightarrow L}\circ T_{L\rightarrow R}\;
                        =\;\mbox{\rm Id}_{{\cal V}_L}
\hspace{1cm}\mbox{and}\hspace{1cm}
   T_{L\rightarrow R}\circ T_{R\rightarrow L}\;
                =\;\mbox{\rm Id}_{{\cal V}_R}\,.
$$
Thus we can define the {\it transition matrix} $T$ to be
$T_{R\rightarrow L}$ and its inverse $T_{L\rightarrow R}$. In terms of
the bases $\{E^L_i\}$ and $\{E^R_i\}$, we may let
$$
 A^L\;=\;\left(\,a^L_{ij}\,\right)_{k_0\times k}\,,\hspace{.3cm}
 A^R\;=\;\left(\,a^R_{ij}\,\right)_{k_0\times k}\,
 \hspace{.3cm}\mbox{and}\hspace{.3cm}
 T\;=\;\left(\,(E^R_i,R^L_j)\,\right)_{k\times k}\,,
$$
then ${\cal Z}_L$
(resp.\ ${\cal Z}_R$) is the subspace in ${\cal V}_L$ (resp.\
${\cal V}_R$) generated by the row vectors of $A^L$ (resp.\ $A^R$) and
the two matrices $A^L$, $A^R$ are related by
$$
A^R\:T\;=\;A^L\,.
$$
The slope condition now reads
$$
 A^R\,(\mbox{\boldmath $v$}_1\,+\, T \mbox{\boldmath $v$}_2)
  \;=\;\mbox{\bf 0} \hspace{1em}\mbox{ for some column vectors
       $\mbox{\boldmath $v$}_1$,
          $\mbox{\boldmath $v$}_2\in {\Bbb R}^k_{>0}$}\,,
$$
where ${\Bbb R}^k_{>0}$ is the strictly positive orthant of
${\Bbb R}^k$. Equivalently,
$$
 \left({\Bbb R}^k_{>0}\,+\,T{\Bbb R}^k_{>0}\right)\,\cap\,
  {\cal U}_R\;\neq\;\emptyset\,.
$$

\bigskip

\noindent{\it Remark 3.1.4.} Let $\{e_j\}$ be the standard basis for
${\Bbb R}^k$; then $Te_j$ lies in the hyperplane
$\{(x_1, \cdots, x_k)^t\,|\,\sum x_i = -1 \}$ for $j=1,\cdots,k$. To
see this, since $Te_j=\sum_i(E^R_i,E^L_j)\,e_i$, it suffices to
show that
$$
\sum_i(E^R_i,E^L_j)\;=\;-1\hspace{1cm}\mbox{for $j=1,\ldots,k$}\,.
$$
Let $\Omega^L_j$ be the left strip associated to $E^L_j$ and
$$
 \cdots\,,\;\Omega^R_{j_{-1}}\,,\;\Omega^R_{j_0}\,,\;
                        \Omega^R_{j_1}\,,\;\Omega^R_{j_2}\,,\;\cdots
$$
be the sequence of right-strips that $\Omega^L_j$ crosses
following the future-direction. (Note that same strip appears in general
more than once in this sequence.) For $s$ negatively (resp.\
positively) large enough, $\Omega^L_j$ must cross both the boundary
components of $\Omega^R_{j_s}$ from the halves causally before
(resp.\ after) the singularities. Thus, if let $c_{sj}$ be the
contribution to $\sum_i(E^R_i,E^L_j)$ at each occurrence of crossing,
then the sequence $\{c_{sj}\}_s$ must be of the form
$$
 \stackrel{0}{\ldots\ldots\ldots}\,,\,-1\,,\,
   \stackrel{0}{\ldots}\,,\,1\,,\,\cdots\cdots\,,\,-1\,,\,1\,,\,
    \stackrel{0}{\ldots}\,,\, -1\,,\stackrel{0}{\ldots\ldots\ldots}\,,
$$
namely, a sequence obtained by inserting 0's to a finite
1, $-1$ alternating sequence beginning and ending with $-1$
({\sc Figure 3\,-\,2}). Consequently,
$$
 \sum_i\,(E^R_i,E^L_j)\;=\;\sum_s\,c_{sj}\;=\;-1
$$
as required. Incidentally, it follows from this that  the
intersection of $\left({\Bbb R}^k_{>0}\,+\,T{\Bbb R}^k_{>0}\right)$ with
$\{(x_1, \cdots, x_k)^t\,|\,\sum x_i = 0 \}$ is
$\mbox{\it Span}\,_{{\supbscriptsizeBbb R}_{>0}}\{e_i+Te_j\}_{ij}$.
\begin{figure}[htbp]
\setcaption{{\sc Figure 3\,-\,2.} The pattern in which a left-strip
     crosses right-strips and the contribution at each occurrence of
     crossing to the sum of pairings.      }
\centerline{\psfig{figure=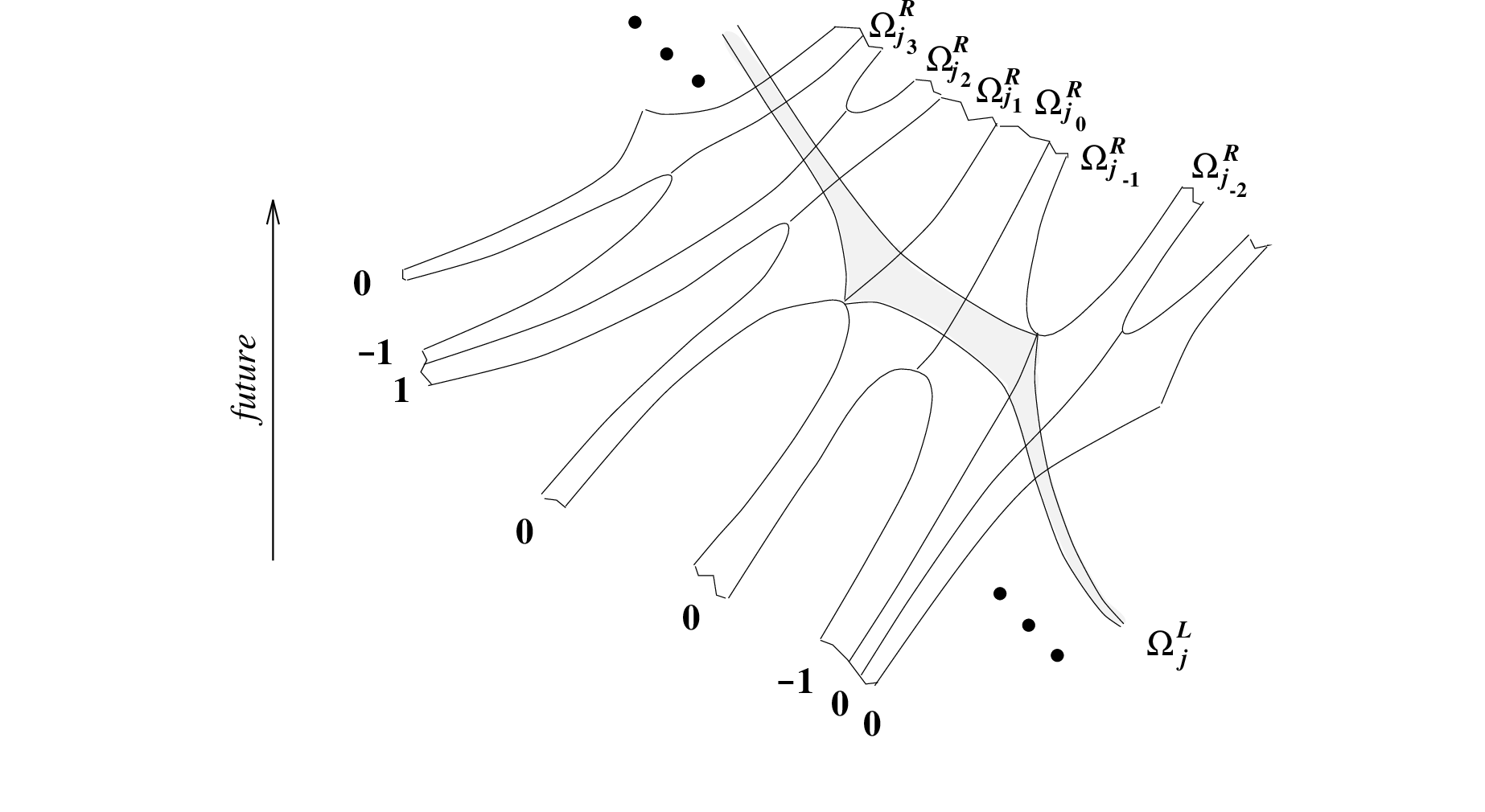,width=13cm,caption=}}
\end{figure}

\bigskip

\subsection{\bf Electrical circuits and examples of
                                               unrectifiability.}
Given an s-d-l Lorentz surface with a simple cut system
$(\Sigma,\{C_{\alpha}\}_{\alpha})$. Recall from Sec.\ 2.3 the network
$\mbox{\it Net}\,(\Sigma,\{C_{\alpha}\}_{\alpha})$ associated to it,
whose edge $E_{\alpha}$ has a {\it favored direction} induced from the
time-orientation of $\Sigma$. Let
$A_{\alpha}=S^1\times [0,2\pi k_{\alpha}]$ be grafted to $\Sigma$ along
$C_{\alpha}$. If the new surface $\Sigma^{\prime}$ is rectifiable to a
Mandelstam diagram $\Xi$ with time function $t$ and characteristic
1-forms $\mu_L$, $\mu_R$ (for $\Sigma$ orientable; otherwise
$(\mu_L,\mu_R)$ is the local splitting of the characteristic bi-valued
1-form $\mu$ for $\Sigma$ non-orientable). Then on $A_{\alpha}$
$$
 \mbox{\raisebox{.3ex}{$\left.\left(
      \int_{\{0\}\times[0,2\pi k_{\alpha}]} dt \right)\,\right/$}
   \raisebox{-.3ex}{$\left(\int_{S^1\times\{0\}}\mu_L\right)$}}\;
  =\; k_{\alpha}\,.
$$
And similarly for $\mu_R$. These relations resemble the {\it Ohm's law}:
$$
 \mbox{\raisebox{.3ex}{$V\,/$}\raisebox{-.3ex}{\,$I$}}\;=\;R \,,
$$
where $V$ is the (electrical) potential difference at the end points of
a conducting rod, $I$ the current through it, and $R$ the total
resistance thereon. The behavior of the transverse measures to null
foliations on $\Sigma$ reminds one the {\it Kirchhoff's first law}
which states that the algebraic sum of the currents at each vertex must
be zero. These considerations suggest one the following approximate
dictionary:

\vspace{.8cm}

\centerline{
\begin{tabular}{lll|lll}   \hline
 && \hspace{-12pt}
    \begin{minipage}[t]{5.5cm}\rule{0ex}{3.5ex}
    \hspace{-.6ex}{\it A grafted s-d-l Lorentz \newline surface
    $\left(\Sigma\,;\{(C_{\alpha},k_{\alpha})\}_{\alpha}\right)$}:
    \rule[-2ex]{0ex}{2ex}
  \end{minipage}  & & &
  \hspace{-12pt}\begin{minipage}[t]{6cm}
    {\it An electrical circuit supported on the network
    $\mbox{\it Net}\,(\Sigma,\{C_{\alpha}\}_{\alpha})$}:
  \end{minipage}  \\ \hline
 &&\begin{minipage}[t]{5cm}
   \hspace{-10pt}$\bullet$\rule{0ex}{3.5ex}
   {\it total left} (or {\it right}) {\it trans-} \newline
   {\it verse measure} $\mu_{\alpha}$ along $C_{\alpha}$
  \end{minipage}  & & &
  \begin{minipage}[t]{6cm}
    \hspace{-10pt}$\bullet$
    {\it current} through edge $E_{\alpha}$
  \end{minipage}  \\
 &&\begin{minipage}[t]{5cm}
    \hspace{-10pt}$\bullet$\rule{0ex}{3ex}
    {\it time-orientation} on $\Sigma$
  \end{minipage}  & & &
  \begin{minipage}[t]{5cm}
     \hspace{-10pt}$\bullet$
     {\it favored direction} of current; \newline
     direction of the edges of \newline
     $\mbox{\it Net}\,(\Sigma,\{C_{\alpha}\}_{\alpha})$
  \end{minipage}  \\
 &&\begin{minipage}[t]{5cm}
     \hspace{-10pt}$\bullet$\rule{0ex}{3ex}
     {\it step} $k_{\alpha}$ of grafting along $C_{\alpha}$
     ($k_{\alpha}$ large)
  \end{minipage}  & & &
  \begin{minipage}[t]{6cm}
     \hspace{-10pt}$\bullet$
     {\it resistance} $R_{\alpha}$ at edge $E_{\alpha}$
  \end{minipage} \\
 &&\begin{minipage}[t]{5cm}
     \hspace{-10pt}$\bullet$\rule{0ex}{3ex}
     {\it time function} on $\Sigma$
  \end{minipage}  & & &
  \begin{minipage}[t]{6cm}
     \hspace{-10pt}$\bullet$
     {\it potential} on
     $\mbox{\it Net}\,(\Sigma,\{C_{\alpha}\}_{\alpha})$
     \rule[-2ex]{0ex}{1ex}
  \end{minipage} \\ \hline
\end{tabular}
} 

\vspace{1cm}

This realization explains why there are {\it unrectifiable} s-d-l
Lorentz surfaces. Suppose we begin with any
$(\Sigma,\{C_{\alpha}\}_{\alpha})$ whose associated electrical
circuit contains a bridge. Then by varying the step $k_{\alpha}$ of
grafting along $C_{\alpha}$, one varies the resistence $R_{\alpha}$ of
the circuit and can manage to force the current through the bridge go in
the unfavored direction. If, in addition, these $k_{\alpha}$ are kept
large enough, then trying to rectify the new s-d-l Lorentz surface
$\Sigma^{\prime}=(\Sigma;\{(C_{\alpha},k_{\alpha})\}_{\alpha})$ will
render some characteristic strip in $\Sigma^{\prime}$ reverse its
original restricted time-orientation. Hence it cannot be rectified. The
following example illustrates this.

\bigskip

\noindent{\bf Example 3.2.1.} Let $\Sigma$ be an s-d-l Lorentz surface
of genus 2 with one past- and one future-end. Let $\{C_{\alpha}\}$ be
a simple cut system as indicated in ({\sc Figure 3\,-\,3}). Assume that
each $C_{\alpha}$ is oriented from right to left so that the transverse
measure $\int_{C_{\alpha}}\mu_R$ is positive for any positive 1-form on
$\Sigma^{\rm ornt}_0/$\raisebox{-.6ex}{${\cal F}_R$}.
\begin{figure}[htbp]
\setcaption{{\sc Figure 3\,-\,3.} An s-d-l Lorentz surface with a
            simple cut system $(\Sigma,\{C_{\alpha}\}_{\alpha})$ and
            its associated electrical circuit. }
\centerline{\psfig{figure=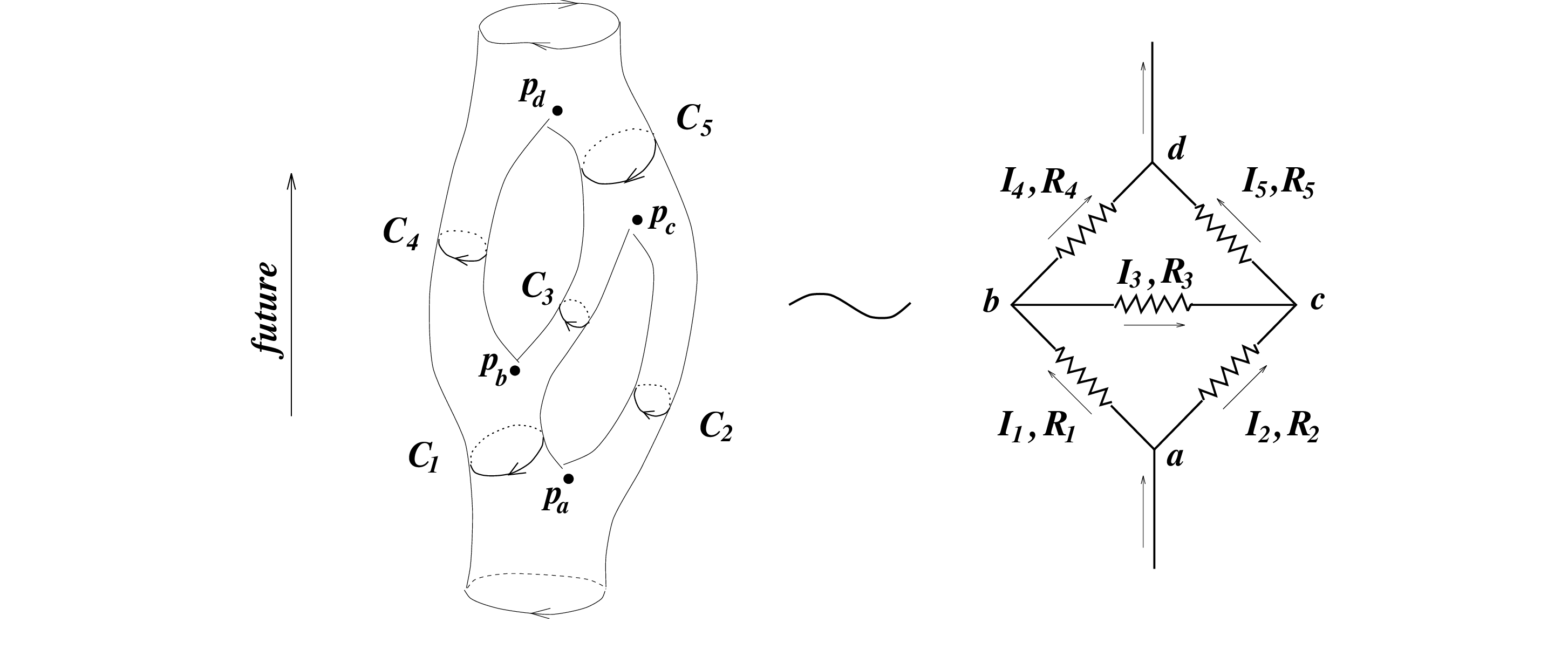,width=13cm,caption=}}
\end{figure}
The associated electrical circuit is then a simple bridge. Recall that
$$
 I_3\;>\;\mbox{(resp.\ $=$, $<$)}\;0\hspace{.5cm}\mbox{if and only if}
 \hspace{.5cm}R_2\,R_4\;>\;\mbox{(resp.\ $=$, $<$)}\;R_1\,R_5\,.
$$

For clarity of observation, let's recast the surface into an immersed
planar domain (using a Morse function and its associated handlebody
decomposition of $\Sigma$). Assume that the basic structures on $\Sigma$
are as indicated in {\sc Figure 3\,-\,4}.
\begin{figure}[htbp]
\setcaption{{\sc Figure 3\,-\,4.} Basic structures of $\Sigma$.}
\centerline{\psfig{figure=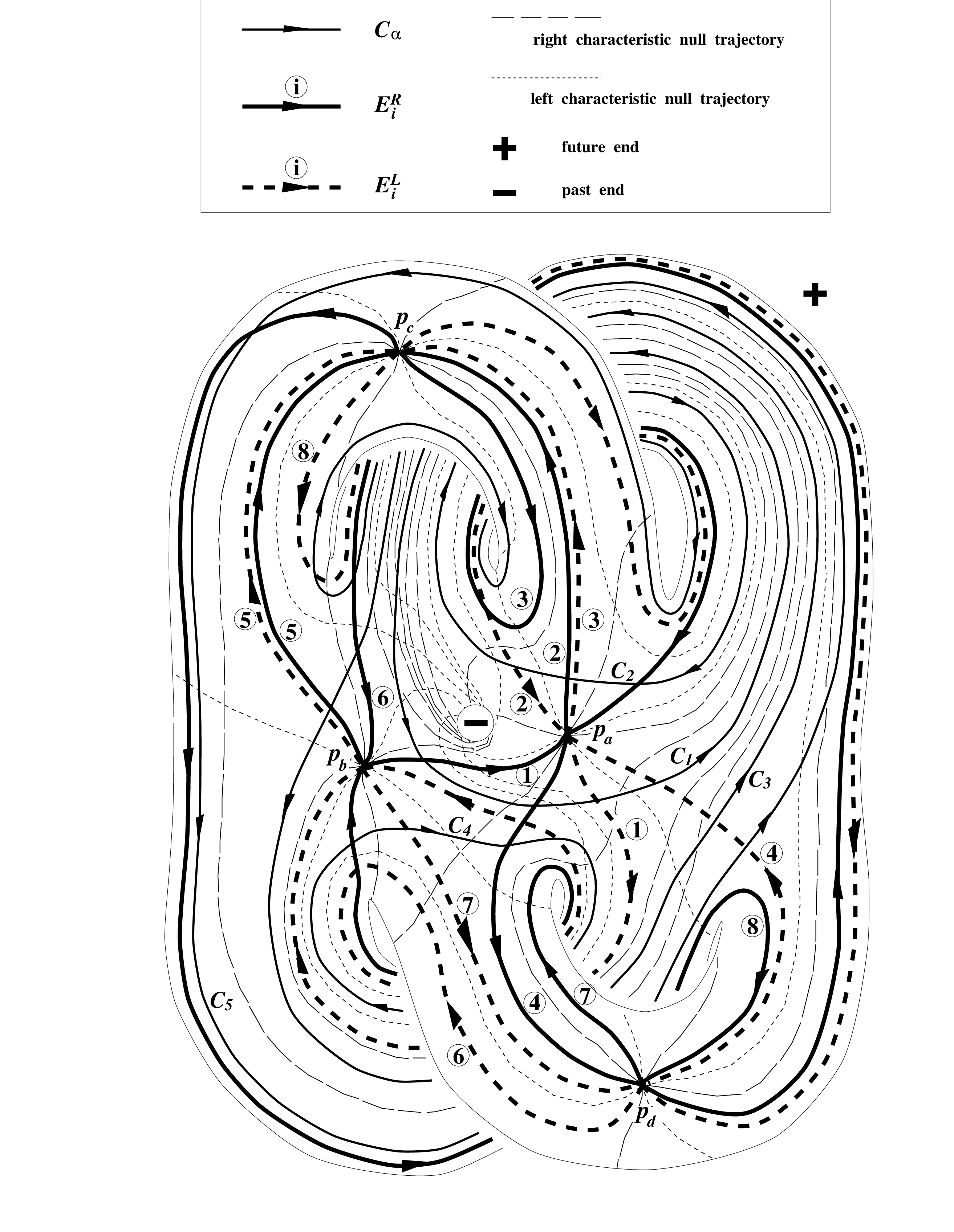,width=13cm,caption=}}
\end{figure}
Up to (topological) automorphism of the surface, the pair of graphs
$(\Gamma_L,\Gamma_R)$ in $\Sigma$ with their edges labelled are shown
in {\sc Figure 3\,-\,5}.
\begin{figure}[htbp]
\setcaption{{\sc Figure 3\,-\,5.} The pair of graphs
   $(\Gamma_L,\Gamma_R)$ in $\Sigma$. The circled number beside an
   edge indicates the labelling of that edge. }
\centerline{\psfig{figure=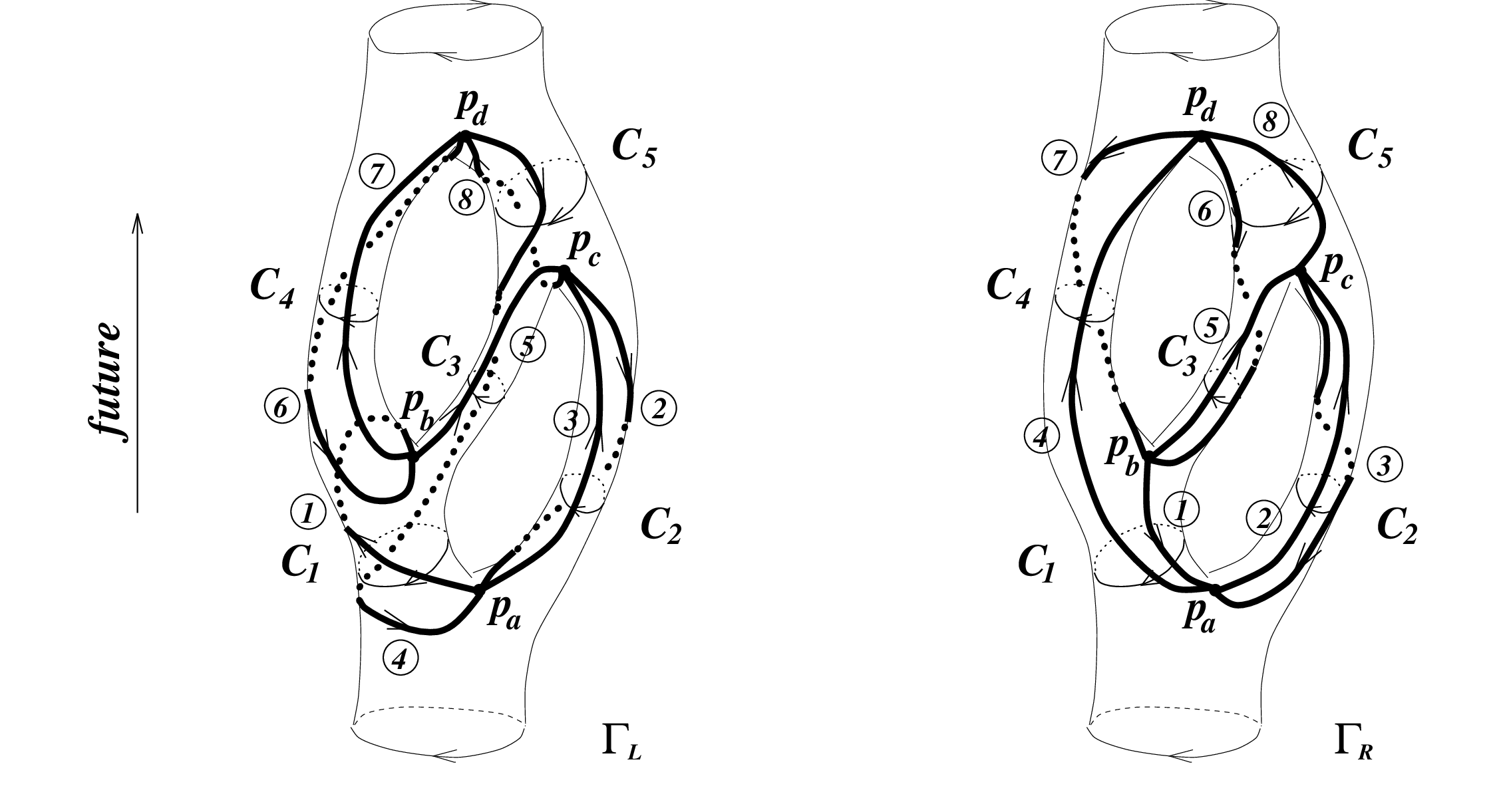,width=13cm,caption=}}
\end{figure}
It follows from this, with careful examination of this pair of
graphs, that the transition matrix $T$ for $\Sigma$ is
$$
 T\;=\;
    \left[ \begin{array}{rrrrrrrr}
         -1 &  0 &  0 &  0 &  0 & -1 & -1 &  0 \\
          0 &  0 &  1 &  0 &  0 &  0 &  0 &  0 \\
          0 & -1 & -2 &  0 &  0 &  0 &  0 &  0 \\
          1 &  0 &  0 &  0 &  0 &  1 &  2 &  0 \\
          0 &  0 &  0 &  0 &  1 &  0 &  0 &  0 \\
          0 &  0 &  0 &  0 & -1 &  0 &  0 & -1 \\
          0 &  0 &  0 &  0 &  0 & -1 & -2 &  0 \\
         -1 &  0 &  0 & -1 & -1 &  0 &  0 &  0
    \end{array} \right]
$$
and the right cut space
${\cal U}_R=\mbox{\it Span}\,_{\supbscriptsizeBbb R}
  \{\mbox{\boldmath $u_a$, $u_b$, $u_c$, $u_d$}\}$, where
$$
\begin{array}{lcrrrrrrrrrl}
 \mbox{\boldmath $u_a$} &= &( &-1, &1, & -1, &1, &0, &0, &0, &0 &)^t \\
 \mbox{\boldmath $u_b$} &= &( & 1, &0, & 0, &0, &1, &-1, &-1, &0 &)^t \\
 \mbox{\boldmath $u_c$} & = &( &0, &-1, &1, &0, &-1, &0, &0, &1 &)^t \\
 \mbox{\boldmath $u_d$} & = &( &0, &0, &0, &-1, &0, &1, &1, & -1 &)^t
\end{array}
$$
with $(\;\cdot\;)^t$ meaning transpose. One can check that, say,
\begin{eqnarray*}
 \lefteqn{ \left(\frac{3}{8}e_1 +e_2 +2e_3 +\frac{1}{2}e_4 +e_5 +e_6
                          +\frac{1}{2}e_7 +\frac{5}{4}e_8\right) }\\
  & & + \left(\frac{1}{4}Te_1 +Te_2 +2Te_3 +Te_4 +Te_5 +Te_6
                                    +\frac{1}{8}Te_7 +Te_8\right)\\
  & & \hspace{-1cm} =\; 3\mbox{\boldmath $u_a$}
                   +2\mbox{\boldmath $u_b$} +\mbox{\boldmath $u_d$}\,.
\end{eqnarray*}
Thus
$\left({\Bbb R}^8_{>0}\,+\,T{\Bbb R}^8_{>0}\right)\,\cap\,
{\cal U}_R\;\neq\;\emptyset$ and $\Sigma$ is rectifiable.

To see how $T$ varies after grafting, one needs also to know both the
right edges $E^R_i$ and the left strips $\Omega^L_j$ that go across a
given $C_{\alpha}$. This can be obtained from {\sc Figure 3\,-\,4}. The
result is listed in the following table:

\vspace{.8cm}

\centerline{
\begin{tabular}{l|rl|l|}
     & \multicolumn{2}{c|}{$E^R_i$ through $C_{\alpha}$
                                           \rule[-2ex]{0ex}{2ex}}
      & \multicolumn{1}{c|}{$\Omega^L_j$ through $C_{\alpha}$} \\ \hline
 \rule{0ex}{3.5ex}
  $C_1\;$ & $\;-E^R_1$, & \hspace{.4em} $E^R_4$
          & $\;\Omega^L_1\,,\; \Omega^L_4\,,\; \Omega^L_5\,,\;
                \Omega^L_6\,,\; \Omega^L_7\,,\; \Omega^L_8\;$  \\
 \rule{0ex}{3ex}
  $C_2\;$ & $\;E^R_2$, & $-E^R_3$
          & $\;\Omega^L_2\,,\; \Omega^L_3\;$ \\
 \rule{0ex}{3ex}
  $C_3\;$ & $\;E^R_5$, & $-E^R_6$
    & $\;\Omega^L_1\,,\; \Omega^L_4\,,\; \Omega^L_5\,,\;
                                                  \Omega^L_8\;$\\
 \rule{0ex}{3ex}
  $C_4\;$ & $\;E^R_4$, & $-E^R_7$
          & $\;\Omega^L_6\,,\; \Omega^L_7\;$  \\
 \rule{0ex}{3ex}
  $C_5\;$ & $\;-E^R_6$, & \hspace{.4em} $E^R_8$
    & $\;\Omega^L_1\,,\; \Omega^L_2\,,\; \Omega^L_3\,,\;
                \Omega^L_4\,,\; \Omega^L_5\,,\; \Omega^L_8\;$   \\
\end{tabular}
} 

\vspace{1cm}

\noindent
The "$-$"-sign before an $E^R_i$ indicates that $E^R_i$ crosses a
$C_{\alpha}$ from the future domain of $C_{\alpha}$ to its past domain.
When no indication (i.e.\ a hidden "+"-sign), it crosses a $C_{\alpha}$
from the past domain of $C_{\alpha}$ to its future domain.
Consequently, after grafted along $C_{\alpha}$ by step $k_{\alpha}$,
the transition matrix $T^{\prime}$ for the new s-d-l Lorentz
surface $\Sigma^{\prime}$ becomes

{\footnotesize
$$
 \hspace{-2cm} T^{\prime}\;=\;
 \left[ \begin{array}{rrrrrrrr}
   -1-2k_1 &  0 &  0 &  -2k_1 &  -2k_1 & -1-2k_1 & -1-2k_1 & -2k_1 \\
          0 &  2k_2 &  1+2k_2 &  0 &  0 &  0 &  0 &  0 \\
          0 & -1-2k_2 & -2-2k_2 &  0 &  0 &  0 &  0 &  0 \\
   1+2k_1 &  0 &  0 & 2k_1 & 2k_1 & 1+2k_1+2k_4 & 2+2k_1+2k_4 & 2k_1 \\
          2k_3&  0 &  0 &  2k_3 & 1+2k_3 &  0 &  0 & 2k_3 \\
   -2k_3-2k_5 & -2k_5 & -2k_5 & -2k_3-2k_5 & -1-2k_3-2k_5
                                             & 0 & 0 & -1-2k_3-2k_5 \\
          0 &  0 &  0 &  0 &  0 & -1-2k_4 & -2-2k_4 & 0 \\
         -1+2k_5 & 2k_5 & 2k_5 & -1+2k_5 & -1+2k_5 &  0 &  0 &  2k_5
    \end{array} \right]\:.
$$     }
As suggested by its associated electrical circuit, assume that
$$
(k_2+2)(k_4+2)\;\ll\;(k_1-2)(k_5-2)\,.
$$
We want to show that $\Sigma^{\prime}$ is then not rectifiable.

Suppose otherwise, then there exist $x_i>0$, $i=1,\ldots,8$, and
$a$, $b$, $c$, $d$ real such that
$$
 \sum_i x_i\, T^{\prime}e_i\;
     <\; a\mbox{\boldmath $u_a$}+b\mbox{\boldmath $u_b$}
                  +c\mbox{\boldmath $u_c$}+d\mbox{\boldmath $u_d$}
$$
(meaning that each component of the former vector is less than the
corresponding one of the latter). Explicitly,
{\footnotesize
$$
 \hspace{-1cm}
 \begin{array}{c}
  (1)\\ (2)\\ (3)\\ (4)\\ (5)\\ (6)\\ (7)\\ (8)
 \end{array}
 \hspace{1cm}\left(
 \begin{array}{r}
  -(x_1+x_6+x_7)-2k_1(x_1+x_4+x_5+x_6+x_7+x_8)\\ x_3+2k_2(x_2+x_3)\\
  -(x_2+2x_3)-2k_2(x_2+x_3)\\
  (x_1+x_6+2x_7)+2k_1(x_1+x_4+x_5+x_6+x_7+x_8)+2k_4(x_6+x_7)\\
  x_5+2k_3(x_1+x_4+x_5+x_8)\\
  -(x_5+x_8)-2k_3(x_1+x_4+x_5+x_8)-2k_5(x_1+x_2+x_3+x_4+x_5+x_8)\\
  -(x_6+2x_7)-2k_4(x_6+x_7)\\
  -(x_1+x_4+x_5)+2k_5(x_1+x_2+x_3+x_4+x_5+x_8)
 \end{array} \right)
 \;<\;\left(
 \begin{array}{r}
   -a+b\\ a-c\\ -a+c\\ a-d\\ b-c\\ -b+d\\ -b+d\\ c-d
 \end{array} \right)\,.
$$   }
That (2), (4), (5), (8) are positive implies that either
(i) $d<c<a\leq b$, or (ii) $d<c<b<a$. Case (i) is ruled out by taking
$(4)+(7)$. For case (ii), let $u=x_1+x_4+x_5+x_8$ (resp.\ $v=x_2+x_3$,
$w=x_6+x_7$) be the total right transverse measure along $C_3$ (resp.\
$C_2$, $C_4$). By considering (3), (4)+(7), (7), and(8), one has
$$
 \begin{array}{rrrrrrrrr}
   0 &< &(u+v)(2k_5-1) &< &c-d &< &b-d &< &2w(k_4+1)\;, \\
   0 &< &(u+w)(2k_1+1) &< &a-b &< &a-c &< &2v(k_2+1)\;.
 \end{array}
$$
This implies that
$$
 [(u+v)(2k_5-1)]\,[(u+w)(2k_1+1)]\;
    <\;[2w(k_4+1)]\,[2v(k_2+1)]\,.
$$
On the other hand, since $u,v,w\,>0$, the assumption that
$(k_2+2)(k_4+2)\;\ll\;(k_1-2)(k_5-2)$ implies the opposite. This leads
to a contradiction; and hence $x_i$ cannot be all positive.
Consequently, $\Sigma^{\prime}$ is {\it not} rectifiable.

\noindent\hspace{14cm}$\Box$

\bigskip

\section{Toward Lorentzian conformal field theories.}
{\small
\begin{quote}
  {\it" "This is a very deep business," $\cdots$. "There are a
  thousand details which I should desire to know before I decide upon
  our course of action. $\cdots$." "}

 \hspace{3cm} --------- $\;$ from $\;$
  \parbox[t]{8cm}{{\it The adventure of the speckled band}, in
   {\sl Adventures of Sherlock Holmes} by Sir A.C.\ Doyle.}
\end{quote}
}

This last section contains discussions on some ingredients of
prospect definitions of Lorentzian conformal field theory (LCFT). It
serves to provoke some thoughts for further investigations and is by no
means complete. We shall give first a proto-definition of LCFT after
Atiyah and Segal ([At1], [At2], [Ge], [Se2], [Se3]; see also [AG-G-M-V],
[F-S], [Mo-S]); and then discuss its refinements. All the Lorentz
surfaces in the discussion are in the coarse category.

\bigskip

\noindent{\bf Proto-definition 4.1 [LCFT].}
A prototype for an abstract (coarse) Lorentzian conformal field theory
after Atiyah and Segal consists of the following data:

\medskip

\noindent\hspace{.5cm}$\bullet$
${\cal C}_{\rm Lorz}$, the category of related geometries:

\medskip

\noindent\hspace{1.2cm}
\begin{minipage}{14cm}
 \begin{quote}
   \hspace{-.8cm}{\it Objects}: An object in ${\cal C}_{\rm Lorz}$ is
      a finite disjoint union of unparametrized circles,
      $\sqcup^{\bullet} S^1$.
      We shall denote the set of objects by
      $\mbox{\it Obj}\,({\cal C}_{\rm Lorz})$. It is isomorphic to
      ${\Bbb N}\cup\{0\}$, the set of nonnegative integers.

   \hspace{-.8cm}{\it Morphisms}: A morphism from
      $\sqcup^m S^1$ to $\sqcup^n S^1$ is a time-oriented s-d-l
      Lorentz surface $\Sigma$ (not necessarily connected or
      orientable) with $m$ past- and $n$ future-ends decorated with an
      ordered collection of $m+n$ parametrized simple spacelike loops
      $C_{\alpha}$ such that each of the first $m$ (resp.\ last $n$)
      loops can be homotoped into a different past- (resp.\ future-)
      end. When there is no risk of confusion, we shall denote a
      morphism $\{\Sigma;(C_1,\ldots,C_{m+n})\}$ simply by $\Sigma$.
      Also we shall denote the set of such by $\mbox{\it Mor}\,(m,n)$
      and their union by $\mbox{\it Mor}\,({\cal C}_{\rm Lorz})$.

   \hspace{-.8cm}{\it Composition of morphisms}: One can compose a
      morphism $\Sigma_2$ from $\sqcup^m S^1$ to
      $\sqcup^n S^1$ to a morphism $\Sigma_1$ from  $\Sigma_2$
      from $\sqcup^l S^1$ to $\sqcup^m S^1$  by sewing
      orderly the last $m$ loops of $\Sigma_1$ to the first $m$ loops of
      $\Sigma_2$ by identifying points of the same parametrization.
      Notice that this determines uniquely a new s-d-l Lorentz surface
      $\Sigma_1\diamond\Sigma_2$.
 \end{quote}
\end{minipage}

\bigskip

\noindent\hspace{.5cm} $\bullet$ \parbox[t]{14cm}{
A functor $Z$ from ${\cal C}_{\rm Lorz}$ to ${\cal C}_{{\rm mod},R}$,
the tensor category of modules over a ring $R$ that satisfies
essentially the following two properties, in addition to some
naturality requirements of the autofunctors on ${\cal C}_{{\rm mod},R}$
induced by $\mbox{\it Diff}\,(S^1)$:  }

\medskip

\noindent\hspace{1.2cm}
\begin{minipage}{14cm}
 \begin{quote}
   \hspace{-.8cm}{\it Multiplicativity}:
      $Z(S^1\sqcup S^1)=Z(S^1)\otimes Z(S^1)$.

   \hspace{-.8cm}{\it Associativity under sewing}:
      $Z(\Sigma_1\diamond\Sigma_2)=Z(\Sigma_2)\circ Z(\Sigma_1)$, up
      to a multiple factor by an element in $R$.
  \end{quote}
\end{minipage}

\noindent\hspace{14cm}$\Box$

\bigskip

\noindent
For string theory, $Z(S^1)$ is in principle the state space for a string
moving in a given target-space.

The above proto-definition for LCFT is a plain parroting from the
Riemannian case. One likes to know if there are refinements that make it
more akin to the nature of s-d-l Lorentz surfaces. To simplify the
argument and make the essential points prominant, we shall restrict
ourselves to oriented Lorentz surfaces for the rest of the discussions.

Let's reflect first on the following question:
\begin{quote}
  {\bf Q.} \parbox[t]{13cm}{\it What could distinguish a would-be
     Lorentzian string theory from Riemannian ones?}
\end{quote}
Let $\Sigma$ be an oriented s-d-l Lorentz surface. Recall from Sec.\ 2
the basic structures associated to $\Sigma$. They either have or suggest
some natural physical interpretations:
\begin{quote}
  \hspace{-23pt}($1^{\circ}$)
    The {\it singular set} $\mbox{\rm Sing}(\Sigma)$: It corresponds to
    the {\it interacting points} of closed strings. By tracing along the
    two sets of characteristic leaves from singularities,
    $\mbox{\rm Sing}(\Sigma)$ provides two collections of {\it labelled
    marked points} on the incoming and outgoing strings. These marked
    points indicate either the prospective or historical interacting
    points on the strings; and their label indicates the type of
    interactions as designated by the index of corresponding
    singularities. The directed graph
    $\mbox{\it Net}\,(\Sigma,\{C_{\alpha}\}_{\alpha})$ associated to
    any simple cut system has $\mbox{\rm Sing}(\Sigma)$ as the vertex
    set. It should be thought of as a {\it Feynman diagram} in the
    space-time depicting the interacting process of particles associated
    to $\Sigma$.

  \hspace{-23pt}($2^{\circ}$)
    The {\it l-c-d tessellation} $\mbox{\rm Tessln}_{\Sigma}$: It gives
    a grid pattern $\mbox{\rm Grid}(\Sigma)$ (the 1-complex in $\Sigma$
    made of characteristic leaves), which suggests a {\it statistical
    mechanical} treatment of interacting strings. It also reminds one
    of {\it adelic string} theories, in which worldsheets for simple
    case could be trees, instead of 2-dimensional manifolds (e.g.\
    [B-F]).

  \hspace{-23pt}($3^{\circ}$)
    {\it Foliations} ${\cal F}_L$ and ${\cal F}_R$: They suggest some
    natural fields and operators in the theory and hint at a connection
    with {\it Connes' non-commutative geometry} (e.g.\ [Co]).
    Furthermore, they provide two ways of identifying incoming strings
    to outgoing strings in a piecewise manner. This relates $\Sigma$ to
    a sequence of {\it bonded directed links}, with adjacent ones
    differing by a simple twist of bond. Thus a connection to knot
    theory is hinted at. (Example 4.2.)
\end{quote}

\bigskip

\noindent{\bf Example 4.2 [Lorentz surfaces and bonded directed links].}
The sequence of bonded directed links associated to the Lorentz surface
depicted in {\sc Figure 3\,-\,4} is illustrated below
({\sc Figure 4\,-\,1}).
\begin{figure}[htbp]
\setcaption{{\sc Figure 4\,-\,1.} The sequence of bonded directed links
associated to $\Sigma$ in {\sc Figure 3\,-\,4}. Notice that twists of
bond resemble splicings of links [Kau].             }
\centerline{\psfig{figure=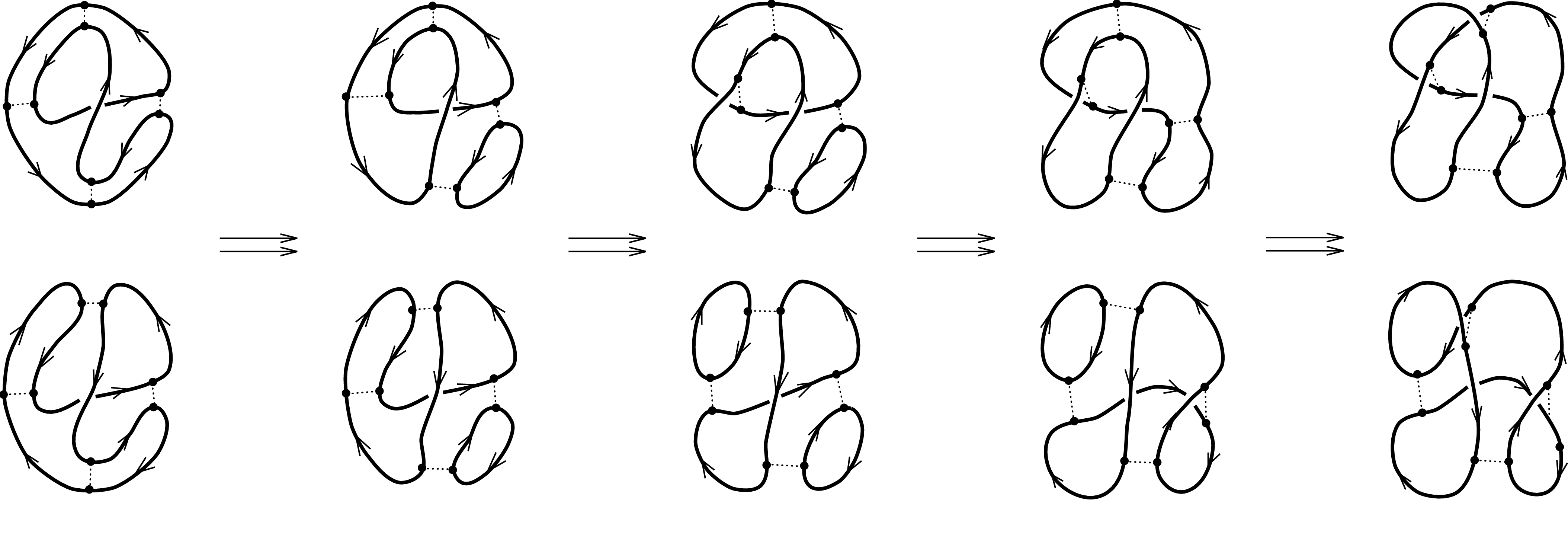,width=13cm,caption=}}
\end{figure}

\noindent\hspace{14cm}$\Box$

\bigskip

In view of these features, the following categories should play roles in
the final picture of LCFT. We list their $\mbox{\it Obj}$ and
$\mbox{\it Mor}$. Composition of morphisms are essentially obtained by
pasting appropriately along the geometric objects involved.

\medskip

\noindent\hspace{.3cm}
(a) \parbox[t]{14cm}{${\cal C}_{\rm t\mbox{-}graph}$
   (cf.\ Item ($1^{\circ}$)): A {\it time-directed} graph
   ({\it t-graph}) is a directed graph that contains no cycles. Any
   graph, each of whose edges is attached to distinct vertices, can be
   made time-directed ({\sc Figure 4\,-\,2}). }

\medskip

\noindent\hspace{1.2cm}
\begin{minipage}{14cm}
 \begin{quote}
   \hspace{-.8cm}{\it Obj} $=\;{\Bbb N}$.

   \hspace{-.8cm}{\it Mor}: A morphism from $m$ to $n$ is a t-graph
     (not necessarily connected) with $m$ incoming and $n$ outgoing
     external edges. These external edges are ordered, first incoming
     ones and then outgoing ones.
 \end{quote}
\end{minipage}

\bigskip

\begin{figure}[htbp]
\setcaption{{\sc Figure 4\,-\,2.} A t-graph interpolates two collections
   of points. It represents a particle interacting process. }
\centerline{\psfig{figure=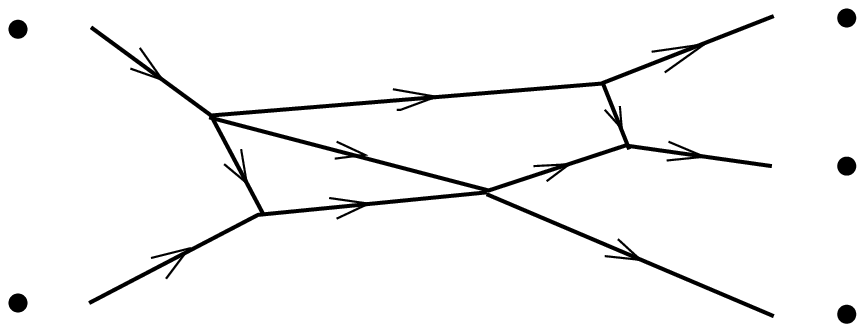,width=13cm,caption=}}
\end{figure}

\bigskip

\noindent\hspace{.3cm}
(b) \parbox[t]{14cm}{${\cal C}_{\rm tes-Lorz}$
  (cf.\ Item ($2^{\circ}$)): A {\it characteristic broken null (CBN-)
  loop} in a Lorentz surface $\Sigma$ is a simple oriented broken null
  loop that lies in $\mbox{\rm Grid}\,(\Sigma)$ and whose arbitrary
  small tubular neighborhood contains a simple spacelike loop. One can
  assign to it a label
  $[i_1,\ldots,i_{2k}]\in
  \mbox{\raisebox{.3ex}{$\left.
    \left({\Bbb Z}-\{0\}\right)^{2k}\hspace{-2pt}\right/$}
               \hspace{-5pt}\raisebox{-.3ex}{${\Bbb Z}_{2k}$} }$
  for some $k$ by counting the signed number of steps it goes following
  the orientation with $+$ for future headed steps and $-$ for past
  headed steps. For the present case, we require that they be oriented
  from left to right relative to the future direction.
  ({\sc Figure 4\,-\,3}).      }

\medskip

\noindent\hspace{1.2cm}
\begin{minipage}{14cm}
 \begin{quote}
   \hspace{-.8cm}{\it Obj} $=\;{\Bbb N}$.

   \hspace{-.8cm}{\it Mor}: A morphism is an oriented s-d-l Lorentz
     surface decorated with an ordered collection of peripheral
     CNB-loops $C_{\alpha}$, one for each end
     (cf.\ ${\cal C}_{\rm Lorz}$). Only $C_{\alpha}$'s that are labelled
     the same are sewable.
 \end{quote}
\end{minipage}

\bigskip

\begin{figure}[htbp]
\setcaption{{\sc Figure 4\,-\,3.} A tessellated oriented s-d-l Lorentz
  surface decorated with CBN-loops. The latter can be cyclically
  labelled.   }
\centerline{\psfig{figure=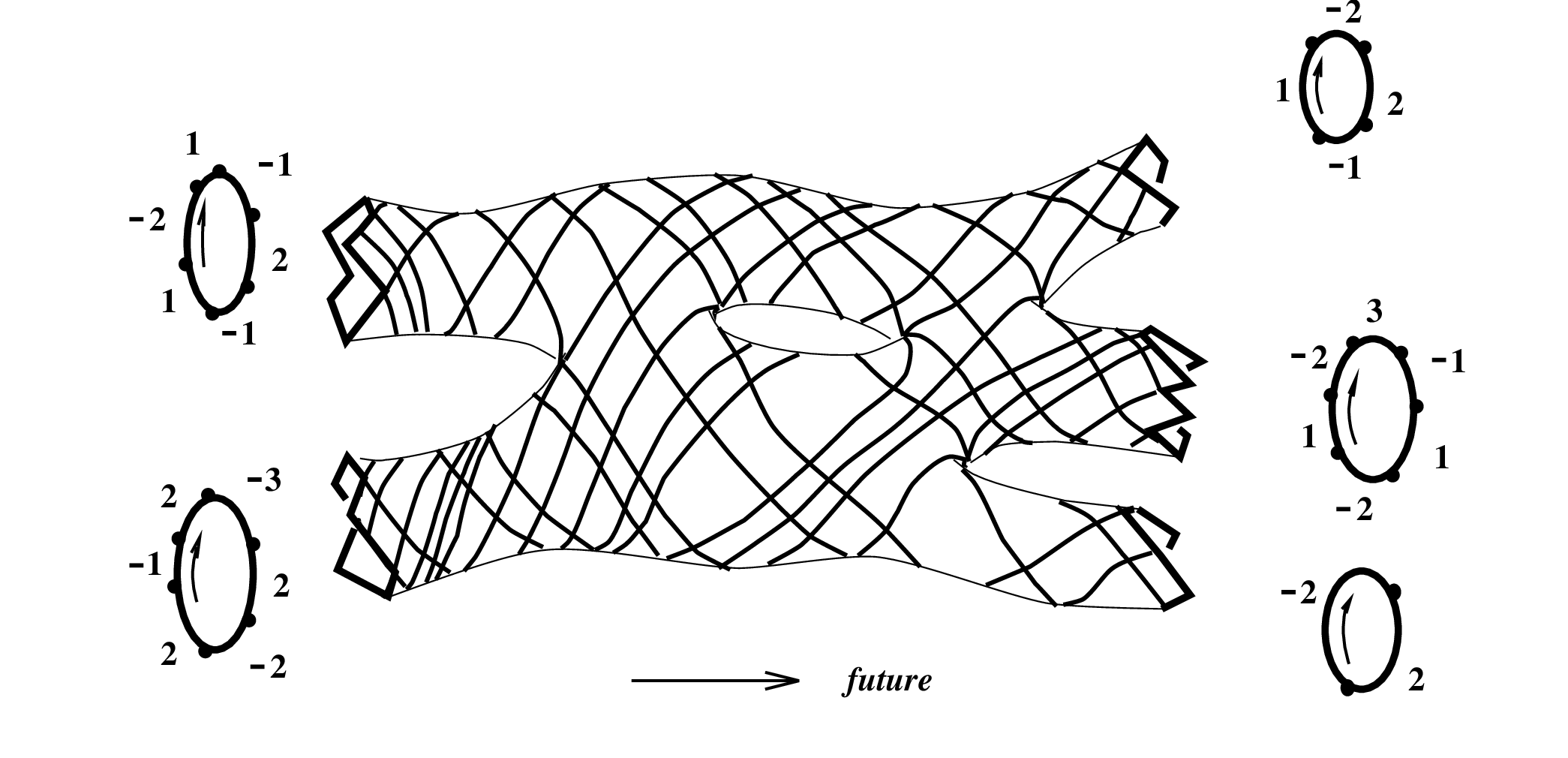,width=13cm,caption=}}
\end{figure}

\bigskip

\noindent\hspace{.3cm}
(c) \parbox[t]{14cm}{
   Category $\mbox{\it B-Link}^{\,\uparrow}$ of {\it bonded d-links}
   (cf.\ Item ($3^{\circ}$)) ({\sc Figures 4\,-\,1} and 4\,-\,4):
   Here a {\it d-link} means a finite disjoint union of directed
   circles. A {\it $k$-bond} is an {\it ordered $k$-prong}. A
   {\it bonded d-link} is a d-link with a finite collection of bonds
   whose ends are attached to the link. There are two kinds of
   {\it simple twists} of bonds: {\it forward}, $\mbox{\it twist}^+$;
   and {\it backward}, $\mbox{\it twist}^-$, as indicated in
   {\sc Figure 4\,-\,4}. Simple twists
   associated to a same bond are inverse of each other; and they
   transform one bonded d-link to another.    }

\medskip

\noindent\hspace{1.2cm}
\begin{minipage}{14cm}
 \begin{quote}
   \hspace{-.8cm}{\it Obj}: The set of bonded d-links.

   \hspace{-.8cm}{\it Mor}: A morphism is a finite sequence of bonded
      d-links obtained by a sequence of simple twists.
 \end{quote}
\end{minipage}

\bigskip

\noindent\hspace{1cm}
\parbox[t]{13.9cm}{It is instructive to think of
$\mbox{\it B-Link}^{\,\uparrow}$ as a dual category of
${\cal C}_{\rm t\mbox{-}graph}$. In some sense,
$\mbox{\it B-Link}^{\,\uparrow}$ and ${\cal C}_{\rm t\mbox{-}graph}$ are
transverse to each other. In the orientable case, the bonded directed
link associated to $\Sigma$ appears as a pair - (left, right) -
(cf.\ Example 4.2).  }

\begin{figure}[htbp]
\setcaption{{\sc Figure 4\,-\,4.} Bonded d-links and simple twists
            of bond. }
\centerline{\psfig{figure=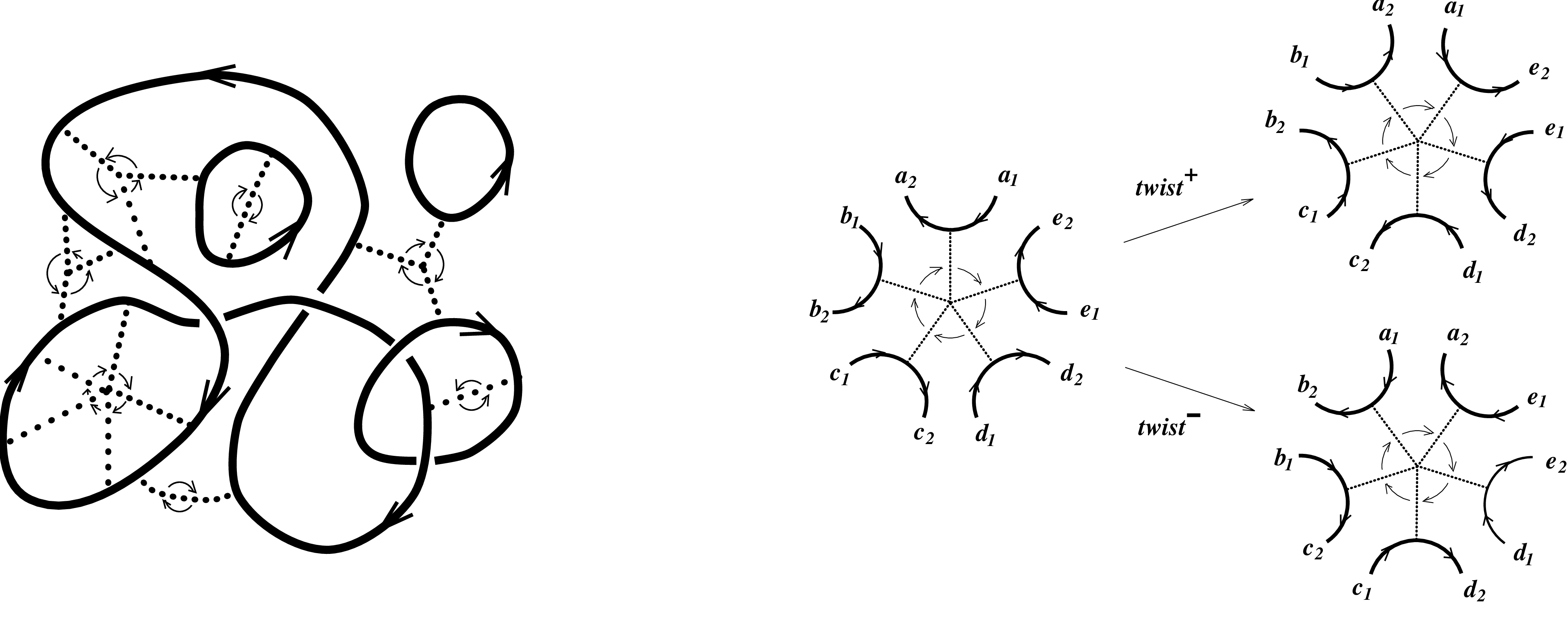,width=13cm,caption=}}
\end{figure}

\bigskip

\noindent\hspace{.3cm}
(d) {\it Extended structures} (cf.\ Items ($1^{\circ}$),
($2^{\circ}$), and ($3^{\circ}$)).

\bigskip

Let us explain Item (d) in brief. Recall that in
{\it extended topological quantum field theories} (ETQFT) at dimension
$d$, one considers generalized path-integrals by associating
{\it higher algebraic structures} to higher co-dimensional manifolds in
a way that satisfies complicated consistency relations due to different
ways of decomposing a manifold (e.g.\ [Fr1], [Fr2], [La] for more
details and terminology). For LCFT that involve only orientable Lorentz
surfaces, the geometric objects involved at various dimensions are:

\vspace{.8cm}

\centerline{
 \begin{minipage}[b]{13.5cm}{
 \begin{tabular}{cccr}
    dimension \rule[-1ex]{0ex}{1ex}
           & & & \multicolumn{1}{c}{geometric object}\\ \hline
    $2$  & & &
       \begin{minipage}[t]{10cm}\rule{0ex}{4ex}{oriented s-d-l Lorentz
          surface $\Sigma$ as in $\mbox{\it Mor}\,({\cal C}_{\rm Lorz})$
          with the decorating loops $C_{\alpha}$ parametrized from
          left to right relative to the future direction }
       \end{minipage} \\
    $1$  & & &
       \begin{minipage}[t]{10cm}\rule{0ex}{4ex}{$S^1$ with either the
          left- or the right-set of marked points labelled with natural
          numbers (cf.\ Item ($1^{\circ}$)) }
       \end{minipage} \\
    $0$  & & &
       \begin{minipage}[t]{10cm}\rule{0ex}{4ex}{coarse conformal
         class $[p]$ of germs of Lorentzian disks around singularity
         $p$}
       \end{minipage}
 \end{tabular}    }
 \end{minipage} \ .
} 

\vspace{1cm}

For the coarse category, $[p]$ is determined by the index of $p$;
hence, the set of geometric objects at dimension zero is parametrized
by ${\Bbb N}$ by assigning $s$ to $[p]$ for $p$ of index $-s$.
For higher dimensions, consider the left sector first. At dimension
one, one concerns about $S^1$ with a left set of labelled marked points
only up to $\mbox{\it Diff}^{\,+}(S^1)$. The set of such classes is
parametrized by
$O^{\hspace{-4pt}\bullet}_L\:=\:\coprod_k\,
           \mbox{\raisebox{.3ex}{${\Bbb N}^k\hspace{-1pt}/$}
               \hspace{-5pt}\raisebox{-.3ex}{\,${\Bbb Z}_k$} }$,
where ${\Bbb Z}_k$ acts on ${\Bbb N}^k$ by cyclically shifting the
coordinates generated by
$(i_1,i_2,\ldots,i_k)\mapsto(i_k,i_1,i_2,\ldots,i_{k-1})$.
Denote an element in $O^{\hspace{-4pt}\bullet}_L$ by
$[i_1,\ldots,i_k]_L$. There are {\it forgetful maps} defined on
$O^{\hspace{-4pt}\bullet}_L$ by deletions of entries. This makes
$O^{\hspace{-4pt}\bullet}_L$ a {\it directed set} by defining
$[j_1,\ldots,j_l]_L \rightarrow [i_1,\ldots,i_k]_L$ if
$[i_1,\ldots,i_k]_L$ is obtainable from $[j_1,\ldots,j_l]_L$ by
a forgetful map. At dimension two, there is a unique
$[i_1,\ldots,i_k]_L$ associated to each $C_{\alpha}$ in
$(\Sigma;(C_{\alpha})_{\alpha})$ obtained by considering the left
characteristic leaves through $C_{\alpha}$ following its orientation
(cf.\ Item ($1^{\circ}$)). And similarly for the right sector.

Thus, as in ETQFT, a prototype for LCFT that takes all dimensions
into account consists of the following data. For simplicity, we
assume all the categories of algebraic structures appearing in the
setting are subcategories in ${\cal C}_{{\rm mod},R}$ :

\medskip

\noindent\hspace{.5cm}$\bullet$
\parbox[t]{14cm}{
   An assignment to each $s\in{\Bbb N}$ a {\it category} $Z(s)$,
   (e.g.\
   $\mbox{\it Rep}\,(\mbox{\rm Sym}_s)$ or
   $\mbox{\it Rep}\,({\Bbb Z}_s)$). The choice of $Z(s)$ should be
   related to the interacting of $s$ strings. }

\bigskip

\noindent\hspace{.5cm}$\bullet$
\parbox[t]{14cm}{
   A {\it rule}
   $$
   {\cal V}\;:\;\coprod_k\,
    \mbox{\raisebox{.3ex}{$\left.
     \left(2^{{\cal C}_{{\rm mod},R}}\right)^k\hspace{-2pt}\right/$}
      \hspace{-5pt}\raisebox{-.7ex}{${\Bbb Z}_k$} }
   \;\longrightarrow\;2^{{\cal C}_{{\rm mod},R}}\,,
   $$
   where ${\Bbb Z}_k$ acts on
   $\left(2^{{\cal C}_{{\rm mod},R}}\right)^k$ by cyclic shift; and an
   assignment to each $[i_1,\ldots,i_k]_L$ in
   $O^{\hspace{-4pt}\bullet}_L$ an element $Z[i_1,\ldots,i_k]_L$ in
   ${\cal V}[Z(i_1),\ldots,Z(i_k)]$. Similarly for the right sector. }

\bigskip

\noindent\hspace{.5cm}$\bullet$
\parbox[t]{14cm}{
   An assignment to each $(\Sigma;(C_{\alpha})_{\alpha})$ a pair of
   {\it homomorphisms} - left and right -, one from a tensor product of
   $Z[i_1,\ldots,i_k]_L$'s related to the incoming $C_{\alpha}$ to that
   related to the outgoing ones and the other similarly for the right
   sector.  }

\bigskip

\noindent\hspace{.5cm}$\bullet$
  {\it Consistency conditions:}

   \begin{quote}
    (i) {\it Naturality from forgetful maps}: Let $I$ indicate the type
    of forgetfulness. The following diagrams commute:
    $$
    \begin{array}{cccl}
    \mbox{\raisebox{.3ex}{$\left.
    \left(2^{{\cal C}_{{\rm mod},R}}\right)^l
    \hspace{-2pt}\right/$}\hspace{-3pt}\raisebox{-.7ex}{${\Bbb Z}_l$}}
    \mbox{\hspace{1cm}} &\longrightrightarrow
    &\mbox{\hspace{1cm}} \mbox{\raisebox{.3ex}{$\left.
    \left(2^{{\cal C}_{{\rm mod},R}}\right)^k
    \hspace{-2pt}\right/$}\hspace{-3pt}\raisebox{-.7ex}{${\Bbb Z}_k$}}
                                                                  & \\
    & & &  \\
    \mbox{\hspace{36pt}$\searrow$} & {\cal V}
         &\mbox{\hspace{-60pt}$\swarrow$}   &  \\
    & & &  \\
    & 2^{{\cal C}_{{\rm mod},R}} & &,
    \end{array}
   $$
   where $l>k$ and "$\longrightrightarrow$" are induced by the forgetful
   maps of deletion of components. And, with abuse of terminology, there
   exists a family of {\it forgetful functors} $F=\{f_I\}_I$ defined on
   $\coprod_k\,{\cal V}\left(
     \mbox{\raisebox{.3ex}{$\left.
     \left(2^{{\cal C}_{{\rm mod},R}}\right)^k\hspace{-2pt}\right/$}
      \hspace{-5pt}\raisebox{-.7ex}{${\Bbb Z}_k$} }\right)$,
   that conjugates $\longrightrightarrow$ with
   $$
    f_I\;:\; {\cal V}\left(
    \mbox{\raisebox{.3ex}{$\left.
     \left(2^{{\cal C}_{{\rm mod},R}}\right)^l
    \hspace{-2pt}\right/$}\hspace{-3pt}
         \raisebox{-.7ex}{${\Bbb Z}_l$}}\right)\; \longrightarrow \;
    {\cal V}\left(\mbox{\raisebox{.3ex}{$\left.
                       \left(2^{{\cal C}_{{\rm mod},R}}\right)^k
    \hspace{-2pt}\right/$}\hspace{-3pt}
                 \raisebox{-.7ex}{${\Bbb Z}_k$}}   \right)
   $$
   for some appropriate $k$, $l$ such that if
   $[j_1,\ldots,j_l]\rightarrow [i_1,\ldots,i_k]$ is of type $I$ then
   $$
    f_I\left({\cal V}[Z(j_1),\ldots,Z(j_l)]\right)\;
                         =\; {\cal V}[Z(i_1),\ldots,Z(i_k)]
   $$
   and
   $$
    f_I(Z[j_1,\ldots,j_l])\; = \;Z[i_1,\ldots,i_k]\,.
   $$
   Similarly for the right sector. Different decompositions of a
   forgetful map are required to lead to the same result.

   (ii) {\it {}From sewings}: Due to the fact that sewing increases
   marked points associated to $C_{\alpha}$, sewing of two composable
   $\Sigma_1$, $\Sigma_2$ now leads to a $\Sigma_1\diamond\Sigma_2$
   whose $Z(\Sigma_1\diamond\Sigma_2)$ has a domain and image modules
   different from the domain of $\Sigma_1$ and the image of $\Sigma_2$
   respectively ({\sc Figure 4\,-\,5}). Consistency conditions arise
   from the fact that a $\Sigma$ could have more than one (though,
   recall that, at most finitely many) non-equivalent simple cut
   system (cf.\ Sec.\ 2.3) and, hence, could admit different sewing
   patterns. $Z(\Sigma)$ should be indifferent of ways of such
   decompositions.
  \end{quote}

\bigskip

\noindent
The whole format extends that in Proto-definition 4.1.

\begin{figure}[htbp]
\setcaption{{\sc Figure 4\,-\,5.} Sewing of $\Sigma_1$ and $\Sigma_2$
  leads to new initial and final collections of circles with marked
  points. --- indicates original characteristic null trajectories; while
  $\cdots$ indicates those new ones due to sewing. Only left sector is
  shown. $\bullet$ =  original marked point;
  $\circ$ = new one due to sewing.  }
\centerline{\psfig{figure=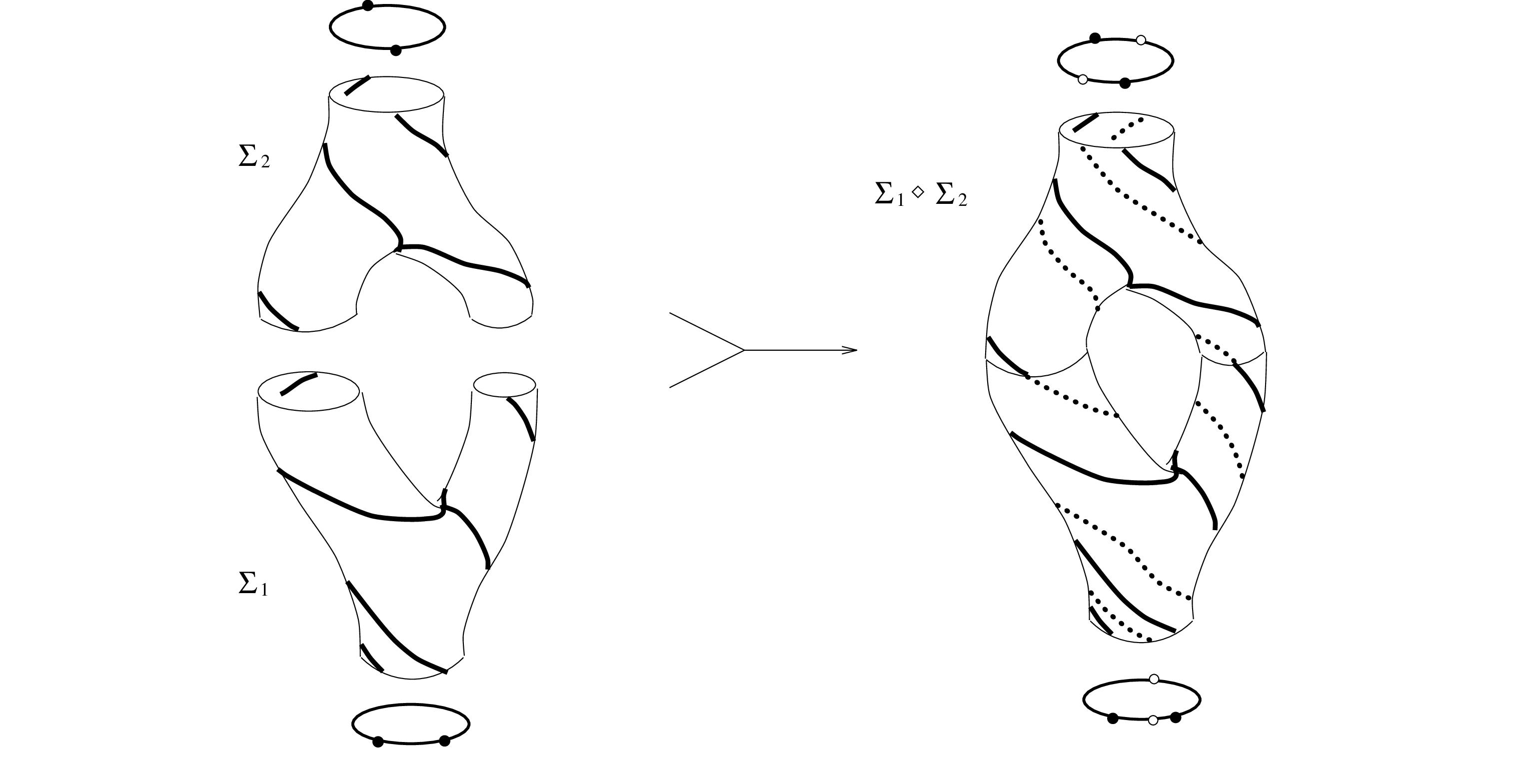,width=13cm,caption=}}
\end{figure}

\bigskip

\noindent
{\it Remark 4.3.} Notice that there are {\it "pinching functors"} from
${\cal C}_{\rm tes-Lorz}$, $\mbox{\it B-Link}^{\,\uparrow}$, and
${\cal C}_{\rm Lorz}$ respectively to ${\cal C}_{\rm t\mbox{-}graph}$.
Physically, ${\cal C}_{\rm t\mbox{-}graph}$ is the most fundamental
category in the theory; all the rest should be its extensions.

\bigskip

\noindent
{\it Remark 4.4.} It should be noted that the space of morphisms in
these categories and natural bundles thereover are among the major
things for study, following the spirit of [F-S] for CFT.

\bigskip

We shall leave more thorough and detailed studies to the future and
conclude the paper here with the wish of rich Lorentzian CFT and
un-Wick-rotated string theory as their Riemannian siblings.

\vspace{4cm}

\begin{flushleft}
{\large\bf Appendix. Quantization of string phase space.}
\end{flushleft}
A most natural quantization for string phase space is through geometric
quantization. Though a complete setting is still beyond grasp at the
moment, some manipulations for the case of finite dimensional phase
spaces go through formally. We shall explore them in this appendix.
We discuss only closed strings and assume that $M$ is Riemannian.

\bigskip

\begin{flushleft}
{\bf Prequantum line bundles.}
\end{flushleft}
Let $\mbox{\rm ev}\,: LM\times S^1 \rightarrow M$ be the
{\it evaluation map} $\mbox{\rm ev}\,(\phi,\sigma)=\phi(\sigma)$. A
$k$-form $\nu$ on $M$ induces a $(k-1)$-form, still denoted by
$\nu$, on $LM$ by setting [Br]
$$
\nu|_{LM}\;=\;\int_{S^1}\,\mbox{\rm ev}^{\ast}\nu|_M\,.
$$
Explicitly, for $Z_i$, $i=1,\cdots, k-1$, in $T_{\phi}LM$,
$$
\nu|_{LM}(Z_1,\ldots,Z_{k-1})\;
  =\;\int_{S^1}\,d\sigma\,\nu|_M(Z_1(\sigma),\ldots,Z_{k-1}(\sigma),
           \phi_{\ast}\partial_{\sigma})\,.
$$
In this way $B$ is regarded as a 1-form on $LM$ and $dB$ a 2-form.
Their pull-back to $LT^{\ast}M$ via projection map shall be denoted the
same. There is also a {\it section-evaluation map}
$\mbox{\rm sev}:S^1\times LT^{\ast}M \rightarrow S^1\times T^{\ast}M$
with $\mbox{\rm sev}\,(\sigma,\gamma)=(\sigma,\gamma(\sigma))$. Recall
$\theta$ the Liouville 1-form on $T^{\ast}M$ and {\boldmath $\theta$}
be the Liouville 1-form on $LT^{\ast}M$. One has
$\mbox{{\boldmath $\theta$}}
    =\int_{S^1}\mbox{\rm sev}^{\ast}d\sigma\wedge\theta$ and
$\mbox{\boldmath $\omega$}=
            \int_{S^1}\mbox{\rm sev}^{\ast}d\sigma\wedge\omega$.
Let $\mbox{\boldmath $\theta$}_B=\mbox{\boldmath $\theta$}+B$,
$\mbox{\boldmath $\omega$}_B=\mbox{\boldmath $\omega$}+dB$, and
${\cal H}_0(\phi,\pi;\sigma)=
  \frac{1}{2}\langle\pi(\sigma),\pi(\sigma)\rangle^{\sim}
  +\frac{1}{2}\langle\phi_{\ast}\partial_{\sigma},
                         \phi_{\ast}\partial_{\sigma}\rangle$.
Then the map $(\phi,\pi)\mapsto (\phi,\pi-B_{\phi})$ is an equivalence
from $(LT^{\ast}M,\mbox{\boldmath $\omega$},{\cal H})$ to
$(LT^{\ast}M,\mbox{\boldmath $\omega$}_B,{\cal H}_0)$ since it pulls
back $\mbox{\boldmath $\theta$}_B$ to {\boldmath $\theta$} and
${\cal H}_0$ to $\cal H$. The string system as given resembles that of
a particle moving in a Riemannian manifold with an external
electromagnetic field.

\bigskip

\noindent
{\it Remark A.1.} Implicit in the validity of the same notation
for a $k$-form on $M$ and its induced $(k-1)$-form on $LM$ is the
commutativity relation:
$$
 d\,\int_{S^1}\,\mbox{\rm ev}^{\ast}\;
       =\; \int_{S^1}\,\mbox{\rm ev}^{\ast}\,d\,,
$$
which follows from the fact that the difference of the two sides is
$\int_{S^1}{\cal L}_{\partial_{\sigma}}$, where $\cal L$ here means
the Lie derivative, and this integral vanishes. (cf.\ Gysin sequence
of sphere-fibration.)

\bigskip

\noindent
{\it Remark A.2.} Observant readers may notice that the
$\mbox{\it Diff}\,(S^1)$-action on
$(LT^{\ast}M,\mbox{\boldmath $\omega$})$ by reparametrization is not
symplectic. This is an obvious defect of the setting.

\bigskip

Recall that a {\it prequantum line bundle} over a symplectic manifold
is a Hermitian line bundle with a connection over that manifold whose
curvature equals the symplectic 2-form up to a conventional factor
($\hbar^{-1}$ in [Wo]). Such a line bundle with connection does not
always exist. When it does, the symplectic manifold is said to be
{\it quantizable}.

\bigskip

\noindent{\bf Assertion A.3 [quantizability].} {\it
The infinite dimensional symplectic manifold
$(LT^{\ast}M,\mbox{\boldmath $\omega$}_B)$ is quantizable.  }

\bigskip

\noindent{\it Reason.} Consider the trivial Hermitian line bundle
${\Bbb L}=LT^{\ast}M\times{\Bbb C}$ over $LT^{\ast}M$. Let
$\gamma_{\tau}$ be a path in $LT^{\ast}M$ and $\gamma_{\tau}(\sigma)$
be its realization in $T^{\ast}M$. Given $z_0\in\Bbb C$, the unique
solution to the first order differential equation
$$
\frac{d}{d\tau}z(\tau)\;=\;
  \frac{i}{\hbar}\int_{S^1}\,d\sigma\, \left\{
    \theta_{\gamma_{\tau}(\sigma)}(\partial_{\tau})\,
       +\,B(\partial_{\tau},\gamma_{\tau\,\ast}\partial_{\sigma})
            \right\}\; \mbox{ with }\;z(0)\,=\,z_0
$$
defines a unique lifting
$\widetilde{\gamma_{\tau}}=(\gamma_{\tau}\,,\,z(\tau))$ of
$\gamma_{\tau}$ in $\Bbb L$ and hence a horizontal distribution
therein. The parallel tranports it generates are unitary due to the
factor $i$. This defines a compatible connection
$\mbox{\boldmath $\nabla$}^B$ in $\Bbb L$ with
$\mbox{\boldmath $\nabla$}^B
               =d-\frac{i}{\hbar}\mbox{\boldmath $\theta$}_B$.
We now check that the curvature of $\mbox{\boldmath $\nabla$}^B$ is
indeed $\hbar^{-1}\mbox{\boldmath $\omega$}_B$.

Let {\boldmath $\alpha$} be the $\Bbb C$-valued connection 1-form in
$\Bbb L$ associated to $\mbox{\boldmath $\nabla$}^B$ and
$\Pr_{\supbscriptsizeBbb C}:\Bbb L\rightarrow\Bbb C$ be the
projection to the $\Bbb C$-component. Then, explicitly,
$$
\mbox{\boldmath $\alpha$}\;=\;
    \Pr\,\hspace{-.3em}_{\supbscriptsizeBbb C\ast}\,
              -\,\frac{i}{\hbar}\mbox{\boldmath $\theta$}_B\,,
$$
where we use the same notation to denote the pullback 1-form in
$\Bbb L$ of $\mbox{\boldmath $\theta$}_B$ and identify the tangent
space of any point in $\Bbb C$ with $\Bbb C$ itself canonically. Let
$\Pr_H$ be the horizontal projection of tangent vectors in $\Bbb L$ to
the horizontal distribution. Then the 2-form
$i\,d\mbox{\boldmath $\alpha$}\circ\Pr_H$ on $\Bbb L$ descends to the
curvature 2-form on $LT^{\ast}M$. On the other hand, observe that
$d\,\Pr_{\supbscriptsizeBbb C\ast}\,=\,0$ due to the fact that
$\Pr_{\supbscriptsizeBbb C}$ is a coordinate
function and hence its differential as a 1-form has to be closed. Thus,
\newline
\hspace{-1.5cm}\parbox{13cm}{
\begin{eqnarray*}
 \lefteqn{i\,d\mbox{\boldmath $\alpha$}\;
     =\;i\,d\,\Pr\,\hspace{-.3em}_{\supbscriptsizeBbb C\ast}\:
               +\:\hbar^{-1}d\mbox{\boldmath $\theta$}_B\;
     =\;\hbar^{-1}\,d\int_{S^1}\,\left\{
       \mbox{\rm sev}^{\ast}d\sigma\wedge\theta \,
          +\,\mbox{\rm ev}^{\ast}B \right\}   }\\
   & & =\;\hbar^{-1}\,\int_{S^1}\,\left\{
         \mbox{\rm sev}^{\ast}d\sigma\wedge\omega\,
                +\,\mbox{ev}^{\ast}dB  \right\}\;
          =\;\hbar^{-1}\mbox{\boldmath $\omega$}_B \,.
\end{eqnarray*}        }\newline
Consequently, $i\,d\mbox{\boldmath $\alpha$}\circ\Pr_H$ descends to
$\hbar^{-1}\mbox{\boldmath $\omega$}_B$ on $LT^{\ast}M$. This concludes
the reason.

\noindent\hspace{14cm}$\Box$

\bigskip

\noindent{\it Remark A.4.} Notice that $\mbox{\boldmath $\nabla$}^B$
is flat along every fiber $T_{\phi}^{\ast}LM$ of $LT^{\ast}M$. Let
{\boldmath $\nabla$} be $\mbox{\boldmath $\nabla$}^B$ with $B=0$; then
the map $(\phi,\pi;z)\mapsto (\phi,\pi-B_{\phi};z)$ gives a
bundle-with-connection isomorphism from
$({\Bbb L},\mbox{\boldmath $\nabla$})$ to
$({\Bbb L},\mbox{\boldmath $\nabla$}^B)$. In general, $\pi_1(M)$ and
$\pi_2(M)$ are non-trivial; and hence there can be non-equivalent
prequantum line bundles over $(LT^{\ast}M,\mbox{\boldmath $\omega$}_B)$.

\bigskip

\begin{flushleft}
{\bf Geometric quantization and string field theory.}
\end{flushleft}
Geometric quantization of the string phase space
$(LT^{\ast}M, \mbox{\boldmath $\omega$}, {\cal H})$ (or its equivalent)
can be regarded as a {\it geometrization of string field theory}.
Sections in $\Bbb L$ are candidates for string fields (or {\it string
wave functions}). An {\it observable} corresponding to a measurable
physical quantity given by a real-valued function $\cal F$ on
$LT^{\ast}M$ is the operator $\widehat{\cal F}$ acting on sections $s$
of $\Bbb L$ by
$$
 \widehat{\cal F}\,s\;=\;
   -i\hbar\mbox{\boldmath $\nabla$}_{X_{\cal F}}s\:+\:{\cal F}s\,,
$$
where $X_{\cal F}$ is the Hamiltonian vector field generated by
$\cal F$. It is the infinitesimal generator for the one-parameter group
action $\hat{\rho}_t$ on sections in $\Bbb L$ by
$$
\hat{\rho}_ts\,(\gamma)\;=\;s(\rho_t\gamma)\,e^{-\frac{i}{\hbar}
  \int_0^t\,dt^{\prime}\,L_{\cal F}}\,,
$$
where $\rho_t$ is the flow generated by $X_{\cal F}$ and
$L_{\cal F}\;=\;\mbox{\boldmath{$\theta$}}(X_{\cal F})\,-\,{\cal F}$
is the Lagrangian of $\cal F$ and the integration
$\int_0^t\,dt^{\prime}$ is taken along the flow $\rho_t$ from $\gamma$
to $\rho_t\gamma$. However, there are subtleties in this naive picture.

{}From the standard geometric quantization ([Wo] for details), one
learns that the {\it polarization} $\cal P$ of $LT^{\ast}M$ by vertical
fibers $T_{\phi}^{\ast}LM$ has to be introduced. Only those sections in
$({\Bbb L},\mbox{\boldmath $\nabla$})$ that are flat along $\cal P$
could be {\it physical}. They are called
{\it $\cal P$-polarized sections} and are string fields that come
from those over $LM$. The {\it canonical line bundle} $K_{\cal P}$
associated to $\cal P$ and its square root
$\delta_{\cal P}=\sqrt{K_{\cal P}}$ also have to be introduced.
Sections $\nu$ in $\delta_{\cal P}$ are {\it half-forms} on $LM$ and one
replaces $\Bbb L$ by ${\Bbb L}_{\cal P}={\Bbb L}\otimes\delta_{\cal P}$.
Since most observables do not preserve $\cal P$, one needs to introduce
a {\it pairing} between $\cal P$-polarized sections $\tilde{s}=\psi\nu$
in ${\Bbb L}_{\cal P}$ and ${\cal P}^{\prime}$-polarized sections
$\tilde{s}^{\prime}=\psi^{\prime}\nu^{\prime}$ in
${\Bbb L}_{{\cal P}^{\prime}}$ for another polarization
${\cal P}^{\prime}$ transverse to $\cal P$. It is defined by
$$
 (\widetilde{s},\widetilde{s}^{\prime})\;=\;
   \int_{LT^{\ast}M}\overline{\psi}\psi^{\prime}\,(\nu,\nu^{\prime})\,
          \mbox{\it vol}_{\mbox{\boldmath $\omega$}}\,,
$$
where $(\nu,\nu^{\prime}) = \sqrt{
       \mbox{\raisebox{.2ex}{${\nu^{\prime}}^2\wedge\bar{\nu}^2/$}}
            \hspace{-3pt}\mbox{\raisebox{-.2ex}{
                $\mbox{\it vol}_{\mbox{\boldmath $\omega$}}$} } }$
and $\mbox{\it vol}_{\mbox{\boldmath $\omega$}}$ is the symplectic
volume form on $LT^{\ast}M$. Such pairing allows one to project
${\cal P}^{\prime}$-polarized sections in ${\Bbb L}_{{\cal P}^{\prime}}$
to $\cal P$-polarized sections in ${\Bbb L}_{\cal P}$. Finally, there
is the {\it metaplectic correction} to give a more coherent treatment of
the half-forms with respect to various polarizations.

Another subtlety arises from {\it symmetries}: (1) the missing but
required symmetry of $\mbox{\it Diff}\,(S^1)$ due to
{\it reparametrizations}; and (2) the manifest {\it conformal} symmetry
of the theory (cf.\ Sec.\ 1). A complete program should contain a
prescription of how to restore the first symmetry and the final
quantities extracted from the setting should be
parametrization-independent. The second symmetry suggests an
extension ${\Bbb L}^{\prime}$ of $\Bbb L$ to include
{\it anti-commuting fields} (ghosts) and a {\it BRST operator} that acts
on sections of ${\Bbb L}^{\prime}$. Only the BRST (co)homology classes
are significant. The Hilbert space $\cal H$ of physical states of the
theory has now a trinity nature: first, it appears usually as a
representation of a graded algebra depending on the target-space $M$;
second, its elements as $\cal P$-polarized sections in
${\Bbb L}^{\prime}_{\cal P}$ should be a generalization of
square-integrable functions in the case of finite dimensional
configuration spaces; and third, these elements are BRST-(co)homology
classes. Unfortunately, not all sutleties are resolvable at the moment.
Nevertheless, the most fundamental object - the Hamiltonian operator on
string fields - can be constructed at the formal level.

\bigskip

\begin{flushleft}
{\bf BKS-construction and the Schr\"{o}dinger equation.}
\end{flushleft}
The string Hamiltonian $\cal H$ is quadratic with respect to the
momentum variable $\pi$; and hence the flow it generates does not
preserve the vertical polarization $\cal P$ in $LT^{\ast}M$. The
{\it Blattner-Kostant-Sternberg-} ({\it BKS-}) {\it construction}
is developed to remedy this ([\'{S}n], [Wo]).

The metric $ds^2$ on $M$ induces a metric on $LM$ via the map
$\mbox{\rm sev}$. Let $\mbox{\it vol}_{LM}$ be the metric volume
form on $LM$ and $\sqrt{\mbox{\it vol}_{LM}}$ be a fixed half-form
associated to $\mbox{\it vol}_{LM}$. A physical section $\tilde{s}$ in
${\Bbb L}_{\cal P}$ can now be written as the pull-back of
$$
\tilde{s}\;=\;\psi\,\sqrt{\mbox{\it vol}_{LM}}
$$
by the projection from $LT^{\ast}M$ to $LM$. Denote the pull-back
section by the same notation. Let $\rho_t$ be the string Hamiltonian
flow on $LT^{\ast}M$ and $\tilde{\rho}_t$ be its induced action on
physical sections in ${\Bbb L}_{\cal P}$ defined by
$$
 \tilde{\rho}_t\,\tilde{s}\;
        =\;\psi_t\,\rho_t^{\ast}\sqrt{\mbox{\it vol}_{LM}}\,,
$$
where
$$
 \psi_t\,(\gamma)\;=\;\psi(\rho_t\gamma)\,
           e^{-\frac{i}{\hbar}\,\int_0^t\,dt^{\prime}\,L_{\cal H}}
$$
with $L_{\cal H}$ the phase-space string Lagrangian. Due to the fact
that the flow $\rho_t$ does not preserve the polarization, the driven
section $\tilde{\rho}_t\,\tilde{s}$ is in general no longer physical.
The pairing between driven and not-driven physical sections,
$\tilde{\rho}_t\tilde{s}$ and $\tilde{s}^{\prime}$ now becomes
$$
 \left(\tilde{\rho}_t\,\tilde{s}\,,\,\tilde{s}^{\prime}\right)\;
   =\; \int_{LT^{\ast}M}\,
     e^{\frac{i}{\hbar}\int_0^t\,dt^{\prime}\,L_{\cal H}}\,
         \overline{\psi\circ\rho_t}\,\psi^{\prime}
    \sqrt{(\rho_t^{\ast}\mbox{\it vol}_{LM},\mbox{\it vol}_{LM})}\,
       \mbox{\it vol}_{\mbox{\boldmath $\omega$}}\,.
$$
One would like to rewrite this integral as an integral over the
configuration space $LM$ of the form:
$$
\int_{LM}\,
  \overline{\left\{\mbox{\rm Id}\,
       -\,\frac{it}{\hbar}{\cal O}_{\cal H}\,
      +\,O(t^2)\right\}\psi}\psi^{\prime}\,\mbox{\it vol}_{LM}\,,
$$
where ${\cal O}_{\cal H}$ depends only on $\cal H$. The string
Hamiltonian operator is then ${\cal O}_{\cal H}$; and the
Schr\"{o}dinger equation reads
$$
 i\hbar\frac{\partial\psi}{\partial t}\;=\;{\cal O}_{\cal H}\psi\,.
$$

Given $t\in{\Bbb R}-\{0\}$. To carry out the above construction, it
turns out more natural to work on the equivalent system
$(LT^{\ast}M, \mbox{\boldmath $\omega$}_B^{(t)}, {\cal H}_0^{(t)})$,
where
$$
\mbox{\boldmath $\omega$}_B^{(t)}\;
   =\;\frac{1}{t}\mbox{\boldmath $\omega$}\,+\,dB
$$
and
$$
 {\cal H}_0^{(t)}(\phi,\pi)\; =\;\frac{1}{2t^2}\int_{S^1}\,d\sigma\,
   \langle \pi,\pi \rangle^{\sim} \,+\,\frac{1}{2}\int_{S^1}\,d\sigma\,
    \langle \phi_{\ast}\partial_{\sigma},
                              \phi_{\ast}\partial_{\sigma} \rangle\,.
$$
The potential 1-form associated to $\mbox{\boldmath $\omega$}_B^{(t)}$
is now
$\mbox{\boldmath $\theta$}_B^{(t)}
 =\frac{1}{t}\mbox{\boldmath $\theta$}+B$.
Assume that $\mbox{\it dim}\,M\geq 3$ and that $\phi$ is generic; hence,
an embedding. Fix a {\it Fermi coordinate system} $x$ [Hi] in a tubular
neighborhood $U$ of $\phi$ in $M$ by choosing an orthonormal frame
$\{e_i\}$ along $\phi$ with $e_1$ the unit tangent vector of $\phi$.
This then induces a trivialization
$\{(\phi^i(\sigma),\pi_i(\sigma))\,|\,\sigma\in [0,2\pi)\}$
of $LT_U^{\ast}M$ . With respect to this, for
$(Y,Z)\in T_{(\phi,\pi)}LT^{\ast}M$ with $Y$ the horizontal component
and $Z$ the vertical component, one has
$$
 d{\cal H}_0^{(t)}|_{(\phi,\pi)}(Y,Z)\;
    =\;\frac{1}{t^2}\int_{S^1}\,d\sigma\,
      \langle Z, \pi(\sigma)\rangle^{\sim}\:-\:
          \int_{S^1}\,d\sigma\, \langle Y,
            \nabla_{\partial_{\sigma}}\partial_{\sigma} \rangle\,.
$$
The correspondence between $T^{\ast}LT^{\ast}M$ and
$T_{\ast}LT^{\ast}M$ induced from $\mbox{\boldmath $\omega$}_B^{(t)}$
is given by
\begin{eqnarray*}
 d\phi^i  & \longrightarrow  & -t\frac{\partial}{\partial\pi_i}\\
 d\pi_i  & \longrightarrow
   & t\frac{\partial}{\partial\phi^i}\:
       -\:t^2\left(i_{\frac{\partial}{\partial\phi^i}}
            i_{\phi_{\ast}\partial_{\sigma}}dB\right)_j\,
               \frac{\partial}{\partial\pi_j}\,;
\end{eqnarray*}
and hence
\begin{eqnarray*}
 \lefteqn{ \left. X_{{\cal H}_0^{(t)}}\right|_{(\phi,\pi)}\;
    =\; \frac{1}{t}\pi^{\sim} \,
           -\,i_{\pi^{\sim}}i_{\phi_{\ast}\partial_{\sigma}} dB \,
        +\, t \left(\nabla_{\phi_{\ast}\partial_{\sigma}}
                     \phi_{\ast}\partial_{\sigma}\right)^{\sim}  }\\
   & & =\; \frac{1}{t}\pi^i\frac{\partial}{\partial\phi^i}\,
     -\,\pi^i \left(i_{\frac{\partial}{\partial\phi^i}}
            i_{\phi_{\ast}\partial_{\sigma}}dB\right)_j
            \frac{\partial}{\partial\pi_j}\,
     +\, t \left( \nabla_{\phi_{\ast}\partial_{\sigma}}
              \phi_{\ast}\partial_{\sigma}\right)^{\sim}_{\;i}
                    \frac{\partial}{\partial\pi_i}\,.
\end{eqnarray*}
We shall denote its horizontal part by $Y_{\cal H}$ and its vertical
part by $Z_{\cal H}$.

The phase $L_{\cal H}$ now becomes
$\mbox{\boldmath $\theta$}_B^{(t)}(X_{{\cal H}_0^{(t)}})
                                          -{\cal H}_0^{(t)}$.
Straightforward computation gives
$$
 L_{{\cal H}_0^{(t)}}(\phi,\pi)\;
   =\; {\cal H}_0^{(t)}(\phi,\pi)\, +\, \int_{S^1}\,d\sigma\, \left\{
     \frac{1}{t} \langle \pi,B_{\phi} \rangle^{\sim}\,-\,
      \langle\phi_{\ast}\partial_{\sigma}\,,\,
                 \phi_{\ast}\partial_{\sigma}\rangle \right\} \,.
$$
Since ${\cal H}_0^{(t)}$ is invariant along the Hamiltonian flow, one
may simply evaluate it at $t^{\prime}=0$; and the phase factor becomes
$$
 e^{\frac{i}{2\hbar t}\int_{S^1}d\sigma\,
                          \langle\pi,\pi\rangle^{\sim}}\cdot
 e^{\frac{it}{2\hbar}\int_{S^1}d\sigma\,
    \langle\phi_{\ast}\partial_{\sigma},
                            \phi_{\ast}\partial_{\sigma}\rangle}\cdot
 e^{\frac{i}{\hbar t}\int_0^t dt^{\prime}\int_{S^1}d\sigma\,
    \langle\pi(\sigma,t^{\prime}),
           B_{\phi}(\sigma,t^{\prime})\rangle^{\sim}}\cdot
 e^{-\frac{i}{\hbar}\int_0^t dt^{\prime}\int_{S^1}d\sigma\,
              \langle \phi_{t^{\prime}\ast}\partial_{\sigma},
                   \phi_{t^{\prime}\ast}\partial_{\sigma}\rangle}\,.
$$
The first factor
$e^{\frac{i}{2\hbar t}\int_{S^1}d\sigma\,
                          \langle\pi,\pi\rangle^{\sim}}$
makes the integral along $T^{\ast}_{\phi}LM$ Gaussian. Since what
matters is the result after taking
$\left.\frac{d}{dt}\right|_{t=o}$, one only needs to expand everything
else in the integrand of $\int_{LT^{\ast}M}(\cdots)$ vertically up to
orders $t$ and $\pi^2$ around the vertical critical set
$\Lambda_c=\left\{\pi=0\right\}$ of ${\cal H}_0^{(t)}$
and then integrate out $\pi$ by applying the stationary phase
approximation formula
\begin{eqnarray*}
 \lefteqn{\left(\frac{1}{2\pi\hbar t}
    \right)^{\frac{1}{2}dim\,L{\supbscriptsizeBbb R}^n}
      \int_{L{\supbscriptsizeBbb R}^n}
   \left[{\cal D}\pi\right]\,e^{\frac{i}{2\hbar t}
      \int_{S^1\times S^1}d\sigma d\sigma^{\prime}\,
         g^{ab}(\sigma)\delta(\sigma-\sigma^{\prime})
        \pi_a(\sigma)\pi_b(\sigma)} {\cal F}(\pi)}\\
    & &\hspace{-1cm} \sim \frac{e^{\frac{i\pi}{4}\,sign\,
          \left(g^{ab}(\sigma)\delta(\sigma-\sigma^{\prime})\right)}}
             {\sqrt{\left|\mbox{\rm det}\left(g^{ab}(\sigma)
                \delta(\sigma-\sigma^{\prime})\right)\right|}}
   \left[\sum_0^{\infty}\frac{(\hbar t)^k}{k!}
            \left(\frac{i}{2}\sum_{a,b}
     \int_{S^1\times S^1}d\sigma d\sigma^{\prime}\,g_{ab}(\sigma)
             \delta(\sigma-\sigma^{\prime})\frac{\delta^2}
       {\delta\pi_a(\sigma)\delta\pi_b(\sigma^{\prime})}
                   \right)^k{\cal F}(\pi)\right]_{\pi=0}\,.
\end{eqnarray*}
The outcome will be an integral over $LM$ and ${\cal O}_{\cal H}$
can thus be obtained.

\bigskip

\begin{flushleft}
{\bf The details.}
\end{flushleft}
In the following computation, $\pr:LT^{\ast}M\rightarrow LM$ is the
cotangent bundle projection map and $T^{\ast}_{\phi}LM$ the fibre at
$\phi$. And we shall denote $\frac{\partial}{\partial x^i}$ by
$\partial_i$.

\bigskip

\noindent
{\bf (a) The phase factor.} First one has
\begin{eqnarray*}
 \lefteqn{ \left.\frac{d}{dt^{\prime}}\right|_{t^{\prime}=0}
   \int_{S^1}d\sigma\, \langle\pi(\sigma,t^{\prime}),
                   B_{\phi}(\sigma,t^{\prime})\rangle^{\sim}\;
    =\; \int_{S^1}d\sigma\, (Y_{\cal H}g^{ij})\pi_i B_{\phi\,j} }\\
  & & \hspace{3em}+\: \int_{S^1}d\sigma
             \langle Z_{\cal H}, B_{\phi} \rangle^{\sim} \:
       +\: \int_{S^1}d\sigma\, \langle \pi,
   i_{Y_{\cal H}}di_{\phi_{\ast}\partial_{\sigma}}B\rangle^{\sim}\,.
\end{eqnarray*}

The first term vanishes since the only possible non-zero
$Y_{\cal H}g^{ij}$ at $t^{\prime}=0$ under Fermi coordinates is
$Y_{\cal H}g^{11}$, in which case
$B_{\phi\,1}=B(e_1,\phi_{\ast}\partial_{\sigma})=0$.
The third term also vanishes since
$\langle \pi,i_{Y_{\cal H}}di_{\phi_{\ast}\partial_{\sigma}}
   B\rangle^{\sim}\, =\,\frac{1}{t}i_{\pi^{\sim}}i_{\pi^{\sim}}
               di_{\phi_{\ast}\partial_{\sigma}}B\,=\,0$.
Hence only the second term remains and
\begin{eqnarray*}
 \lefteqn{ \frac{i}{\hbar t}\int_0^t dt^{\prime}\,
   \int_{S^1} d\sigma \langle \pi(\sigma,t^{\prime}),
                 B_{\phi}(\sigma,t^{\prime}) \rangle^{\sim}\;
      =\; \frac{i}{\hbar}
               \int_{S^1} d\sigma \langle \pi,B_{\phi} \rangle^{\sim}}\\
  & & \hspace{3em} +\: \frac{it}{2\hbar} \int_{S^1} d\sigma\, B\left(
       -\,(i_{\pi^{\sim}} i_{\phi_{\ast}\partial_{\sigma}} dB)^{\sim}\:
        +\:t\nabla_{\phi_{\ast}\partial_{\sigma}}
             \phi_{\ast}\partial_{\sigma}
              \,,\,\phi_{\ast}\partial_{\sigma} \right) \:+\: O(t^2)\\
  & & =\; \frac{i}{\hbar}
          \int_{S^1} d\sigma\,\langle \pi, B_{\phi} \rangle^{\sim}\:
           +\: O(t\pi, t^2)\,.
\end{eqnarray*}

Next, by first variation,
\begin{eqnarray*}
\lefteqn{-\frac{i}{\hbar}\int_0^t dt^{\prime}\int_{S^1}d\sigma\,
              \langle \phi_{t^{\prime}\ast}\partial_{\sigma},
                     \phi_{t^{\prime}\ast}\partial_{\sigma}\rangle}\\
 & & =\;-\frac{it}{\hbar}\int_{S^1}d\sigma\,
        \langle\phi_{\ast}\partial_{\sigma},
                      \phi_{\ast}\partial_{\sigma}\rangle\,
      +\,\frac{it^2}{\hbar}\int_{S^1}d\sigma\,
            \langle\nabla_{\phi_{\ast}\partial_{\sigma}}
                        \phi_{\ast}\partial_{\sigma},
                            Y_{\cal H}(\sigma,t)\rangle\,+\,O(t^2) \\
 & & =\;-\frac{it}{\hbar}\int_{S^1}d\sigma\,
        \langle\phi_{\ast}\partial_{\sigma},
                      \phi_{\ast}\partial_{\sigma}\rangle\,
         +\, O(t\pi, t^2)\,.
\end{eqnarray*}

Altogether and explicitly in local trivialization,
\begin{eqnarray*}
\lefteqn{ e^{\frac{it}{2\hbar}\int_{S^1}d\sigma\,
                      \langle\phi_{\ast}\partial_{\sigma},
                            \phi_{\ast}\partial_{\sigma}\rangle}\cdot
   e^{\frac{i}{\hbar t}\int_0^t dt^{\prime}\int_{S^1}d\sigma\,
        \langle\pi(\sigma,t^{\prime}),
          B_{\phi}(\sigma,t^{\prime})\rangle^{\sim}}\cdot
   e^{-\frac{i}{\hbar}\int_0^t dt^{\prime}\int_{S^1}d\sigma\,
              \langle \phi_{t^{\prime}\ast}\partial_{\sigma},
                    \phi_{t^{\prime}\ast}\partial_{\sigma}\rangle}}\\
 & & =\; \left[1\,+\,\frac{i}{\hbar}\int_{S^1}d\sigma\,
                            B_{\phi}^i(\sigma)\pi_i(\sigma)\,
       -\,\frac{1}{2\hbar^2}\int_{S^1\times S^1}d\sigma^1 d\sigma^2\,
         B_{\phi}^i(\sigma^1)B_{\phi}^j(\sigma^2)
         \pi_i(\sigma^1)\pi_j(\sigma^2)\right.\\
 & & \hspace{3em} \left.\,
                   -\,\frac{it}{2\hbar}\int_{S^1}d\sigma\,
         \langle\phi_{\ast}\partial_{\sigma},
         \phi_{\ast}\partial_{\sigma}\rangle\,+\,O(t\pi,t^2)\right]\,.
\end{eqnarray*}

\bigskip

\noindent{\bf (b) The} $\psi\circ\rho_t$ {\bf part.} Regard
$d\psi|_{\phi}$ as a complex-valued 1-form along $\phi$ in $M$. Let
$\mbox{\rm grad}\,\psi$ be the vector field on $LM$ with
$\left.\mbox{\rm grad}\,\psi\right|_{\phi}$ the metric equivalent of
$d\psi|_{\phi}$. Then
\begin{eqnarray*}
\lefteqn{(\psi\circ\rho_t)|_{T^{\ast}_{\phi}LM}(\pi)\;=\;
  \psi\circ\rho_t(\phi,\pi)\;=\;\psi\left(\pr\circ\rho_t(\phi,\pi)\right)
    \;=\;e^{tY_{\cal H}}\psi(\phi)}\\
 & & =\;\psi(\phi)\,+\,tY_{\cal H}\psi(\phi)\,+\,\frac{t^2}{2}
       Y_{\cal H}Y_{\cal H}\psi(\phi)\,+\,O(\pi^3)     \\
 & & =\;\psi(\phi)\,+\,t\int_{S^1}d\sigma\,
     \left.\langle Y_{\cal H},\mbox{\rm grad}\,\psi
                                      \rangle\right|_{\phi(\sigma)}\,
     +\,\frac{t^2}{2}\int_{S^1}d\sigma\,
       \left.\langle \nabla_{Y_{\cal H}}Y_{\cal H},
                \mbox{\rm grad}\,\psi\rangle\right|_{\phi(\sigma)}\\
 & & \hspace{3cm}+\,\frac{t^2}{2}\int_{S^1}d\sigma\,
        \left.\langle Y_{\cal H},
          \nabla_{Y_{\cal H}}\mbox{\rm grad}\,\psi
                      \rangle\right|_{\phi(\sigma)}\,+\,O(\pi^3)\,.
\end{eqnarray*}
The term
\begin{eqnarray*}
\lefteqn{ \frac{t^2}{2}\int_{S^1}d\sigma\,
       \left.\langle \nabla_{Y_{\cal H}}Y_{\cal H},
             \mbox{\rm grad}\,\psi\rangle\right|_{\phi(\sigma)} }\\
 & & =\;\frac{t^2}{2}\int_{S^1}d\sigma\,\left.
     \langle \left(Y_{\cal H} Y_{\cal H}^r\right)\partial_r\,
      +\,Y_{\cal H}^r Y_{\cal H}^s \nabla_{\partial_r}\partial_s\,,\,
                  \mbox{\rm grad}\,\psi\rangle\right|_{\phi(\sigma)}\,.
\end{eqnarray*}
Observe that
$$
 Y_{\cal H} Y_{\cal H}^r\;
  =\;\left.\frac{d}{dt^{\prime}}\right|_{t^{\prime}=0}
     \pi^r(\sigma,t^{\prime})
$$
is the vertical component $Z_{\cal H}$ of $X_{\cal H}$ up to metrical
dual; thus
$\frac{t^2}{2}\int_{S^1}d\sigma\,\left.
     \langle \left(Y_{\cal H} Y_{\cal H}^r\right)\partial_r\,
              \mbox{\rm grad}\,\psi\rangle\right|_{\phi(\sigma)}$
is of order $O(t^2\pi, t^3)$. Consequently,
\begin{eqnarray*}
\lefteqn{ (\psi\circ\rho_t)|_{T^{\ast}_{\phi}LM}(\pi)}\\
 & & =\;\psi(\phi)\,+\,\int_{S^1}d\sigma\,
     \left.\pi(\mbox{\rm grad}\,\psi)\right|_{\phi(\sigma)}\,
       +\,\frac{1}{2}\int_{S^1}d\sigma\,
          \left.\langle \nabla_{\partial_1}\partial_i \,,\,
          \mbox{\rm grad}\,\psi\rangle
                  \right|_{\phi(\sigma)}\pi^1\pi^i\\
 & & \hspace{3em}
       +\,\frac{1}{2}\int_{S^1}d\sigma\, \langle \pi^{\sim}\,,
         \nabla_{\pi^{\sim}}\mbox{\rm grad}\,\psi\rangle_{\phi(\sigma)}
                              \,+\,O(t^2\pi, t^3)\,.
\end{eqnarray*}

\bigskip

\noindent{\bf (c) The volume factor}
$\sqrt{(\rho_t^{\ast}\mbox{\it vol}_{LM},\mbox{\it vol}_{LM})}${\bf .}
In terms of the coordinates $x$, one may write locally and formally that
$$
\begin{array}{lll}
 \mbox{\it vol}_{LM} &= & \sqrt{\mbox{\rm det}{\cal O}_g}
   \curlywedge_{(i,\sigma)\in\{1,\cdots,n\}\times S^1}dx^i(\sigma)\\
 \mbox{\it vol}_{\mbox{\boldmath $\omega$}} &=
  & \left(
  \frac{1}{2\pi\hbar}\right)^{dim\,L{\supbscriptsizeBbb R}^n}\cdot
    \curlywedge_{(i,\sigma)\in\{1,\cdots,n\}\times S^1}
       \left(dp_i(\sigma)\wedge dx^i(\sigma)\right)\,,
\end{array}
$$
where ${\cal O}_g$ is the linear operator on $T_{\cal U}LM$
defined by
$$
\xi^{i}(\sigma)\left.\partial_i\right|_{x(\sigma)}
\longmapsto\delta^{ir}g_{rs}(x(\sigma))\xi^s(\sigma)
           \left.\partial_i\right|_{x(\sigma)}
$$
and the curly wedge $\curlywedge$ represents a formal continuous wedge
product.

Denote $\pr\circ\rho_t(\phi,\pi)$ in coordinates by $x^i(\sigma,t)$.
Then
$$
 x^i(\sigma,t)\;=\;x^i(\sigma)\,+\,
     t\frac{\delta{\cal H}}{\delta p_i(\sigma)}\,+\,O(t^2)\,.
$$
Hence
\begin{eqnarray*}
 \lefteqn{dx^i(\sigma,t)\;=\;dx^i(\sigma)\,+\,
    t\int_{S^1}d\sigma^1\, \left.
      \frac{\delta^2{\cal H}}{\delta p_j(\sigma^1)
       \delta p_i(\sigma)}\right|_{(\phi,\pi)} dp_j(\sigma^1)  }\\
  & & \hspace{9em} +\,  t\int_{S^1}d\sigma^1\,\left.
   \frac{\delta^2{\cal H}}{\delta x^j(\sigma^1)\delta p_i(\sigma)}
          \right|_{(\phi,\pi)} dx^j(\sigma^1)\,+\,O(t^2)
\end{eqnarray*}
and
\begin{eqnarray*}
\lefteqn{\left(\rho_t^{\ast}\mbox{\it vol}_{LM}\right)\curlywedge
\mbox{\it vol}_{LM}}\\
 & & =\;\sqrt{\mbox{\rm det}\,{\cal O}_g(\pr\circ\rho_t(\phi,\pi))}
 \sqrt{\mbox{\rm det}\,{\cal O}_g(\phi)}\,
  \left(\curlywedge_{(i,\sigma)\in\{1,\cdots,n\}\times S^1}
     dx^i(\sigma,t)    \right)\curlywedge\left(
      \curlywedge_{(i,\sigma)\in\{1,\cdots,n\}\times
                                           S^1}dx^i(\sigma)\right)\\
 & & =\;\sqrt{\mbox{\rm det}\,{\cal O}_g(\pr\circ\rho_t(\phi,\pi))}
          \sqrt{\mbox{\rm det}\,{\cal O}_g(\phi)}\,
    \;\mbox{\rm det}\left(t\left.
     \frac{\delta^2{\cal H}}{\delta p_j(\sigma^1)
       \delta p_i(\sigma)}\right|_{(\phi,\pi)}\right)
     \curlywedge_{j,\sigma^1} (dp_j(\sigma^1)\wedge dx^j(\sigma^1))\\
 & & =\;\sqrt{\mbox{\rm det}\,{\cal O}_g(\pr\circ\rho_t(\phi,\pi))}
          \sqrt{\mbox{\rm det}\,{\cal O}_g(\phi)}\,
   \;\mbox{\rm det}
       \left(t\,g^{ij}(x(\sigma))\delta(\sigma-\sigma^1)\right)\,
      \curlywedge_{j,\sigma^1} (dp_j(\sigma^1)\wedge dx^j(\sigma^1))\\
 & & =\;(2\pi\hbar)^{dim\,L{\supbscriptsizeBbb R}^n}
      \sqrt{\mbox{\rm det}\,{\cal O}_g(\pr\circ\rho_t(\phi,\pi))}
      \sqrt{\mbox{\rm det}\,{\cal O}_g(\phi)}\,
     \;\mbox{\rm det}\left(t\,{\cal O}_g^{-1}\right)\,
                            \mbox{\it vol}_{\mbox{\boldmath $\omega$}}\\
 & & =\;(2\pi\hbar t)^{dim\,L{\supbscriptsizeBbb R}^n}
       \sqrt{  \mbox{\rm det}\,{\cal O}_g(\pr\circ\rho_t(\phi,\pi))\,
             \cdot\,\mbox{\rm det}\,{\cal O}_g^{-1}(\phi)   }\,
                        \mbox{\it vol}_{\mbox{\boldmath $\omega$}}\,.
\end{eqnarray*}

To get the expansion at $(\phi,\pi)$, we need the following digression.
Recall that, with respect to the normal coordinate system $y$ at a point
$q$ in $M$, one has
$$
 g_{ab}(y)\;=\;g_{ab}(q)\,-\,\frac{1}{3}R_{acbd}y^c y^d\,+\,o(|y|^2)\,,
$$
where $R$ is the curvature tensor evaluated at $q$.
Observe that for $q$ at loop $\phi$, the Fermi coordinates $x$
and the normal coordinates $y$ around $q$ satisfies
$$
(y^1,y^2,\cdots,y^n)\;=\;(x^1-x^1(q),x^2,\cdots,x^n)\,+\,o(|y|)
$$
since the induced map of the coordinate transformation, say from $x$
to $y$, on the tangent space is the identity map at $q$. Consequently,
in coordinates $x$,
\begin{eqnarray*}
\lefteqn{ g_{ij}(x(\sigma,t))\;=\;g_{ij}(x(\sigma,0)) }\\
 & & \hspace{3em} -\,\frac{t^2}{3}
     R\left(\left.\partial_i\right|_{x(\sigma,0)}\, ,\,
          Y_{\cal H}(\sigma,0)\,,\,
             \left.\partial_j\right|_{x(\sigma,0)}\,,\,
                 Y_{\cal H}(\sigma,0)\right)\,+\,O(t^3)  \\
 & & =\;g_{ij}(x(\sigma,0))\,-\,\frac{1}{3}
        R\left(\left.\partial_i\right|_{x(\sigma,0)}\, ,\,
        \pi^{\sim}\,,\,\left.\partial_j\right|_{x(\sigma,0)}\,,\,
                    \pi^{\sim}\right)\,+\,O(t^3)\,.
\end{eqnarray*}

With the above curvature term denoted by
$R(\partial_i,\pi^{\sim},\partial_j,\pi^{\sim})$, one has in terms of
local distributions
\begin{eqnarray*}
\lefteqn{\mbox{\rm det}\,{\cal O}_g(\pr\circ\rho_t(\phi,\pi))\,
                    \cdot\,\mbox{\rm det}\,{\cal O}_g^{-1}(\phi)}\\
 & & = \;\mbox{\rm det}\left(g_{ij}(x(\sigma,t))
       \delta(\sigma^1-\sigma^2)\right)\,\cdot\, \mbox{\rm det}
        \left(g^{kl}(x(\sigma))\delta(\sigma^3\,-\,\sigma^4)\right)\\
 & & =\;\mbox{\rm det}\left(\left(g_{ij}(x(\sigma^1,0))\,-\,\frac{1}{3}
             R\left(\partial_i,\pi^{\sim},\partial_j,
                   \pi^{\sim}\right)\,+\,O(t^3)\right)
                               \delta(\sigma^1-\sigma^2)\right)\\
 & & \hspace{3cm}\cdot\,\mbox{\rm det}\left(
           g^{kl}(x(\sigma))\delta(\sigma^3-\sigma^4) \right)\\
 & & =\;\mbox{\rm det}\left(\mbox{\rm Id}\,
          -\,\frac{1}{3}{\cal O}_{\rm Riemann}(
                  \pi^{\sim},\pi^{\sim})\, +\,O(t^3)\right)\\
 & & =\;1\, -\,\frac{1}{3}\,\mbox{\rm tr}\,{\cal O}_{\rm Riemann}(
                   \pi^{\sim},\pi^{\sim})\,+\,O(t^3)\,,
\end{eqnarray*}
where $\mbox{\rm Id}$ is the identity map at $T_{\phi}LM$ and
$$
\begin{array}{ccccl}
 {\cal O}_{\rm Riemann}(\pi^{\sim},\pi^{\sim}) &:
                      &T_{\phi}LM &\longrightarrow &T_{\phi}LM\\
  & &\xi &\longmapsto
  & R\left(\,\cdot\,,\pi^{\sim},\xi,\pi^{\sim}\right)^{\sim}
\end{array}
$$
with "$^{\sim}$" representing the metrically equivalent vector field
to a 1-form along $\phi$. Formally,
$$
 \mbox{\rm tr}\,{\cal O}_{\rm Riemann}(\pi^{\sim},\pi^{\sim})\;
    =\;\int_{T_{\phi}LM}\left[{\cal D}\xi\right]\,
                \langle \xi\,,\,{\cal O}_{\rm Riemann}\xi\rangle\;
    =\;\int_{S^1}\,d\sigma\, {\cal O}_{\rm Ric}(\phi(\sigma))
                                  (\pi^{\sim},\pi^{\sim})\,,
$$
where ${\cal O}_{\rm Ric}(\phi(\sigma))$ is the local density functional
of $\mbox{\rm tr}\,{\cal O}_{\rm Riemann}$ along $S^1$. Thus
$$
 \sqrt{\mbox{\rm det}\,{\cal O}_g(\pr\circ\rho_t(\phi,\pi))\,
     \cdot\,\mbox{\rm det}\,{\cal O}_g^{-1}(\phi) }\;
    =\;1\,-\,\frac{1}{6}\,\int_{S^1}d\sigma\,
       ({\cal O}_{\rm Ric}(\phi(\sigma))^{rs}\,
            \pi_r(\sigma)\pi_s(\sigma)\, +\,O(t^2)
$$
and
$$
\sqrt{\left(
      \rho_t^{\ast}\mbox{\it vol}_{LM},\mbox{\it vol}_{LM}\right)}\;
  =\;(2\pi\hbar t)^{\frac{1}{2}dim\,L{\supbscriptsizeBbb R}^n}\,\cdot\,
     \left(1\,-\,\frac{1}{12}\,\int_{S^1}d\sigma\,
      ({\cal O}_{\rm Ric}(\phi(\sigma))^{rs}\,
         \pi_r(\sigma)\pi_s(\sigma)\, +\,O(t^2)\right)\,.
$$

\bigskip

\noindent{\bf (d) All together.} Putting all these expansions together
and extracting terms of type $1$, $t$, and $\pi^2$ from the product,
one obtains
\begin{eqnarray*}
\lefteqn{ \int_{LT^{\ast}M}\,e^{\frac{i}{\hbar}\int_0^t\,dt^{\prime}\,
     L_{\cal H}}\,\overline{\psi\circ\rho_t}\,\psi^{\prime}
       \sqrt{(\rho_t^{\ast}\mbox{\it vol}_{LM},
       \mbox{\it vol}_{LM})}\,
                  \mbox{\it vol}_{\mbox{\boldmath $\omega$}}}\\
 & & =\;\int_{LM}\, \left(\frac{1}{2\pi\hbar}
               \right)^{dim\,L{\supbscriptsizeBbb R}^n}\,
    [{\cal D}x_{\sigma}]\, \int_{T_{x_{\sigma}}LM}\, \left(\frac{1}{t}
                 \right)^{dim\,L{\supbscriptsizeBbb R}^n}\,
     [{\cal D}\pi_{\sigma}]\,
        e^{\frac{i}{2\hbar t}\int_{S^1}d\sigma\,
           \langle\pi(\sigma),\pi(\sigma)\rangle^{\sim}}\cdot\\
 & & \hspace{0.7cm} \left[\,1\,+\,\frac{i}{\hbar}\int_{S^1}d\sigma\,
                           B_{\phi}^i(\sigma)\pi_i(\sigma)\,
       -\,\frac{1}{2\hbar^2}\int_{S^1\times S^1}d\sigma^1 d\sigma^2\,
         B_{\phi}^i(\sigma^1)B_{\phi}^j(\sigma^2)
         \pi_i(\sigma^1)\pi_j(\sigma^2)\right.\\
 & & \hspace{3em}\left. -\,\frac{it}{2\hbar}\int_{S^1}d\sigma\,
                \langle\phi_{\ast}\partial_{\sigma},
                        \phi_{\ast}\partial_{\sigma}\rangle\,
                                 +\,O(t^2,t\pi)\,\right]\cdot\\
 & & \hspace{0.7cm} \left[\,
      \bar{\psi}(\phi)\,+\,\int_{S^1}d\sigma\,
       \left.\pi(\sigma)(\mbox{\rm grad}\,\bar{\psi})
                                          \right|_{\phi(\sigma)}\,
        +\,\frac{1}{2}\int_{S^1}d\sigma\,
        \left.\langle \nabla_{\partial_1}
              \partial_i  \,,\,\mbox{\rm grad}\,\bar{\psi}\rangle
                  \right|_{\phi(\sigma)}
           \pi^1(\sigma)\pi^i(\sigma)\right.\\
 & & \hspace{3em} \left. +\,\frac{1}{2}\int_{S^1}d\sigma\,
      \langle \pi^{\sim}\,,\nabla_{\pi^{\sim}}\mbox{\rm grad}\,
       \bar{\psi}\rangle_{\phi(\sigma)}\,+\,O(t^3)\, \right]\cdot\\
 & & \hspace{0.7cm}
       \left[\, (2\pi\hbar t)^{\frac{1}{2}dim\,
                        L{\supbscriptsizeBbb R}^n}\,\cdot\,
         \left(1\,-\,\frac{1}{12}\,\int_{S^1}d\sigma\,
           {\cal O}_{\rm Ric}(\phi(\sigma))^{rs}\,
                            \pi_r(\sigma)\pi_s(\sigma)\,
                             +\,O(t^2)\right) \right]\,\psi^{\prime}\\
 & & =\;\int_{LM}[{\cal D}x_{\sigma}]\,
       \left(\frac{1}{2\pi\hbar t}\right)^{\frac{1}{2}dim\,
                                        L{\supbscriptsizeBbb R}^n}
        \int_{T_{x_{\sigma}}LM}[{\cal D}\pi_{\sigma}]\,
                e^{\frac{i}{2\hbar t}\int_{S^1}d\sigma\,
           \langle\pi(\sigma),\pi(\sigma)\rangle^{\sim}}\cdot\\
 & & \hspace{0.7cm}
      \left[\,\bar{\psi}(\phi) -\frac{it}{2\hbar}\,\bar{\psi}(\phi)\,
        \int_{S^1}d\sigma\,\langle\phi_{\ast}\partial_{\sigma},
                            \phi_{\ast}\partial_{\sigma}\rangle\,
      +\,\frac{1}{2}\int_{S^1}d\sigma\,
       \langle \pi^{\sim}\,,\nabla_{\pi^{\sim}}\mbox{\rm grad}\,
                        \bar{\psi}\rangle_{\phi(\sigma)}\,\right.\\
 & & \hspace{1cm} -\,\frac{1}{12}\,\bar{\psi}(\phi)\,\int_{S^1}d\sigma\,
       \left.{\cal O}_{\rm Ric}\right|_{\phi}(\pi^{\sim},\pi^{\sim})  \\
 & & \hspace{2cm} +\,\frac{1}{2}\int_{S^1}d\sigma\,
          \left.\langle \nabla_{\partial_1}
              \partial_i  \,,\,\mbox{\rm grad}\,\bar{\psi}\rangle
          \right|_{\phi(\sigma)}\pi^1(\sigma)\pi^i(\sigma) \\
 & & \hspace{3cm}  +\, \frac{i}{\hbar}\int_{S^1}d\sigma\,
                           B_{\phi}(\pi^{\sim})\cdot
             \int_{S^1}d\sigma\,
         \left.\pi(\sigma)(\mbox{\rm grad}\,
                               \bar{\psi})\right|_{\phi(\sigma)}\\
 & & \hspace{2cm}   -\,\frac{1}{2\hbar^2}\,\bar{\psi}(\phi)\,
        \left[ \int_{S^1}d\sigma\,B_{\phi}(\pi^{\sim})\right]^2\,
  \left. +\,\mbox{\raisebox{0ex}[1.8ex][1.5ex]{(the irrelevent rest)}}
                                         \right]\, \psi^{\prime}\,.
\end{eqnarray*}

Let $\mbox{\it Sec}_{\cal P}({\Bbb L})$ be the space of
$\cal P$-polarized sections in $\Bbb L$. Define the following operators
from
$(T_{\ast}LM\otimes T_{\ast}LM)\otimes\mbox{\it Sec}_{\cal P}({\Bbb L})$
to $\mbox{\it Sec}_{\cal P}({\Bbb L})$:

\vspace{-2.5em}
\begin{eqnarray*}
 \lefteqn{   }\\
  & & {\cal O}_{d^2}(\eta,\xi)\,\psi \;\;\;
       =\; \int_{S^1\times S^1}d\sigma_1 d\sigma_2\,\langle
        \eta|_{\phi(\sigma_1)}\,,
         \nabla_{\xi|_{\phi(\sigma_2)}}
          \mbox{\rm grad}\, \psi\rangle|_{\phi(\sigma_2)} \,,  \\
  & & {\cal O}_{B,d}(\eta,\xi)\,\psi \;\,
        =\; \int_{S^1}d\sigma\, B_{\phi}(\eta)\:
                  \int_{S^1}d\sigma\, d\psi (\xi) \,,\\
  & & {\cal O}_{B^2}(\eta,\xi)\,\psi \;\;
       =\; \psi\,\int_{S^1}d\sigma\,B_{\phi}(\eta)\:
         \int_{S^1}d\sigma\,B_{\phi}(\xi) \,,
                               \hspace{3em} \mbox{\rm and} \\
  & & {\cal O}_{h,d}(\eta,\xi)\,\psi \;\:
       =\; \int_{S^1\times S^1}d\sigma_1 d\sigma_2\,
         \left.\langle \nabla_{\partial_1}
          \partial_i  \,,\,\mbox{\rm grad}\,\psi\rangle
           \right|_{\phi(\sigma_1)}
            \eta^1|_{\phi(\sigma_1)}\xi^j|_{\phi(\sigma_2)}\,;
\end{eqnarray*}
and their trace $\mbox{\rm Tr}$
$$
 \left.\mbox{\rm Tr}\,{\cal O}\right|_{\phi}\;
   =\; \int_{S^1\times S^1} d\sigma_1 d\sigma_2\,
       {\cal O}_{ij}|_{(\phi(\sigma_1),\phi(\sigma_2))}
          g^{ij}(\phi(\sigma_1))\delta(\sigma_1-\sigma_2)\,,
$$
where ${\cal O}_{ij}|_{(\phi(\sigma_1),\phi(\sigma_2))}$ are components
of the density of ${\cal O}$ at $\phi$ along $S^1\times S^1$ with
respect to the Fermi coordinates. Then, after applying the Gaussian
integration, one has
\begin{eqnarray*}
\lefteqn{ \int_{LT^{\ast}M}\,e^{\frac{i}{\hbar}\int_0^t\,dt^{\prime}\,
     L_{\cal H}}\,\overline{\psi\circ\rho_t}\,\psi^{\prime}
       \sqrt{(\rho_t^{\ast}\mbox{\it vol}_{LM},
       \mbox{\it vol}_{LM})}\,
                  \mbox{\it vol}_{\mbox{\boldmath $\omega$}}}\\
 & & \sim\; \int_{LM}[{\cal D}x_{\sigma}]\,
      \frac{e^{\frac{i\pi}{4}\,sign\,
        \left(g^{ab}(\sigma)\delta(\sigma-\sigma^{\prime})\right)}}
           {\sqrt{\left|\mbox{\rm det}\left(g^{ab}(\sigma)
              \delta(\sigma-\sigma^{\prime})\right)\right|}}\,
      \left\{\bar{\psi}\,-\,
                     \left(\frac{it}{\hbar}\right)\,
     \left[ \frac{1}{2}\,\int_{S^1}d\sigma\,
               \langle\phi_{\ast}\partial_{\sigma},
                   \phi_{\ast}\partial_{\sigma}\rangle\right.\right.\\
 & & \hspace{6em}
       -\,\frac{{\hbar}^2}{2}\,\mbox{\rm Tr}\,{\cal O}_{d^2}\,
       -\,i{\hbar}\, \mbox{\rm Tr}\,{\cal O}_{B,d}\,
       +\,\frac{1}{2}\,\mbox{\rm Tr}\,{\cal O}_{B^2}\\
 & & \hspace{9em}
       -\,\frac{{\hbar}^2}{2}\, \mbox{\rm Tr}\,{\cal O}_{h,d}\,
       +\,\frac{{\hbar}^2}{12}\, \mbox{\rm Tr}\,{\cal O}_{\rm Ric}\,
    \left. \mbox{\raisebox{0ex}[1.8ex][1.5ex]{}} \right]_{\phi}\,
                 \cdot\,\bar{\psi} \left.    +\,O(t^2)\,
       \mbox{\raisebox{0ex}[1.8ex][1.5ex]{}}\right\}\,\psi^{\prime}\,.
\end{eqnarray*}

The factor
{\small
$\left. e^{\frac{i\pi}{4}\,sign\,
  \left(g^{ab}(\sigma)\delta(\sigma-\sigma^{\prime})\right)}\right/$
 \hspace{-2ex}
    \raisebox{-.6ex}{$\sqrt{\left|\mbox{\rm det}\left(g^{ab}(\sigma)
  \delta(\sigma-\sigma^{\prime})\right)\right|}$} }
should be absorbed into the pairing by the further metaplectic
correction, which we won't discuss. Hence
\begin{eqnarray*}
 \lefteqn{ {\cal O}_{\cal H}\;
   =\; -\,\frac{{\hbar}^2}{2}\,\mbox{\rm Tr}\,\left( {\cal O}_{d^2}\,
         +\, \frac{2i}{\hbar}\,{\cal O}_{B,d}\,
           -\, \frac{1}{{\hbar}^2}\, {\cal O}_{B^2}    \right) }\\
 & & \hspace{3em} +\,\frac{1}{2}\,\int_{S^1}d\sigma\,
           \langle\phi_{\ast}\partial_{\sigma},
            \phi_{\ast}\partial_{\sigma}\rangle\,
      -\,\frac{{\hbar}^2}{2}\, \mbox{\rm Tr}\,{\cal O}_{h,d}\,
      +\,\frac{{\hbar}^2}{12}\, \mbox{\rm Tr}\,{\cal O}_{\rm Ric}\,,
\end{eqnarray*}
and the Schr\"{o}dinger equation follows. Notice that the part
$\mbox{\rm Tr}\,({\cal O}_{d^2}+\frac{2i}{\hbar}{\cal O}_{B,d}
                                 -\frac{1}{{\hbar}^2}{\cal O}_{B^2})$
resembles $(\nabla-\frac{i}{\hbar}B)(\nabla-\frac{i}{\hbar}B)$ and
should be regarded as the Laplacian operator on sections of
${\Bbb L}$. The only term in ${\cal O}_{\cal H}$ that does not have
a parallel to the case for a point-like charged particle is
$-\frac{{\hbar}^2}{2}\mbox{\rm Tr}\,{\cal O}_{h,d}$, which one may call
the {\it holonomy term}. The restriction of this term to a geodesic
loop $\phi$ in $M$ with trivial holonomy is zero.

\bigskip

\noindent
{\it Remark A.5.} Presumably, the Feynman propagator for string field
theory can also be worked out in the BKS method [Wo]. What is presented
in this Appendix is the easy part. The hard part is to introduce further
some regularizations to make physical sense of these traces. That we
haven't yet succeeded.

\end{document}